\documentclass{article}
\usepackage{graphicx} % Required for inserting images
\usepackage{url}
\usepackage{jcappub}
\usepackage{float}
\usepackage{hyperref}
\usepackage{multirow}
\usepackage{tabularx}
\usepackage{array}
\usepackage{adjustbox}
\usepackage[most]{tcolorbox}
\usepackage{bm}
\usepackage{subcaption} % in preamble
\usepackage{enumitem}
\usepackage{amssymb}
\usepackage{pifont}

\title{The Primordial power spectrum from the largest to smallest CMB scales}
\author[a,b]{Debabrata Chandra}
\author[a,b,c]{Dhiraj Kumar Hazra}
\author[d,e]{Arman Shafieloo}
\author[f]{Tarun Souradeep}
\affiliation[a]{The Institute of Mathematical Sciences, CIT Campus, Chennai 600 113, India}
\affiliation[b]{Homi Bhabha National Institute, Training School Complex, Anushakti Nagar, Mumbai 400094, India}
\affiliation[c]{INAF/OAS Bologna, Osservatorio di Astrofisica e Scienza dello Spazio, Area della ricerca
CNR-INAF, via Gobetti 101, I-40129 Bologna, Italy}
\affiliation[d]{Korea Astronomy and Space Science Institute, Daejeon 34055, South Korea}
\affiliation[e]{University of Science and Technology, Daejeon 34113, South Korea}
\affiliation[f]{Astronomy and Astrophysics Group, Raman Research Institute, Bangalore, India}
\emailAdd{debabratac@imsc.res.in}
\emailAdd{dhiraj@imsc.res.in}
\emailAdd{shafieloo@kasi.re.kr}
\emailAdd{tarun@rri.res.in}

\abstract{
We reconstruct the primordial power spectrum (PPS) across the full range of cosmological scales accessible to Cosmic Microwave Background (CMB) observations. Using the Modified Richardson--Lucy algorithm, we perform free-form reconstructions through deconvolution of Planck PR3 and PR4, Atacama Cosmology Telescope (ACT) DR6, and South Pole Telescope 3G (SPT-3G) D1 data. Across different regularization schemes, we find no evidence for significant deviations from a power-law primordial spectrum. The reconstructed spectra show a strong correlation between Planck and ACT, even at the level of localized features over their overlapping range, $0.07 < k < 0.14\,\mathrm{Mpc}^{-1}$, demonstrating consistency between the two observations. Using a complementary parametric Bayesian reconstruction, we find that the previously discussed preference for a blueward tilt in the ACT data at small scales is preferred only at the $1\sigma$ level.
}

\date{\today}

\begin{document}

\maketitle

\section{Introduction}
Cosmic Microwave Background (CMB) observations are the best probes to constrain the initial condition for the generation of perturbations in the early Universe. The Planck CMB mission~\cite{Planck:2018jri,Planck:2018vyg,Planck:2019nip} has provided us with a highly precise map of the CMB sky. Recently, ACT and SPT two ground based CMB missions have also released their latest data, ACT-DR6~\cite{AtacamaCosmologyTelescope:2025blo} and SPT-3G D1~\cite{SPT-3G:2025bzu}, respectively. These three datasets, while providing substantial overlap also complement each other in cosmological scales in both temperature and polarization. This has enabled us to probe the primordial universe across the widest possible scale from CMB anisotropy. Planck being a space-based mission probes full CMB sky and precisely measures a widest angular scale about an arcmin scale. On large to intermediate angular scales, {\it Planck} remains the most powerful probe, owing to its nearly full-sky coverage and cosmic-variance-limited temperature measurements, whereas at smaller angular scales, ground-based observations provide superior sensitivity and resolution. The ACT-DR6 mission covers $45 \%$ of the sky. The SPT-3G D1 survey deeply probes the Main field covering only about $4 \%$ of the total CMB sky. In this region the foreground contamination is less which enables SPT to stringently measure EE and TE polarization spectra over $\ell = 1800-4000$ and $\ell = 2200-4000$, respectively.
These missions complement each other in sky-coverage, sensitivity and resolution that allows us to have a precise measurement of the CMB anisotropy across a wide scale.

The cosmic structures observed today originated from the quantum fluctuations generated during inflation. From CMB anisotropy maps, we measure the statistical properties of these primordial fluctuations, which are well-explained by a power-law spectrum parametrized with amplitude $A_s$, and spectral index $n_s$. We have detected $A_s$ and $n_s$ with strong statistical significance from different CMB missions~\cite{Planck:2018jri,AtacamaCosmologyTelescope:2025blo,SPT-3G:2025bzu}) and it strongly confirms a nearly scale invariant (slightly red tilted) primordial power spectrum (PPS). Simple single field slow-roll inflation with a canonical kinetic energy term predict nearly scale invariant power-law PPS. However, certain local and non-local outliers in the CMB data may hint towards the departure from single field slow-roll inflation. Such outliers may be addressed by the presence of sharp features in the inflationary potential~\cite{Adams:2001vc,Starobinsky:1992ts}, by the presence of sudden bend in the track of inflaton field in multi-field inflation~\cite{Gao:2012uq,Achucarro:2010da}, or by the presence of abrupt change in the sound speed of inflaton fields~\cite{Miranda:2012rm,Bean:2008na}, all such physical scenarios introduce oscillatory signals in the PPS. Additionally, there are other models for instance, brane inflation~\cite{Bean:2008na}, axion monodromy inflation~\cite{Flauger:2009ab}, where we find that the oscillatory behaviour of inflationary dynamical parameters like slow-roll parameters, sound speed gives rise to logarithmic oscillations in the PPS. However, apart from oscillatory signals there exist other models which can generate bumps in the PPS~\cite{Barnaby:2009mc,Romano:2008rr}. Thus, any decisive identification or a conclusive ruling out of any such signal in the PPS would allow us to decipher the physics of inflationary paradigm. Along with this, detecting any such features has far-reaching consequences, like resolving tensions that are persisting among different cosmological surveys probing early and late universe~\cite{Hazra:2022rdl,DiValentino:2021izs,Abdalla:2022yfr}. Thus, the primordial power spectrum serves a crucial role in unveiling the physics of early universe. 

Several competing algorithms can be found in the literature that allow us to constrain PPS from CMB observations. It is possible to broadly categorize these methodologies into two distinct classes. In the first class, in a bottom-up approach we consider different templates of the PPS allowing for departures from the minimal power-law PPS. In bottom-up approach, different physically/phenomenologically motivated templates including oscillatory features~\cite{Planck:2018jri,Ichiki:2009xs,Meerburg:2011gd,Meerburg:2013cla,Meerburg:2013dla,Meerburg:2015owa,Fergusson:2014hya,Fergusson:2014tza,Hu:2014hra}, bump features~\cite{Nerval:2026iev,Raffaelli:2026bhk}, cutoff model~\cite{Planck:2015sxf, Sinha:2005mn, Upadhyay:2026unf}, modified spectral tilt~\cite{Planck:2015sxf,Planck:2018jri,Joy:2010ea}, etc. have been studied. Besides template-based approaches, another widely used strategy is to represent deviations from a power-law primordial power spectrum using a set of basis functions defined at discrete knots. The spectrum between these knots is then reconstructed using suitable interpolation schemes~\cite{Bridle:2003sa,Bridges:2008ta,Sealfon:2005em,Verde:2008zza,Peiris_2010}, with some analyses also allowing the knot positions to vary~\cite{V_zquez_2012,Aslanyan:2014mqa,Abazajian:2014tqa,Handley:2019fll,Planck:2015sxf,Planck:2018jri}. The statistical significance of the reconstructed features can subsequently be assessed through Bayesian model comparison or, alternatively, through a principal-component analysis based on the Fisher information matrix~\cite{Leach:2005av}.
The second category can be referred to as the top-down approach. In top-down approach the strategy is to reconstruct the non-parametric PPS directly from the observed data. Some such methods include using cosmic inversion method~\cite{Matsumiya:2001xj,Matsumiya:2002tx,Kogo:2003yb} where PPS is reconstructed by directly solving the cosmological perturbation equations, using wavelet basis functions to probe local features~\cite{Mukherjee:2003yx, Mukherjee:2005dc, Shafieloo:2006hs}, implementing likelihood maximization method controlled with some penalty factor operating either directly on the function~\cite{Tegmark:2002cy} or its derivative of different orders~\cite{Planck:2013jfk,Planck:2018jri,Planck:2015sxf,Tocchini-Valentini:2004kwg,Tocchini-Valentini:2005mzh,Nagata:2008zj,Hunt:2013bha,Hunt:2015iua,Gauthier_2012} to reconstruct free-form PPS. Other reconstruction methods such as the wave-number binning method~\cite{Wang:1998gb,Hannestad:2000pm,Hannestad:2003zs,Elgaroy:2001wu}, and the Singular Value Decomposition method~\cite{Nicholson_2010} are also there in the literature. The method we adopt in this work, the Modified Richardson-Lucy algorithm~\cite{Shafieloo:2003gf,Shafieloo:2006hs,Shafieloo:2007tk,Nicholson_2009,Gibelyou:2010qe,Hamann_2010,Hazra:2013xva,Hazra:2013eva,Hazra:2014jwa,Chandra:2021ydm} also comes under the umbrella of top-down category.

The Richardson-Lucy algorithm~\cite{Richardson:1972hli,Lucy:1974yx} is widely used in astronomy in performing image analysis. For an observed image where the point spread function of the imaging system is known the Richardson-Lucy algorithm can iteratively reconstruct the true object. In articles \cite{10.1093/mnras/265.1.145,10.1093/mnras/267.2.323}, the Richardson-Lucy algorithm was first introduced in cosmology however, article \cite{Shafieloo:2003gf} first implemented it for CMB analysis. After this, several new features have been added to this algorithm to make it more robust, which include introducing a new factor to incorporate uncertainties associated with $C_{\ell}$, and adding various regularization schemes to tame the noise over fitting issue. We will refer to this new updated algorithm as the Modified Richardson-Lucy algorithm (MRL).  

In this article, we intend to investigate the behaviour of the primordial power spectrum from CMB data. In this work, we will address following major directions.
\begin{itemize}

\item  Model independent reconstructions of the primordial power spectrum combining different datasets that covers the widest CMB scales of interest.
%\item  Are primordial power spectrum reconstructed from different datasets consistent?
\item  Primordial features that are supported by multiple datasets.
\item  Consistency between different CMB observations.
%\item  Additionally, we will also explore the internal consistency of diffrent Planck data releases.
\end{itemize}

\section{Methodology and datasets: free-form statistics}
In this section, we discuss the application of the deconvolution principle in probing the behaviour of the primordial power spectrum directly from the CMB angular power spectrum data. Here, we leverage the Modified Richardson-Lucy deconvolution algorithm in reconstructing the primordial power spectrum, and the entire scheme adopted in this work is elaborated subsequently.
\subsection{Strategy}
\begin{figure}[ht]
\centering
\includegraphics[width=\textwidth, keepaspectratio]{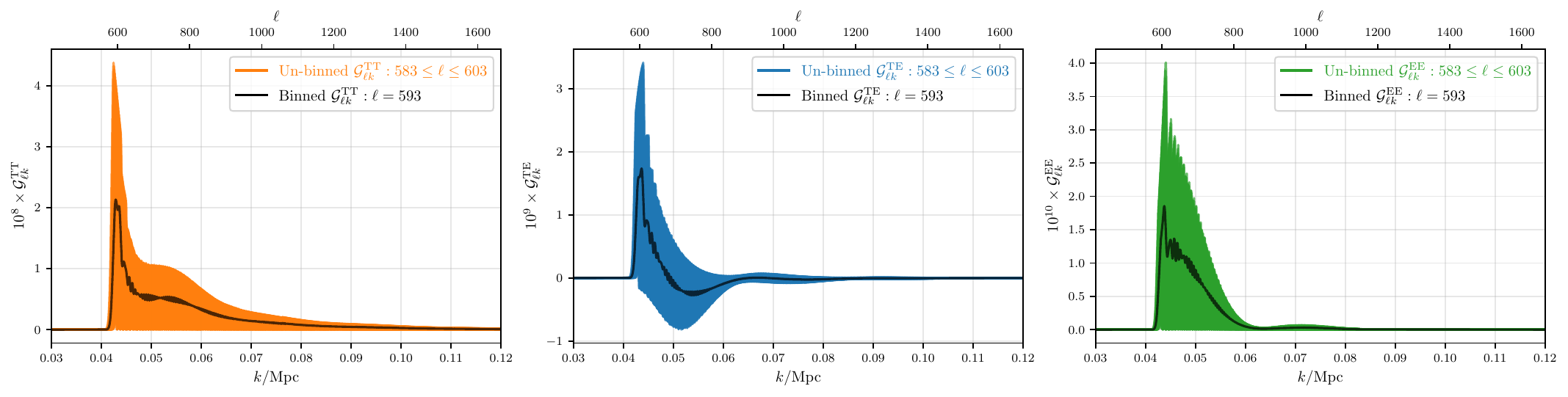}
\caption[]{This figure depicts the radiative-transport kernels $\mathcal{G}^{XY}_{\ell k}$ across a multipole band ($583 \leq \ell \leq 603$) and corresponding bin centre, $\ell = 593$ for TT, TE, and EE spectra.}
\label{fig:glk}
\end{figure}

The CMB physics tells us how the primordial power spectrum gives rise to CMB angular power spectrum through cosmological evolution. The relation that connects the primordial power spectrum ($\mathcal{P}_{\mathcal{R}}(k)$) with CMB angular power spectrum ($\mathcal{C}_\ell^{XY}$) appears as follows:
\begin{equation}
\mathcal{C}_\ell^{XY} = 4\pi \int \frac{dk}{k} \, \mathcal{P}_{\mathcal{R}}(k)\, \Delta_\ell^{X}(k) \, \Delta_\ell^{Y}(k),
\label{eq:ClXY}
\end{equation}
here, $ \Delta_\ell^{X}(k)$ refers to the transfer function encapsulating the physics of CMB photons described by the Boltzmann equation. The superscript indices X and Y in Eq. (\ref{eq:ClXY}) stands for CMB temperature anisotropy ($T$) or E-mode polarization ($E$). 
We recast the integral expression of Eq. (\ref{eq:ClXY}) into a discretized version in wave-number space ($k$): 
\begin{equation}
\mathcal{C}_\ell^{XY} = \sum_k \, \mathcal{G}^{XY}_{\ell k} \mathcal{P}_k.
\label{eq:disClXY}
\end{equation}
we can rewrite Eq. (\ref{eq:disClXY}) in a matrix form as
\begin{equation}
\mathcal{C} = \mathcal{G} \mathcal{P}.
\label{eq:mat_Cl}
\end{equation}
The radiative-transport kernel $\mathcal{G}^{XY}_{\ell k}$ defined in Eq. (\ref{eq:disClXY}), contains contributions from both the transfer functions $ \Delta_\ell^{X}(k)$ and the integration weight $\frac{dk}{k}$ from Eq. (\ref{eq:ClXY}). We show $\mathcal{G}^{XY}_{\ell k}$ of temperature and polarization spectra for a multipole band, $583 \leq \ell \leq 603$, and for the corresponding bin centre, $\ell = 593$ in Fig. \ref{fig:glk}.
The Relation (\ref{eq:disClXY}) signifies that in $\mathcal{C}_\ell^{XY}$, $\mathcal{P}_{\mathcal{R}}(k)$ captures all the information of primordial perturbations, and $\mathcal{G}^{XY}_{\ell k}$ describes the cosmological evolution dictated by the cosmological parameters ($H_0, \, \tau, \, \Omega_bh^2, \, \Omega_ch^2, \, \Omega_k, \, \sum m_{\nu}, \,$ etc.). Consequently, we can separate out the information of primordial physics and the late time physics distinctly, and it turns out that extracting the information of $\mathcal{P}_{\mathcal{R}}(k)$ becomes a deconvolution problem. The $\mathcal{G}^{XY}_{\ell k}$ matrix exhibits a highly oscillatory behaviour and it is numerically singular~\cite{Nicholson_2010}, consequently forbidding us from directly evaluating $\mathcal{G}^{-1}$. Moreover, the Relation (\ref{eq:disClXY}) has severe degeneracy because the dimension of $\ell$ (number of rows in the $\mathcal{G}^{XY}_{\ell k}$ matrix) is much less than the dimension of $k$ (number of columns in the $\mathcal{G}^{XY}_{\ell k}$ matrix) which makes the system underdetermined.
To circumvent this inversion problem we use the Modified Richardson-Lucy deconvolution method, which we elaborate below.

\subsection{Modified Richardson-Lucy deconvolution algorithm}
The Richardson-Lucy (RL) algorithm is a well explored device in the field of image reconstruction~\cite{Richardson:1972hli,Lucy:1974yx,10.1093/mnras/265.1.145,10.1093/mnras/267.2.323}. However, its application in reconstructing the primordial power spectrum from observed CMB angular power spectrum data has been first explored in article~\cite{Shafieloo:2003gf}. 

The Richardson-Lucy (RL) algorithm states that the primordial power spectra obtained at $i^{th}$ and $(i+1)^{th}$ iterations, $\mathcal{P}^{i}(k)$ and $\mathcal{P}^{i+1}(k)$, respectively, obey following recurrence relation: 
\begin{equation}
\mathcal{P}^{i+1}_k = \mathcal{P}^{i}_k + \mathcal{P}^{i}_k \times \sum_{\ell} \, \tilde{\mathcal{G}}_{\ell k} \, \left( \frac{\mathcal{C}_\ell^{\mathrm{Data}} - \mathcal{C}_\ell^{\mathrm{Theory},i}}{\mathcal{C}_\ell^{\mathrm{Theory},i}}  \right).
\label{eq:rl}
\end{equation}

With every iteration the algorithm updates $\mathcal{P}^{i}(k)$ to reduce the residual between the theoretical and observed angular power spectra. In Eq. (\ref{eq:rl}), $\mathcal{C}_\ell^{\mathrm{Theory},i}$ and $\tilde{\mathcal{G}}_{\ell k}$ are defined by
\begin{equation}
\mathcal{C}_\ell^{\mathrm{Theory},i} = \sum_k \, \mathcal{G}_{\ell k} \mathcal{P}^i_k,
\label{eq:ithClTh}
\end{equation}
\begin{equation}
\tilde{\mathcal{G}}_{\ell k} = \frac{\mathcal{G}_{\ell k}}{\sum_{\ell^\prime} \, \mathcal{G}_{\ell^\prime k}}.
\label{eq:norm_glk}
\end{equation}

In Eq. (\ref{eq:rl}), the RL algorithm does not take into account the uncertainty associated with $\mathcal{C}_\ell^{\mathrm{Data}}$, and causing it to generate features that are actually the consequence of over-fitting the noise present in the data. This limitation was addressed in article~\cite{Shafieloo:2003gf} by introducing a "$\bm {\tanh}$" term to weight the updates according to the data uncertainties. To work with binned and un-binned data and to use the full covariance matrix instead of only diagonal elements further modifications are made in the RL method. After all these improvements and modifications Eq. (\ref{eq:rl}) takes the following final form:
\begin{equation}
\mathcal{P}^{i+1}_k = \mathcal{P}^{i}_k + \mathcal{P}^{i}_k \times \sum_{\ell} \, \tilde{\mathcal{G}}_{\ell k} \, \left( \frac{\mathcal{C}_\ell^{\mathrm{Data}} - \mathcal{C}_\ell^{\mathrm{Theory},i}}{\mathcal{C}_\ell^{\mathrm{Theory},i}}  \right)\tanh^2 \left[\mathcal{E}_{\ell}^i \left(\mathcal{C}_\ell^{\mathrm{Data}} - \mathcal{C}_\ell^{\mathrm{Theory},i} \right) \right],
\label{eq:mrl}
\end{equation}
which we call the Modified Richardson-Lucy (MRL) deconvolution algorithm. The expression $\mathcal{E}_{\ell}^i$ in Eq. (\ref{eq:mrl}) is the error matrix associated with the multipole $\ell$, which we evaluate using the relation:
\begin{equation}
\mathcal{E}_{\ell}^i = \sum_{\ell^\prime} \mathrm{Cov}^{-1}_{\ell,\ell^\prime} \left(\mathcal{C}_{\ell^\prime}^{\mathrm{Data}} - \mathcal{C}_{\ell^\prime}^{\mathrm{Theory},i} \right).
\label{eq:err_mat}
\end{equation}
In this work, we do not consider cross-covariances among the TT, TE, and EE datasets when computing $\mathcal{E}_{\ell}$ (\ref{eq:err_mat}). It is important to mention here that when we have binned $\mathcal{C}_{\ell}^{\mathrm{Data, \, binned}}$ data we compute binned $\tilde{\mathcal{G}}_{lk}^{\mathrm{Binned}} $, and for un-binned $\mathcal{C}_{\ell}^{\mathrm{Data, \, un-binned}}$ data the un-binned $\tilde{\mathcal{G}}_{lk}^{\mathrm{Un-binned}}$ is computed. In our analysis, we use the publicly available Boltzmann solver code, namely, CAMB \cite{Lewis:1999bs} to compute  $\mathcal{C}_{\ell}^{\mathrm{Theory}}$ and $\mathcal{G}_{lk}$ for a given cosmological model. For the MRL reconstruction the converged solution $\mathcal{P}^{\mathrm{Converged}}_k$ is independent of the choice of initial power spectrum $\mathcal{P}^0_k$~\cite{Shafieloo:2003gf}, which is in our case the standard power-law power spectrum:
\begin{equation}
\mathcal{P}^0_k = A_s \left(\frac{k}{k_{\mathrm{pivot}}} \right)^{n_s-1},
\label{eq:dis_base_pk}
\end{equation}
where $A_s$ and $n_s$ are the amplitude and spectral index, respectively, evaluated at the pivot scale, $k_{\mathrm{pivot}}=0.05 \, \mathrm{Mpc}^{-1}$~\cite{Planck:2018jri,Planck:2018vyg,Planck:2019nip}. The MRL method tries to reduce the difference between $\mathcal{C}_\ell^{\mathrm{Data}} $ and $ \mathcal{C}_\ell^{\mathrm{Theory},i}$ with each passing iteration, as a result it starts fitting the noise after sufficiently many iterations. Thus, it is important to carefully terminate the process once the adopted stopping-criterion is reached.  

\subsection{Regularization schemes}\label{regularization_schemes}
The working principle of the MRL algorithm is to find solutions of $\mathcal{P}_k$ that provide better fit to the observed data. However, the very design of the algorithm allows it to generate solutions having unphysical sharp oscillatory features by over-fitting the noise with increasing iterations. To alleviate the contributions coming from noise over-fitting and have physically plausible $\mathcal{P}_k$s, we allow the MRL solutions to pass through regularization process~\cite{Sohn:2022jsm}. In this work, we employ three different regularization schemes:

\begin{itemize}
    \item \textbf{Gaussian regularization (GR):}
    
In this regularization recipe, we perform a convolution operation on the MRL reconstructed power spectra with a Gaussian kernel and normalize the result to obtain the final smoothed spectrum $\mathcal{P}_k^{\mathrm{GR}}$. Mathematically, it takes the following expression in discrete k-space:  
\begin{equation}
\mathcal{P}_k^{\mathrm{GR}} =
\frac{
\sum_{k^\prime} \mathcal{P}_{k^\prime} \exp\left[
- \left( \frac{\log k^\prime - \log k}{\Sigma_{\mathrm{GR}}} \right)^2
\right]
}{
\sum_{k^\prime} \exp\left[
- \left( \frac{\log k^\prime - \log k}{\Sigma_{\mathrm{GR}}} \right)^2
\right]
};
\label{eq:gauss_reg}
\end{equation}
also note that the convolution is carried out in logarithmic space with adjustable kernel width $\Sigma_{\mathrm{GR}}$. This regularization prescription is highly efficient in suppressing noise-induced features but with a limitation that it imposes uniform smoothing at all scales controlled by a single parameter $\Sigma_{\mathrm{GR}}$.  
    \item \textbf{Diffusive regularization (DR):}

To circumvent the limitation of the Gaussian regularization, and to have a more physically realistic reconstruction by preserving most of the major features together with effectively regularizing unwanted over-fitting, the diffusive regularization method has been introduced in Ref.~\cite{Sohn:2022jsm}. The proposal of this improved regularization method is to introduce a double derivative term $\mathcal{D}^2_{k k^\prime}$ in Eq. (\ref{eq:mrl}) by taking inspiration from the heat diffusion equation, 
\begin{equation}
\frac{\partial f}{\partial t}
= \kappa \frac{\partial^2 f}{\partial x^2},
\label{eq:diffusion}
\end{equation}
as the name of the method itself suggests, which finally gives rise to the equation:
\begin{equation}
\mathcal{P}^{i+1}_k = \mathcal{P}^{i}_k + \mathcal{P}^{i}_k \times \sum_{\ell} \, \tilde{\mathcal{G}}_{\ell k} \, \left( \frac{\mathcal{C}_\ell^{\mathrm{Data}} - \mathcal{C}_\ell^{\mathrm{Theory},i}}{\mathcal{C}_\ell^{\mathrm{Theory},i}}  \right)\tanh^2 \left[\mathcal{E}_{\ell}^i \left(\mathcal{C}_\ell^{\mathrm{Data}} - \mathcal{C}_\ell^{\mathrm{Theory},i} \right) \right] + \kappa_{\mathrm{DR}} \sum_{k^\prime} \, \mathcal{D}^2_{k k^\prime} \mathcal{P}^{i}_{k^\prime}.
\label{eq:regmrl}
\end{equation}
Introduction of the $\mathcal{D}^2_{k k^\prime}$ term (discretized version of $\frac{\partial^2}{\partial x^2}$ in $k$-space) in Eq. (\ref{eq:regmrl}) causing the MRL algorithm to "diffuse away" any features that are very sharp at each iteration, while finding solutions for $\mathcal{P}_k$ that are giving better fit to the data. The factor $\kappa_{\mathrm{DR}}$ in Eq. (\ref{eq:regmrl}) acts as a regulator which controls the amount of smoothing during reconstruction. By setting $\kappa_{\mathrm{DR}}=0$, we can get back to our typical MRL results obtained from Eq. (\ref{eq:mrl}).
    \item \textbf{Total variation regularization (TVR):}
    
The total variation regularization is a well studied method, and it has been widely used to analyse images. In this method, the regularization process is governed by a penalty term given by
\begin{equation}
\int dx \, \left| \frac{\partial f}{\partial x} \right|,
\label{eq:pen_lasso}
\end{equation}
which we see in the case of LASSO regression. To incorporate this regularization into the MRL algorithm, we add the term: $\kappa \sum_{k^\prime} \, \mathcal{D}_{k k^\prime} \, \mathrm{sign}\left(\sum_{k^{\prime \prime}} \, \mathcal{D}_{k^\prime k^{\prime\prime} } \mathcal{P}^{i}_{k^{\prime\prime}} \right)$ in Eq. (\ref{eq:mrl}), which leads to
\begin{align}
\mathcal{P}^{i+1}_k = \mathcal{P}^{i}_k + \mathcal{P}^{i}_k \times \sum_{\ell} \, \tilde{\mathcal{G}}_{\ell k} \, \left( \frac{\mathcal{C}_\ell^{\mathrm{Data}} - \mathcal{C}_\ell^{\mathrm{Theory},i}}{\mathcal{C}_\ell^{\mathrm{Theory},i}}  \right)\tanh^2 \left[\mathcal{E}_{\ell}^i \left(\mathcal{C}_\ell^{\mathrm{Data}} - \mathcal{C}_\ell^{\mathrm{Theory},i} \right) \right] \notag \\
+ \, \kappa_{\mathrm{TVR}} \sum_{k^\prime} \, \mathcal{D}_{k k^\prime} \, \mathrm{sign}\left(\sum_{k^{\prime \prime}} \, \mathcal{D}_{k^\prime k^{\prime\prime} } \mathcal{P}^{i}_{k^{\prime\prime}} \right).
\label{eq:mrltv}
\end{align}
Here $\kappa_{\mathrm{TVR}}$ plays the same role as $\kappa_{\mathrm{DR}}$ plays in the case of diffusive regularization. The $\mathrm{sign}(x)$ function appearing in Eq. (\ref{eq:mrltv}) takes the value $1 \, (-1)$ for all positive (negative) values of $x$ and for $x=0$ it is "$0$".
\end{itemize}
For further details on these regularization schemes in the context of the MRL reconstruction one can refer to article~\cite{Sohn:2022jsm}.

\subsection{Data concatenation}
%In this work, we use both temperature and polarization data. in order to reconstruct the primordial power spectrum from the whole dataset, which includes TT, TE, and EE all three angular power spectrum data. To achieve this, we construct a concatenated data matrix with all these datasets we construct a full data vector with TT, TE, and EE angular power spectrum data to use 
In this work, we use both temperature and polarization measurements by concatenating TT, TE and EE angular power spectra into a single data vector, 
\begin{equation}
\mathcal{C}_L^{\mathrm{Total}} =
\begin{bmatrix}
\mathcal{C}^{TT}_{\ell_{\min}} & \cdots & \mathcal{C}^{TT}_{\ell_{\max}} &
\gamma\, \mathcal{C}^{TE}_{\ell_{\min}} & \cdots & \gamma\, \mathcal{C}^{TE}_{\ell_{\max}} &
\gamma^2\, \mathcal{C}^{EE}_{\ell_{\min}} & \cdots & \gamma^2\, \mathcal{C}^{EE}_{\ell_{\max}}
\end{bmatrix}_{\!L}
\label{eq:cl_total_mat}
\end{equation}.
The concatenated angular power spectrum matrix (\ref{eq:cl_total_mat}) too obeys Eq. (\ref{eq:disClXY}), thus we can write
\begin{equation}
\mathcal{C}_\ell^{\mathrm{Total}} = \sum_k \, \mathcal{G}^{\mathrm{Total}}_{lk} \mathcal{P}_k,
\label{eq:dis_Cl_tot_eq}
\end{equation}
%since for a given $\mathcal{P}_k$ Eq. (\ref{eq:disClXY}) is valid for each of TT, TE, and EE spectrum,
where $\mathcal{G}_{Lk}^{\mathrm{Total}}$ is given by
\begin{equation}
\mathcal{G}_{Lk}^{\mathrm{Total}} =
\begin{bmatrix}
\mathcal{G}^{TT}_{\ell_{\min} k_{\min}} & \cdots & \cdots & \mathcal{G}^{TT}_{\ell_{\min} k_{\max}} \\
\vdots & \vdots & \ddots & \vdots \\
\mathcal{G}^{TT}_{\ell_{\max} k_{\min}} & \cdots & \cdots & \mathcal{G}^{TT}_{\ell_{\max} k_{\max}} \\ \\
\gamma\, \mathcal{G}^{TE}_{\ell_{\min} k_{\min}} & \cdots & \cdots & \gamma\, \mathcal{G}^{TE}_{\ell_{\min} k_{\max}} \\
\vdots & \vdots & \ddots & \vdots \\
\gamma\, \mathcal{G}^{TE}_{\ell_{\max} k_{\min}} & \cdots & \cdots & \gamma\, \mathcal{G}^{TE}_{\ell_{\max} k_{\max}} \\ \\
\gamma^2\, \mathcal{G}^{EE}_{\ell_{\min} k_{\min}} & \cdots & \cdots & \gamma^2\, \mathcal{G}^{EE}_{\ell_{\min} k_{\max}} \\
\vdots & \vdots & \ddots & \vdots \\
\gamma^2\, \mathcal{G}^{EE}_{\ell_{\max} k_{\min}} & \cdots & \cdots & \gamma^2\, \mathcal{G}^{EE}_{\ell_{\max} k_{\max}} \\
\end{bmatrix}_{\!Lk}.
\label{eq:glk_total_mat}
\end{equation}
%In Eqs. (\ref{eq:cl_total_mat}) and (\ref{eq:glk_total_mat}) the $\ell_{\min}$ and $\ell_{\max}$ for TT, TE, and EE datasets are subjected to the choice of data that we consider and also the choice of multipole range made during reconstruction.
In Eq. (\ref{eq:cl_total_mat}), we multiply $\mathcal{C}^{TE}_{\ell}$ and $\mathcal{C}^{EE}_{\ell}$ data with factors $\gamma$ and $\gamma^2$, respectively when concatenating with $\mathcal{C}^{TT}_{\ell}$ data, to impose a relative weight among TT, TE, and EE datasets. Accordingly, we consider the parameters $\gamma$ and $\gamma^2$ in concatenated $\mathcal{G}_{Lk}^{\mathrm{Total}}$ matrix as well to satisfy the Eq. (\ref{eq:dis_Cl_tot_eq}). We also use the tuning parameter $\gamma$ consistently when computing the full covariance matrix for the concatenated dataset.

\subsection{Datasets}\label{subsec:datasets}
We use the temperature and polarization data from the latest data releases of ACT and SPT, and Planck PR3 and PR4 likelihoods to conduct a free-form reconstruction of the primordial power spectrum employing the MRL algorithm. %We have already mentioned in the introduction that in this analysis, we use datasets from two different CMB missions from their latest data release, namely, Atacama Cosmology Telescope (ACT) mission and South Pole Telescope (SPT) mission to do a free-form reconstruction of the primordial power spectrum.

\begin{itemize}
    \item \textbf{ACT-DR6:}
    
    In this work, we use the temperature and polarization band power spectrum data from the latest $6^{th}$ data release of ACT mission~\cite{AtacamaCosmologyTelescope:2025blo} \footnote{\url{https://lambda.gsfc.nasa.gov/product/act/act_dr6.02/}}. In the DR6 release, ACT provides power spectrum data for two different binning schemes with minimum bin widths ($\Delta \ell_{\mathrm{min}}$) of 20 and 50; the actual bin width increases with multipole. ACT uses 50 bin width data for the likelihood analysis, and data with $\Delta \ell_{\mathrm{min}}=20$ are used in the ACT paper~\cite{AtacamaCosmologyTelescope:2025blo} for plotting. %ACT uses 20 and 50 bin-width datasets To perform likelihood analysis ACT uses dataset with $\Delta \ell_{\mathrm{min}}=50$, and for plotting purpose the ACT paper~\cite{AtacamaCosmologyTelescope:2025blo} uses the data with $\Delta \ell_{\mathrm{min}}=20$. %In the DR6 data, ACT provides two different binning schemes with minimum bin widths ($\Delta \ell_{\mathrm{min}}$) of 50 and 20.  %The dataset having the bin-width of 50 is used for the likelihood analysis. However, in the ACT paper~\cite{AtacamaCosmologyTelescope:2025blo} they used the power spectra data having the bin-width of 20 for the plotting purpose.
    For the MRL reconstruction, we use the dataset of bin size, $\Delta \ell_{\mathrm{min}}=20$. From ACT, we have TT, TE, and EE data for the multipole ($\ell$) range of $[593, 8319]$. However, in this work, we consider the temperature and polarization angular power spectrum data only up to maximum multipole $(\ell_{\mathrm{max}})$ 4810, because for high $\ell$s the CMB power becomes low due to Silk damping phenomenon and also the foreground power starts dominating the CMB signal for scales $\ell > 4000$ ~\cite{AtacamaCosmologyTelescope:2025blo}. %As a result, in our MRL reconstruction with the ACT-DR6 data, we consider the multipole range $[593, 4810]$ for both temperature (TT) and polarization (TE/EE) data.
    ACT measures three different frequency bands with band centres 98, 150, and 220 GHz covering $45 \%$ of the sky. However, after taking into account the contaminations from different galactic and extragalactic sources only $ 25\%$  of the sky contributes to the final analysis. %it utilizes $ 25\%$  of the sky for the final analysis after masking contaminations arising from different galactic and extragalactic sources.
    ACT provides us with the data for auto and cross frequency band power spectra, and associated covariance matrices, using which we prepare the coadded TT, TE, and EE band power spectra and corresponding covariance matrices (here the cross terms are also present along with diagonal elements). We use our coadded ACT data in this work in reconstructing the free-form PPS employing the MRL algorithm.

    \item \textbf{SPT-3G D1:}
    
    Along with ACT-DR6 data, we also use the latest SPT-3G D1 data~\cite{SPT-3G:2025bzu} \footnote{\url{https://lambda.gsfc.nasa.gov/product/spt/spt_3gd1/}} in this reconstruction analysis. The SPT-3G D1 data come from the measurement of the Main field of the sky, a portion which covers about $4 \% $ of the whole CMB sky. The SPT-3G D1 survey observes CMB temperature and E-mode polarization fields across three frequencies of 95, 150, and 220 GHz. We use in this work both the temperature and polarization data from the SPT-3G D1 survey. SPT-3G D1 uses same bin-size ($\Delta \ell)$ of 50 across the entire multipole range for TT, TE, and EE band powers. The SPT-3G D1 survey provides the TT band power data for the multipole range of $ 400 < \ell < 3000$, whereas for the TE/EE mode we have band power data for the multipole range of $400 < \ell < 4000$. %SPT mission the $\ell$ range of TT data is $< 3000$ (to be specific, $\ell_{\mathrm{max}} =2974.46$) and for TE and EE polarization modes $\ell_{\mathrm{max}}$ extends up to $4000$, specifically, $3974.56$ and $3974.49$, respectively. 
    However, in reconstructing with SPT data, we truncate TE and EE data at $\ell_{\mathrm{max}} < 3000$, since we have TT data only up to $\ell < 3000$. We use here the cleaned combined TT, TE, and EE band power data, prepared from the multifrequency data employing minimum-variance method and subtracting contributions coming from nuisance and foreground. %SPT-3G D1 data provides us with the minimum-variance foreground- and nuisance-subtracted cleaned combined TT, TE, and EE band power spectra prepared from the multifrequency band power data.

   \item \textbf{CamSpec-NPIPE:} 

   We also reconstruct PPS from Planck PR4 CamSpec-NPIPE data~\cite{Rosenberg:2022sdy}. We use a cleaned coadded unbinned angular power spectrum ($C_{\ell}$) data from Planck PR4 CamSpec-NPIPE likelihood. This data release of Planck utilizes $ 80 \%$ of the sky map. We have CamSpec-NPIPE TT, TE, and EE data over multipole ranges of $[2,2500]$, $[30,2500]$, and $[30,2000]$, respectively. We use this full multipole range when we reconstruct only from Planck PR4 CamSpec-NPIPE data. However, we use a truncated CamSpec-NPIPE data when combining with ACT-DR6 and SPT-3G D1 datasets. To combine CamSpec data with ACT and SPT data, we consider each dataset only for that $\ell$ range where they have relatively lower uncertainties in measuring $C_{\ell}$s. When combining with ACT-DR6 data we find that CamSpec imparts tighter uncertainties on $C^{TT}_{\ell}$, $C^{TE}_{\ell}$, and $C^{EE}_{\ell}$ data over multipole ranges $[2-1086]$, $[30-981]$, and $[30-792]$, respectively, and for rest of the overlapping multipole region ACT performs better than CamSpec. Similarly, within the overlapping region of CamSpec and SPT-3G D1 data, across multipoles $[2-2049]$, $[30-1399]$, and $[30-1049]$, CamSpec has higher signal to noise ratio than SPT for $C^{TT}_{\ell}$, $C^{TE}_{\ell}$, and $C^{EE}_{\ell}$, respectively. Accordingly, we truncate CamSpec data to reconstruct PPS from CamSpec and ACT/SPT combined dataset. 
    
\end{itemize}

\subsection{Statistical significance}\label{subsec:stat_sig}
We assess the MRL reconstructions with a goodness of fit statistic and a simulation-based error analysis, which we describe subsequently.
\begin{enumerate}
    \item \textbf{Goodness of fit:}
    
    We evaluate the chi-square per datum to quantify the change in fit relative to the observed data obtained from the MRL solution at each iteration. The reduced chi-squared statistic is defined as follows: 
    \begin{equation}
    \tilde{\chi}^2_{\mathrm{XY}} \equiv \frac{1}{n_{\mathrm{XY}}} 
    \sum_{\ell,\ell'} 
    \Delta \mathcal{C}^{\mathrm{XY}}_{\ell} \, 
    \mathrm{Cov}^{-1}_{\ell,\ell'} \, 
    \Delta \mathcal{C}^{\mathrm{XY}}_{\ell'},
    \label{eq:reduce_chi2}
\end{equation}
where the parameter $n_{\mathrm{XY}}$ represents the total number of data points, and $\Delta \mathcal{C}^{\mathrm{XY}}_{\ell}$ is the residual given by $\mathcal{C}^{\mathrm{Data}}_{\ell}-\mathcal{C}^{\mathrm{Recons}}_{\ell}$. We use $\tilde{\chi}^2_{\mathrm{XY}}$ as an estimation of the goodness of fit to the data. For all our reconstructions, we compute the change in the chi-square values obtained for the reconstructed and power-law power spectra at each iteration: 
    \begin{equation}
    \Delta\tilde{\chi}^2_{\mathrm{XY}}=\tilde{\chi}^2_{\mathrm{XY,~recons}}-\tilde{\chi}^2_{\mathrm{XY,~power-law}}. 
    \label{eq:residual_chi2}
    \end{equation}
    
    \item \textbf{Error analysis:} 
    
    We conduct a statistical hypothesis test between the baseline power-law power spectrum (\ref{eq:dis_base_pk}) and the MRL reconstructions using diffusive and total variation regularization methods. In our analysis, the null hypothesis $H_0$ states that the observed CMB data corroborate the power-law power spectrum, and our alternative hypothesis $H_1$ states that the observed CMB data rejects power-law $\mathcal{P}_k$ and we need features in the primordial power spectrum to explain the observed data. To test our hypothesis, we conduct the $p$-value analysis at each $k$ demonstrating the statistical significance of any feature in the reconstructions showing a deviation from the power-law form. We evaluate pointwise $p$-values in $k$ using 1000 simulated $\mathcal{C}_{\ell}$ realizations drawn from a multivariate Gaussian distribution with a power-law mean and the covariance matrix of the data, and then from these realizations we reconstruct 1000 primordial $\mathcal{P}_k$s using the same MRL algorithm. We compare the $\mathcal{P}_k$s reconstructed from actual data with these 1000 simulated $\mathcal{P}_k$s defining the $68 \%$, $95 \%$, and $99 \%$ confidence intervals (hereafter CIs). %This $p$-value statistics allows us to find whether any feature at any $k$ is rejecting the power-law behaviour with certain CIs. 
    We also evaluate the ratio of $\Delta \mathcal{P}(k)$ vs $\sigma(k)$ for the simulated $\mathcal{P}_k$s, where $\Delta \mathcal{P}(k)$ stands for the residual reconstructed power spectrum with respect to the median power spectrum obtained from simulations, and $\sigma(k)$ is the standard deviation of the simulated power spectra at each $k$.

\end{enumerate}

\section{Result and analysis: free-form statistics}\label{sec:result_nonpar}
In this section, we discuss our results on the MRL reconstructions obtained for different datasets taken into consideration in this work. We reconstruct for each dataset four free-form power spectra ($\mathcal{P}_k$) using methods outlined in \texttt{Regularization schemes} section. We refer to the power spectra reconstructed using Eqs. (\ref{eq:mrl}), (\ref{eq:gauss_reg}), (\ref{eq:regmrl}), and (\ref{eq:mrltv}) as \texttt{MRL}, \texttt{MRL-GR}, \texttt{MRL-DR}, and \texttt{MRL-TVR} throughout this article.
% \begin{itemize}
%     \item First, we reconstruct employing the MRL algorithm illustrated in Eq. (\ref{eq:mrl}), which we refer with the term \texttt{MRL} throughout this article.   
%     \item After reconstruction using Eq. (\ref{eq:mrl}), we further use the Gaussian regularization shown in Eq. (\ref{eq:gauss_reg}) for smoothing. We call this smoothed power spectrum as \texttt{MRL-GR} in this entire article. %We assume the value of the Gaussian kernel width $\Sigma_{\mathrm{GR}}=0.03$ in all our analysis. 
%     \item Next, we reconstruct the free-form $\mathcal{P}_k$ for the \texttt{Diffusive regularization} method by applying Eq. (\ref{eq:regmrl}). We label this reconstructed power spectrum as \texttt{MRL-DR} in this work. % We set the value of the tuning parameter for the diffusive regularization $\kappa_{\mathrm{DR}}=0.01$ in all our reconstructions.
%     \item Lastly, using the \texttt{Total variation regularization} method illustrated in Eq. (\ref{eq:mrltv}), we reconstruct primordial $\mathcal{P}_k$, which is designated as \texttt{MRL-TVR} throughout this article. %For this regularization method, we adopt the value of regularisation parameter $\kappa_{\mathrm{TVR}}=0.008$ throughout this work. 
% \end{itemize}
 The adopted values of the smoothing parameters in Eqs. (\ref{eq:gauss_reg}), (\ref{eq:regmrl}), and (\ref{eq:mrltv}) are $\Sigma_{\mathrm{GR}}=0.03$, $\kappa_{\mathrm{DR}}=0.01$, and $\kappa_{\mathrm{TVR}}=0.008$, respectively, for all reconstructions. To show the correspondence between $\ell$ and $k$ in all the power spectrum plots, we use the Limber approximation, $\ell \approx k \eta_{\mathrm{rec}}$, where $\eta_{\mathrm{rec}}$ is the conformal distance to the last scattering surface. 
%Now, let us discuss the results obtained from our reconstruction analysis for different datasets. and for ACT and SPT data individually as well as in combination with CamSpec data. To reconstruct PPS for the combined datasets of ACT and CamSpec, and SPT and CamSpec, we consider the best-fit baseline $\Lambda CDM$ cosmology obtained from Planck PR4 CamSpec-NPIPE data as background cosmology.

\subsection{Reconstruction from CamSpec data}\label{subsec:rec_cam}
\begin{figure}
    \centering

    \begin{subfigure}{0.7\textwidth}
        \centering
        \includegraphics[width=\textwidth]{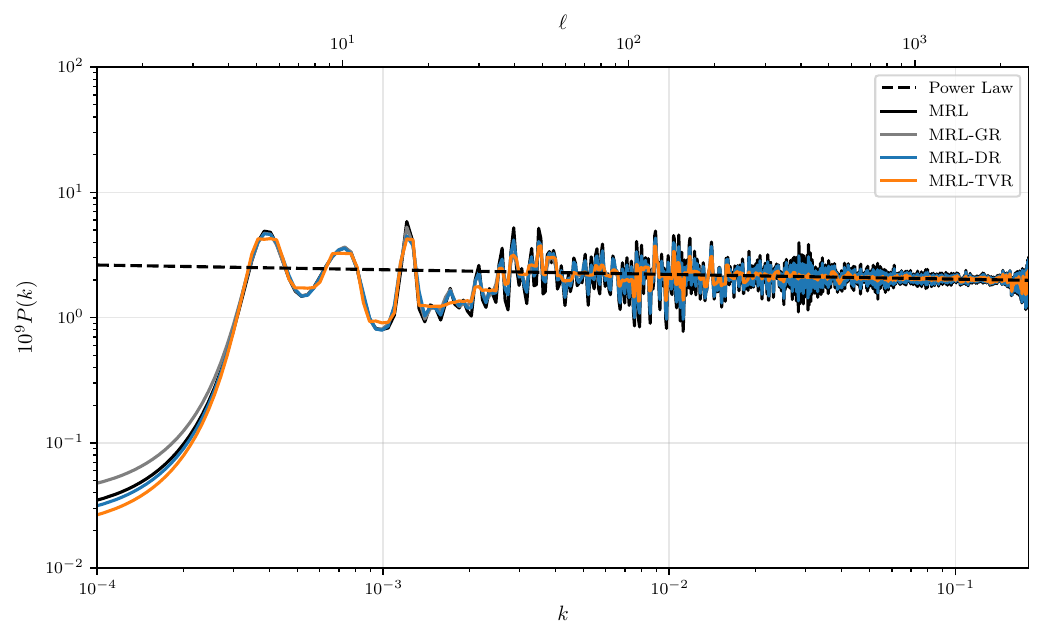}
        \caption{}
        \label{fig:Cams_CamsBF-1a}
    \end{subfigure}

%    \vspace{0.5cm} % space between figures

    \begin{subfigure}{0.7\textwidth}
        \centering
        \includegraphics[width=\textwidth]{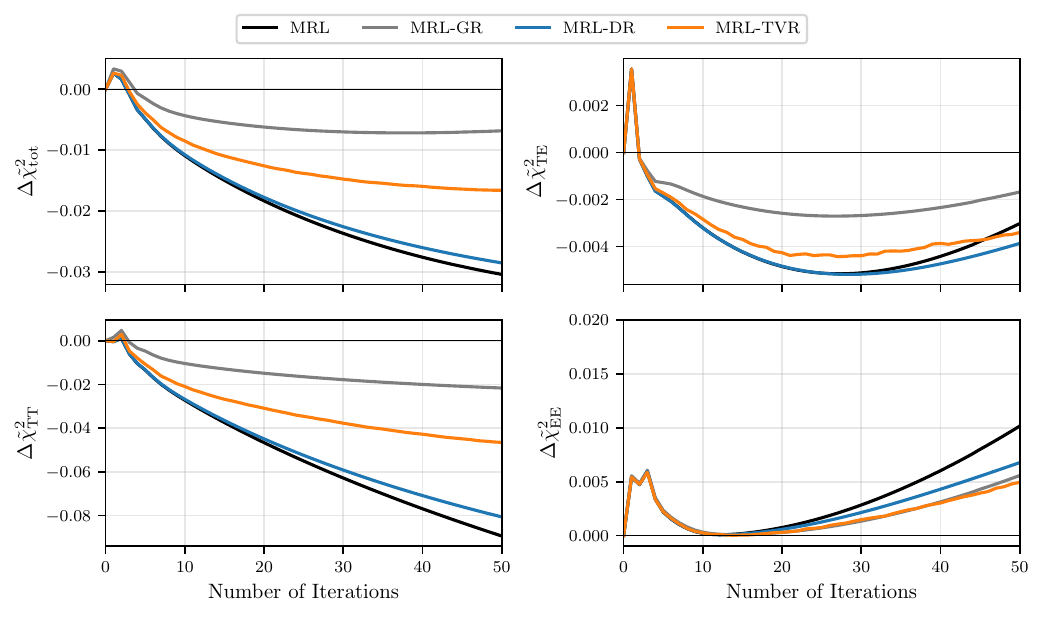}
        \caption{}
        \label{fig:Cams_CamsBF-1b}
    \end{subfigure}

    \caption{(a) Primordial power spectra that minimize the total chi-square reconstructed employing the MRL algorithm with different regularization methods. The black dashed line represents the nearly scale-invariant power-law power spectrum. The range of $k$ in this plot is $[10^{-4} - 0.180]$. (b) Improvements obtained from reconstructed PPS in the chi-squared fit to data with every iteration for combined, TT, TE, and EE spectra. These results have been obtained from Planck PR4 CamSpec-NPIPE data based on Planck PR4 CamSpec-NPIPE best-fit $\Lambda CDM$ background cosmology.}
    \label{fig:Cams_CamsBF-1}
\end{figure}

\begin{figure}
    \centering

    \begin{subfigure}{0.7\textwidth}
        \centering
        \includegraphics[width=\textwidth]{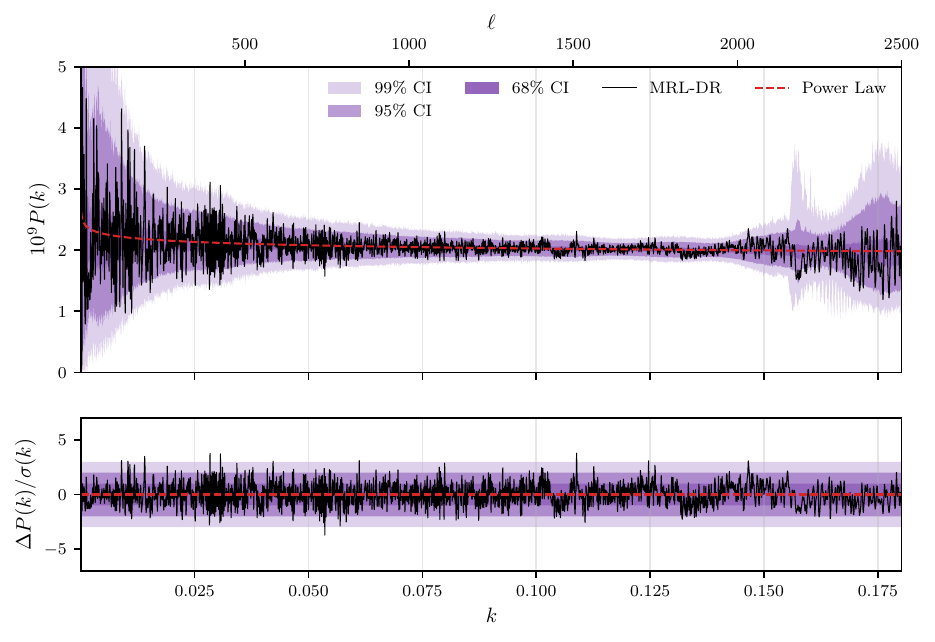}
        \caption{}
        \label{fig:Cams_CamsBF-2a}
    \end{subfigure}

%    \vspace{0.5cm} % space between figures

    \begin{subfigure}{0.7\textwidth}
        \centering
        \includegraphics[width=\textwidth]{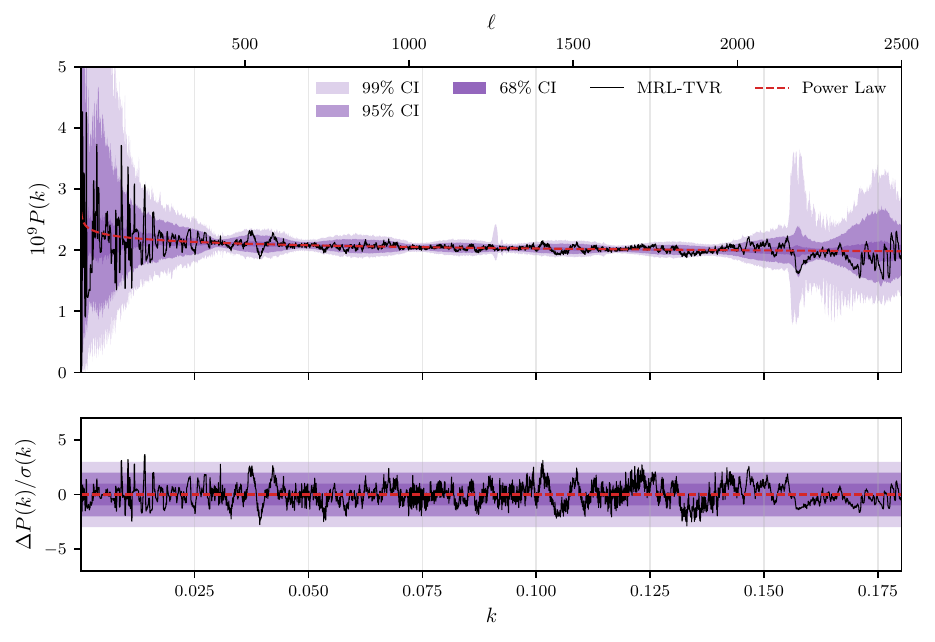}
        \caption{}
        \label{fig:Cams_CamsBF-2b}
    \end{subfigure}

    \caption{(a) Simulation based error analysis on the power spectrum reconstructed using diffusive regularization method (MRL-DR). (b) Simulation based error analysis on the power spectrum reconstructed using total variation regularization method (MRL-TVR). Results in both panels are for Planck PR4 CamSpec-NPIPE data and Planck PR4 CamSpec-NPIPE $\Lambda CDM$ best-fit cosmology. The range of $k$ in both panels is $[10^{-4} - 0.180]$.}

    \label{fig:Cams_CamsBF-2}
\end{figure}

This reconstruction is based on Planck PR4 CamSpec-NPIPE data. We use the $\Lambda CDM $ best-fit cosmology as background cosmology in this reconstruction, evaluated for Planck PR4 CamSpec-NPIPE high-$\ell$ TTTEEE likelihood~\cite{Rosenberg:2022sdy} in combination with Commander low-$\ell$ and SimAll-EE likelihoods applying the \texttt{BOBYQA} minimization method~\cite{Cartis_2021,cartis2018improvingflexibilityrobustnessmodelbased,Powell2009TheBA}. The evaluated best-fit values of the cosmological parameters are listed in Table~\ref{tab:bf_cosmo_all}. The reconstructed power spectra are shown in Fig.~\ref{fig:Cams_CamsBF-1a} for each smoothing technique defined in Section~\ref{regularization_schemes}. The $\mathcal{P}_k$s that minimize the change in chi-square for combined data ($\Delta\tilde{\chi}^2_{\mathrm{tot}}$) are shown in Fig.~\ref{fig:Cams_CamsBF-1a}. The obtained $\Delta\tilde{\chi}^2$ improvements for combined, TT, TE, and EE spectra are exhibited in Fig.~\ref{fig:Cams_CamsBF-1b}. We perform the simulation-based error analysis described in Section~\ref{subsec:stat_sig} for this dataset to estimate the significance of the observed features in the reconstructed power spectra. We present our results obtained from the error analysis for diffusive and total variation regularization methods in Fig.~\ref{fig:Cams_CamsBF-2}, which show that the reconstructions are consistent with the power-law power spectrum with no features.

\subsection{Reconstruction from ACT data}\label{subsec:rec_act}
In this section, we present and discuss the results of our reconstructions with ACT-DR6 data. We reconstruct the primordial power spectrum from ACT-DR6 data for the following three best-fit baseline $\Lambda CDM$ background cosmologies~(Table \ref{tab:bf_cosmo_all}): ACT-DR6, Planck PR3 2018, and Planck PR4 CamSpec-NPIPE. Now, let us discuss each scenario separately in subsequent sections. 
% \begin{itemize}
%     \item First, we reconstruct PPS from the ACT data based on the best-fit baseline $\Lambda CDM$ cosmology obtained from the ACT data (Table \ref{tab:bf_cosmo_all}). %The best-fit values of the cosmological parameters for the baseline $\Lambda CDM$ model is tabulated
%     \item Next, the best-fit baseline $\Lambda CDM$ cosmology (Table \ref{tab:bf_cosmo_all}) obtained from the Planck 2018 PR3 data has been assumed as the background cosmology in reconstructing PPS from the ACT data. 
%     \item Finally, we adopt the best-fit baseline $\Lambda CDM$ cosmology obtained from the Planck PR4 CamSpec-NPIPE data (Table \ref{tab:bf_cosmo_all}) as the background cosmology in reconstructing PPS from the ACT data.
% \end{itemize}

%\begin{itemize}
%    \item \textbf{Reconstruction from ACT data:}

\begin{table*}[t]
\centering
\caption{Best-fit cosmological parameters for the baseline $\Lambda CDM$ model obtained from different CMB datasets.}
\label{tab:bf_cosmo_all}
\begin{adjustbox}{max width=\textwidth}
\renewcommand{\arraystretch}{1.25}
\begin{tabular}{|l|c|c|c|c|c|c|}
\hline
\textbf{Dataset} &
$\mathbf{100\,\Omega_{\mathrm b}h^2}$ &
$\mathbf{100\,\Omega_{\mathrm c}h^2}$ &
$\mathbf{H_0}$ &
$\mathbf{100\,\tau_{\mathrm{reio}}}$ &
$\mathbf{10^9A_{\mathrm s}}$ &
$\mathbf{n_{\mathrm s}}$ \\
\hline\hline

ACT-DR6 &
2.259704 &
12.396975 &
66.045020 &
5.671529 &
2.120059 &
0.967403 \\
\hline

SPT-3G D1 &
2.227060 &
12.263763 &
66.246902 &
5.009288 &
2.120750 &
0.953242 \\
\hline

Planck PR3 2018 &
2.237737 &
12.010350 &
67.321780 &
5.430138 &
2.100428 &
0.965892 \\
\hline

Planck PR4 CamSpec--NPIPE &
2.216025 &
11.984220 &
67.180180 &
5.234995 &
2.081448 &
0.962474 \\
\hline

\end{tabular}
\end{adjustbox}
\end{table*}

\subsubsection{Case I: ACT background cosmology}\label{subsubsec:case1_act}
\begin{figure}
    \centering

    \begin{subfigure}{0.7\textwidth}
        \centering
        \includegraphics[width=\textwidth]{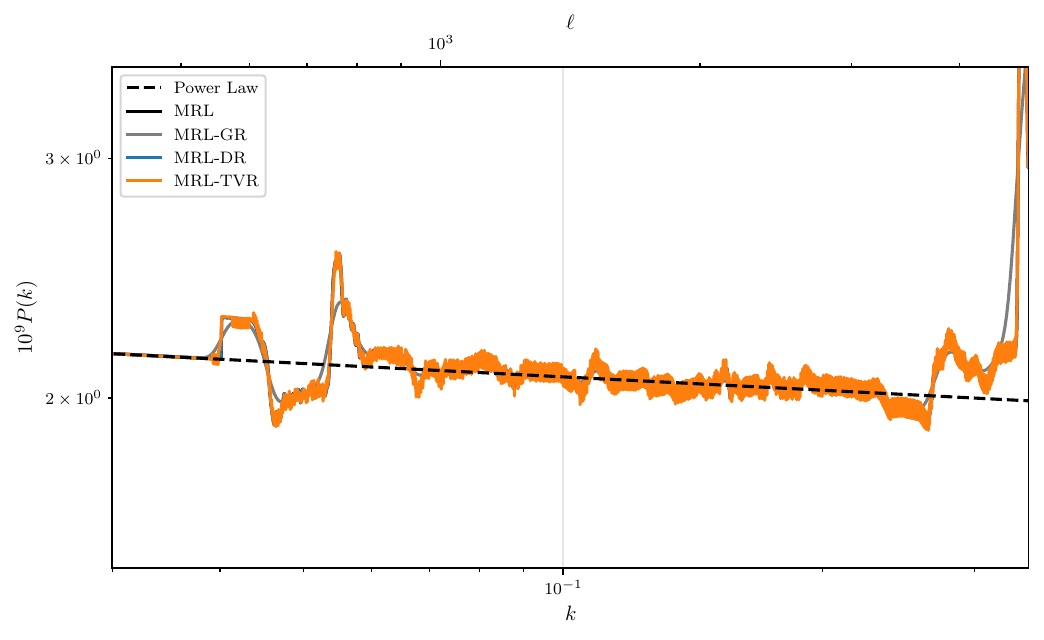}
        \caption{}
        \label{fig:COADACT_ACTBF-1a}
    \end{subfigure}

%    \vspace{0.5cm} % space between figures

    \begin{subfigure}{0.7\textwidth}
        \centering
        \includegraphics[width=\textwidth]{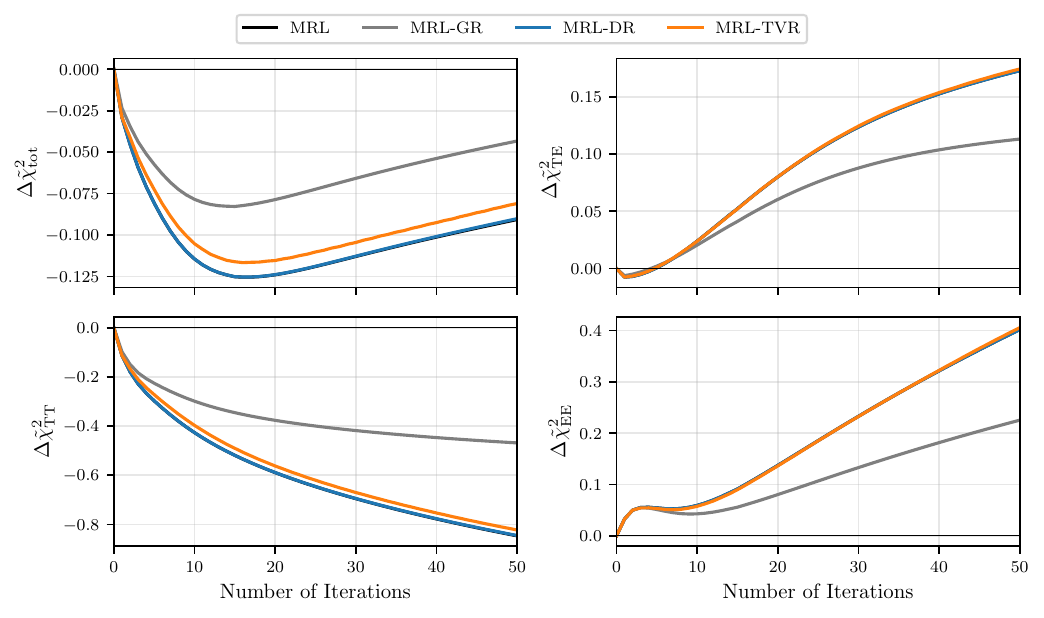}
        \caption{}
        \label{fig:COADACT_ACTBF-1b}
    \end{subfigure}

    \caption{(a) Primordial power spectra that minimize the total chi-square reconstructed employing the MRL algorithm with different regularization methods. The black dashed line represents the nearly scale-invariant power-law power spectrum. The range of $k$ in this panel is $[0.03 - 0.35]$. (b) Improvements obtained from reconstructed PPS in the chi-squared fit to data with every iteration for combined, TT, TE, and EE spectra. These results have been obtained from ACT data based on ACT best-fit $\Lambda CDM$ background cosmology.}
    \label{fig:COADACT_ACTBF-1}
\end{figure}

\begin{figure}
    \centering

    \begin{subfigure}{0.7\textwidth}
        \centering
        \includegraphics[width=\textwidth]{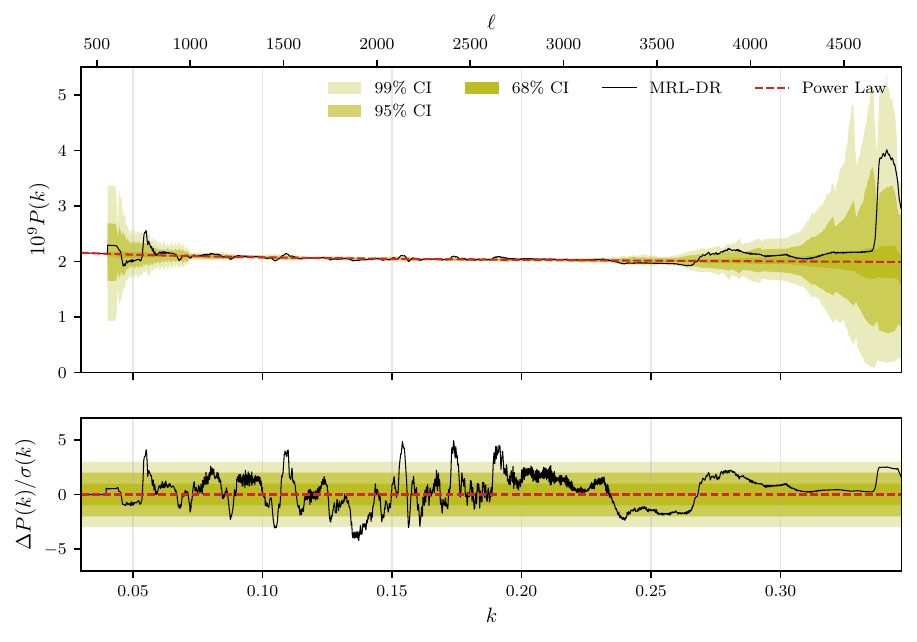}
        \caption{}
        \label{fig:COADACT_ACTBF-2a}
    \end{subfigure}

%    \vspace{0.5cm} % space between figures

    \begin{subfigure}{0.7\textwidth}
        \centering
        \includegraphics[width=\textwidth]{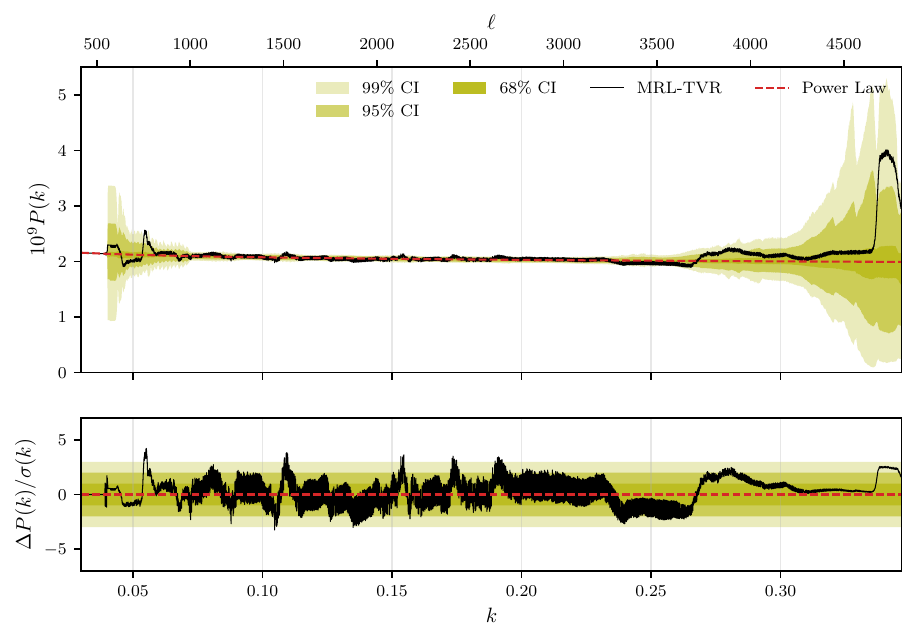}
        \caption{}
        \label{fig:COADACT_ACTBF-2b}
    \end{subfigure}

    \caption{(a) Simulation based error analysis on the power spectrum reconstructed using diffusive regularization method (MRL-DR). (b) Simulation based error analysis on the power spectrum reconstructed using total variation regularization method (MRL-TVR). Results in both panels are for ACT data and ACT $\Lambda CDM$ best-fit cosmology. The range of $k$ in both panels is $[0.03 - 0.35]$.}
    \label{fig:COADACT_ACTBF-2}
\end{figure}

%We now discuss our results for ACT data. 
%We present and discuss the results on the MRL reconstruction obtained for the ACT-DR6 data considering the ACT-DR6 best-fit background cosmology.
In Figs.~\ref{fig:COADACT_ACTBF-1} and~\ref{fig:COADACT_ACTBF-2}, we show the results of the MRL reconstructions from ACT data considering ACT-DR6 best-fit background cosmology. The best-fit values of the baseline $\Lambda CDM$ parameters are obtained from the minimization result released by ACT-DR6 along with their MCMC chains~\footnote{\url{https://lambda.gsfc.nasa.gov/product/act/act_dr6.02/act_dr6.02_chains_prod_table.html}}. We tabulate these values in Table~\ref{tab:bf_cosmo_all}, which we use to compute $\mathcal{P}^0_k$ and $\mathcal{G}^{\mathrm{Total}}_{lk}$ defined in Eqs. (\ref{eq:dis_base_pk}) and (\ref{eq:dis_Cl_tot_eq}), respectively. In Fig.~\ref{fig:COADACT_ACTBF-1a}, we show the reconstructed power spectra for different smoothing methods at the iterations which minimize the change in total reduced chi-square ($\Delta\tilde{\chi}^2_{\mathrm{tot}}$). The results of the goodness-of-fit analysis are presented in Fig.~\ref{fig:COADACT_ACTBF-1b} for combined, TT, TE, and EE spectra. %Each plot in Fig.~\ref{fig:COADACT_ACTBF-1b} is demonstrating how the improvement in reduced chi-square is changing with number of iterations. We have presented the improvement in reduced chi-square compared to the baseline $\mathcal{P}^0_k$, where $\Delta\tilde{\chi}^2_{\mathrm{tot}}$, $\Delta\tilde{\chi}^2_{\mathrm{TT}}$, $\Delta\tilde{\chi}^2_{\mathrm{TE}}$, and $\Delta\tilde{\chi}^2_{\mathrm{EE}}$ are actually presenting the improvement achieved from the combined, TT-only, TE-only, and EE-only datasets. From Fig. \ref{fig:COADACT_ACTBF-1b}, we find that all four methods are improving the fit for the TT data all the way up to 50 iterations. In contrast, for EE data we are not having any improvement to the fit, however for TE data it is initially showing some improvement to the fit but with increased iterations the $\Delta\tilde{\chi}^2_{\mathrm{TE}}$ is getting worse. The $\Delta\tilde{\chi}^2_{\mathrm{tot}}$ is showing improvements during initial iterations reaching minimum value after around 16 iterations, and then started worsening the fit to the combined data with further increased iteration. %Consequently, we have considered presenting reconstructed $\mathcal{P}_k$s after 16 iterations.
In Fig. \ref{fig:COADACT_ACTBF-2}, we show results of the simulation-based error analysis conducted for the MRL reconstructed $\mathcal{P}_k$s. %We perform the error analysis only for MRL-DR and MRL-TVR power spectra, which have been exhibited in the top (\ref{fig:COADACT_ACTBF-2a}) and bottom (\ref{fig:COADACT_ACTBF-2b}) plots, respectively. At the bottom of both Figs. (\ref{fig:COADACT_ACTBF-2a} and \ref{fig:COADACT_ACTBF-2b}), we plot the ratio of $\Delta \mathcal{P}(k)$ vs $\sigma(k)$. %Here $\Delta \mathcal{P}(k)$ stands for the residual reconstructed power spectrum with respect to the median power spectrum obtained from simulations, and $\sigma(k)$ is the standard deviation computed from simulated power spectra at each $k$.
It can be seen from the residual ratio plots in Figs. \ref{fig:COADACT_ACTBF-2a} and \ref{fig:COADACT_ACTBF-2b} that there are a few features lying outside the $99 \%$ CI, at scales $k \approx 0.06 \, \mathrm{Mpc}^{-1}$, $k \approx 0.11 \, \mathrm{Mpc}^{-1}$, $k \approx 0.155 \, \mathrm{Mpc}^{-1}$, $k \approx 0.175 \, \mathrm{Mpc}^{-1}$, and $k \approx 0.19 \, \mathrm{Mpc}^{-1}$ which survive in both regularization methods. Thus, their origin could be physical or statistical rather than reconstruction artifact. We will see whether we recover them in the primordial power spectra reconstructed from SPT data in Section \ref{subsubsec:case1_spt}. %We can observe from residual ratio plots of both MRL-DR and MRL-TVR that a few possible features are visible at scales $k \approx 0.06 \, \mathrm{Mpc}^{-1}$, $k=0.13 \, \mathrm{Mpc}^{-1}$, $k=0.19 \, \mathrm{Mpc}^{-1}$ and $k=0.28 \, \mathrm{Mpc}^{-1}$, however the features around $k=0.13 \, \mathrm{Mpc}^{-1}$ and $k=0.19 \, \mathrm{Mpc}^{-1}$ are less prominent in case of MRL-TVR. Only the features seen around $k=0.06 \, \mathrm{Mpc}^{-1}$ and $k=0.28 \, \mathrm{Mpc}^{-1}$ are equally prominent in both the cases of MRL-DR and MRL-TVR. 
Apart from these locations for the rest of the scales, we find no statistically significant features in the PPS. Besides, at $k \approx  0.33-0.34 \, \mathrm{Mpc}^{-1}$, a localized bump-like feature is visible, which apparently does not comply with the power-law power spectrum but error analysis shows it also sits well inside the $99 \%$ CIs. However, to explore the statistical significance of this feature in a robust manner we conduct a Bayesian analysis with a double-tilt power-law power spectrum in Section \ref{sec:result_par}. %The statistical significance of this feature is a matter of scrutiny.

\subsubsection{Case II: Planck PR3 background cosmology}\label{subsubsec:case2_act}
\begin{figure}
    \centering

    \begin{subfigure}{0.7\textwidth}
        \centering
        \includegraphics[width=\textwidth]{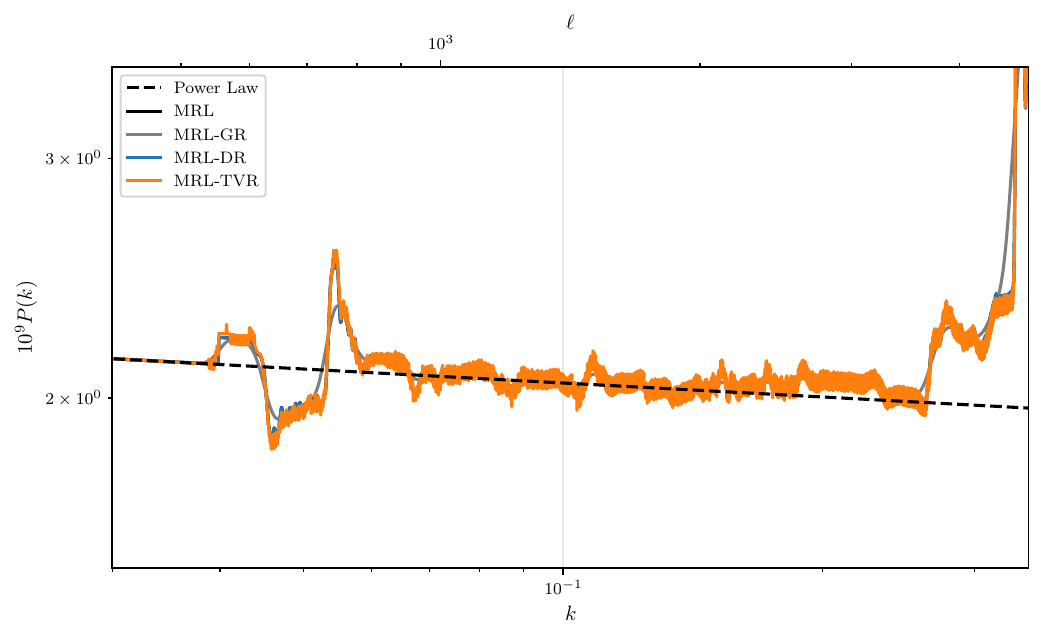}
        \caption{}
        \label{fig:COADACT_P18BF-1a}
    \end{subfigure}

%    \vspace{0.5cm} % space between figures

    \begin{subfigure}{0.7\textwidth}
        \centering
        \includegraphics[width=\textwidth]{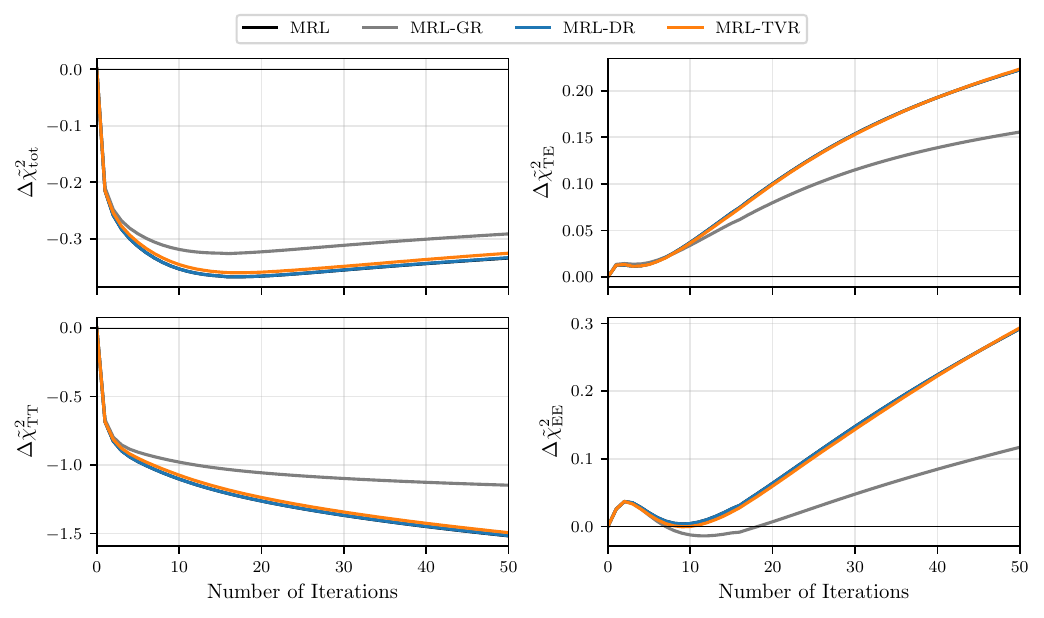}
        \caption{}
        \label{fig:COADACT_P18BF-1b}
    \end{subfigure}

    \caption{(a) Primordial power spectra that minimize the total chi-square reconstructed employing the MRL algorithm with different regularization methods. The black dashed line represents the nearly scale-invariant power-law power spectrum. The range of $k$ in this panel is $[0.03 - 0.35]$. (b) Improvements obtained from reconstructed PPS in the chi-squared fit to data with every iteration for combined, TT, TE, and EE spectra. These results have been obtained from ACT data based on Planck PR3 best-fit $\Lambda CDM$ background cosmology.}
    
    \label{fig:COADACT_P18BF-1}
\end{figure}

\begin{figure}
    \centering

    \begin{subfigure}{0.7\textwidth}
        \centering
        \includegraphics[width=\textwidth]{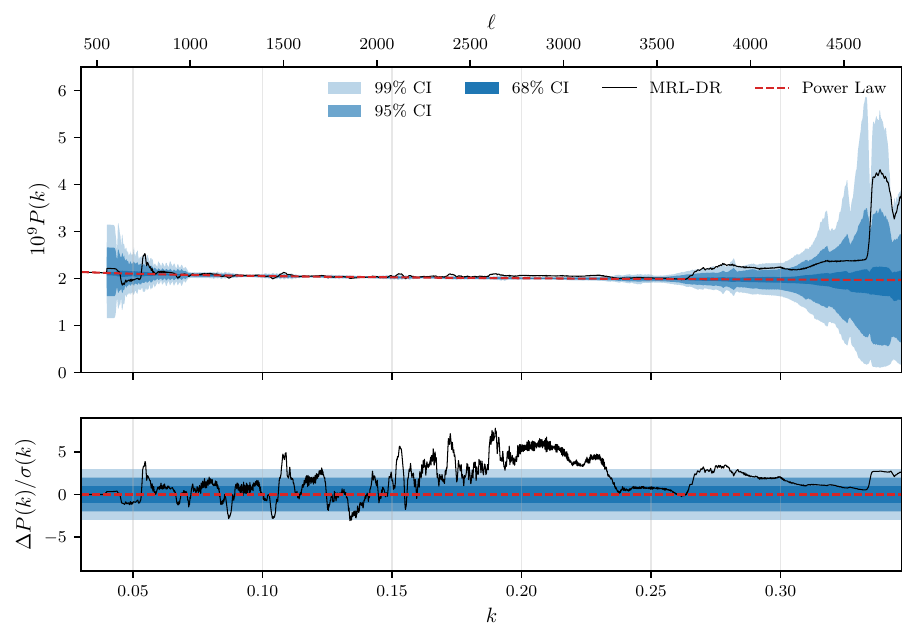}
        \caption{}
        \label{fig:COADACT_P18BF-2a}
    \end{subfigure}

%    \vspace{0.5cm} % space between figures

    \begin{subfigure}{0.7\textwidth}
        \centering
        \includegraphics[width=\textwidth]{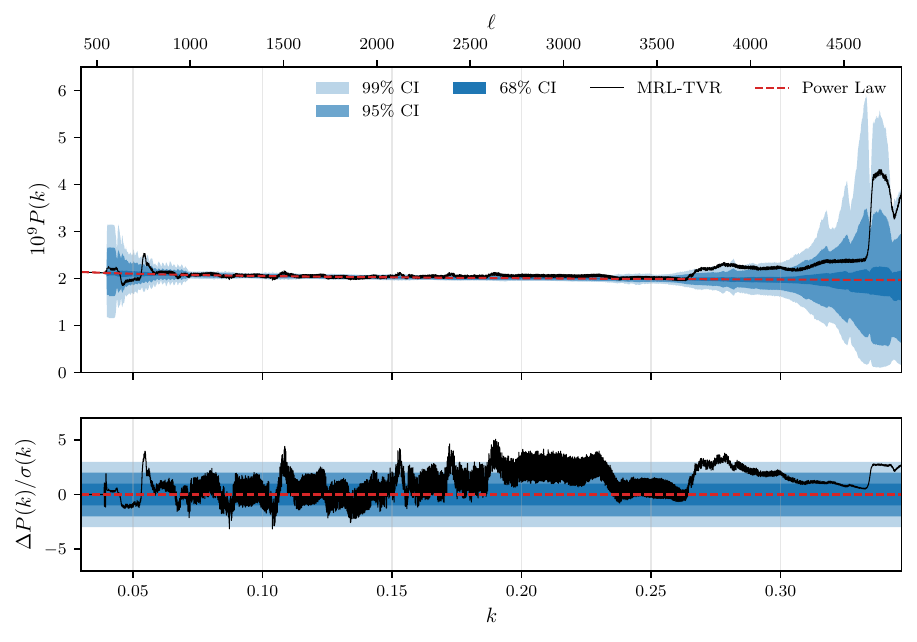}
        \caption{}
        \label{fig:COADACT_P18BF-2b}
    \end{subfigure}

    \caption{(a) Simulation based error analysis on the power spectrum reconstructed using diffusive regularization method (MRL-DR). (b) Simulation based error analysis on the power spectrum reconstructed using total variation regularization method (MRL-TVR). Results in both panels are for ACT data and Planck PR3 $\Lambda CDM$ best-fit cosmology. The range of $k$ in both panels is $[0.03 - 0.35]$.}

    \label{fig:COADACT_P18BF-2}
\end{figure}

%In the last section, we have discussed in detail about the MRL reconstructed power spectra for the ACT data and ACT best-fit background cosmology. 
This section discusses the second analysis with ACT data, which is based on the best-fit baseline $\Lambda CDM$ background cosmology inferred from Planck 2018 PR3 data. The best-fit values adopted in this analysis are listed in Table~\ref{tab:bf_cosmo_all}. We obtain these values from the MCMC chains released by the Planck collaboration\footnote{\url{https://pla.esac.esa.int/pla/\#cosmology}}. The results of this reconstruction are shown in Figs.~\ref{fig:COADACT_P18BF-1} and~\ref{fig:COADACT_P18BF-2}. Fig.~\ref{fig:COADACT_P18BF-1a} shows the MRL reconstructed power spectra for different regularization methods discussed in Section~\ref{regularization_schemes}. Depending upon the regularization methods, we get minimum chi-squares at different iterations, which can be seen from Fig.~\ref{fig:COADACT_P18BF-1b}. The reconstructed power spectra that are minimizing the $\Delta\tilde{\chi}^2_{\mathrm{tot}}$ are depicted in Fig.~\ref{fig:COADACT_P18BF-1a}. We can see from Fig.~\ref{fig:COADACT_P18BF-1a} that at $k \approx 0.20 \, \mathrm{Mpc}^{-1}$ there is a deviation from the baseline power-law power spectrum. The statistical significance of this deviation is estimated with the error-analysis, the results of which are shown in Fig.~\ref{fig:COADACT_P18BF-2}. Fig.~\ref{fig:COADACT_P18BF-2} shows that for the MRL-TVR spectrum (\ref{fig:COADACT_P18BF-2b}) the departure lies within $2-3 \sigma$, whereas for the MRL-DR spectrum (\ref{fig:COADACT_P18BF-2a}) the deviation is greater.
For ACT-DR6 best-fit background cosmology (Fig.~\ref{fig:COADACT_ACTBF-2}), we do not see such deviation.
%Such deviation is not seen in the case of ACT-DR6 best-fit background cosmology (see Fig. \ref{fig:COADACT_ACTBF-2}),
Thus, it becomes a diagnostic test to probe inconsistency between two different datasets. This shows that ACT-DR6 data are not consistent with Planck 2018 PR3 background cosmology. In Section~\ref{sec:result_par}, we constrain this disagreement between ACT and Planck by conducting a Bayesian analysis with a double-tilt power-law power spectrum. %It would allow us to measure the presence of any possible departure from the single power-law behaviour of the primordial power spectrum in terms of change in the spectral tilt and allow us to quantify the inconsistency between the ACT-DR6 and Planck 2018 PR3 datasets. 

\subsubsection{Case III: Planck PR4 background cosmology}\label{subsubsec:case3_act}
\begin{figure}
    \centering

    \begin{subfigure}{0.7\textwidth}
        \centering
        \includegraphics[width=\textwidth]{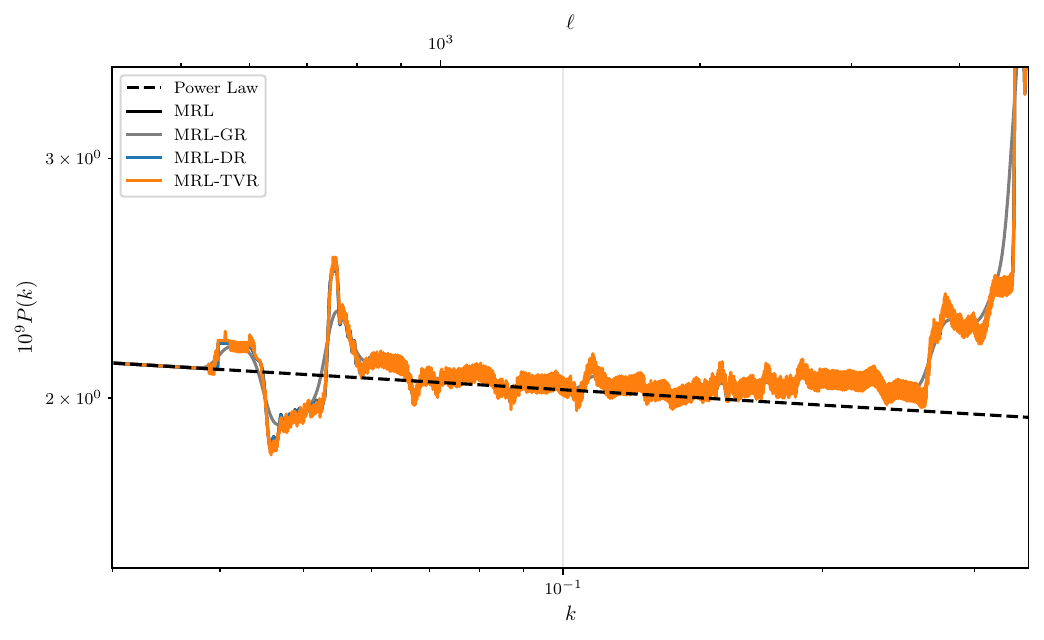}
        \caption{}
        \label{fig:COADACT_CamsBF-1a}
    \end{subfigure}

%    \vspace{0.5cm} % space between figures

    \begin{subfigure}{0.7\textwidth}
        \centering
        \includegraphics[width=\textwidth]{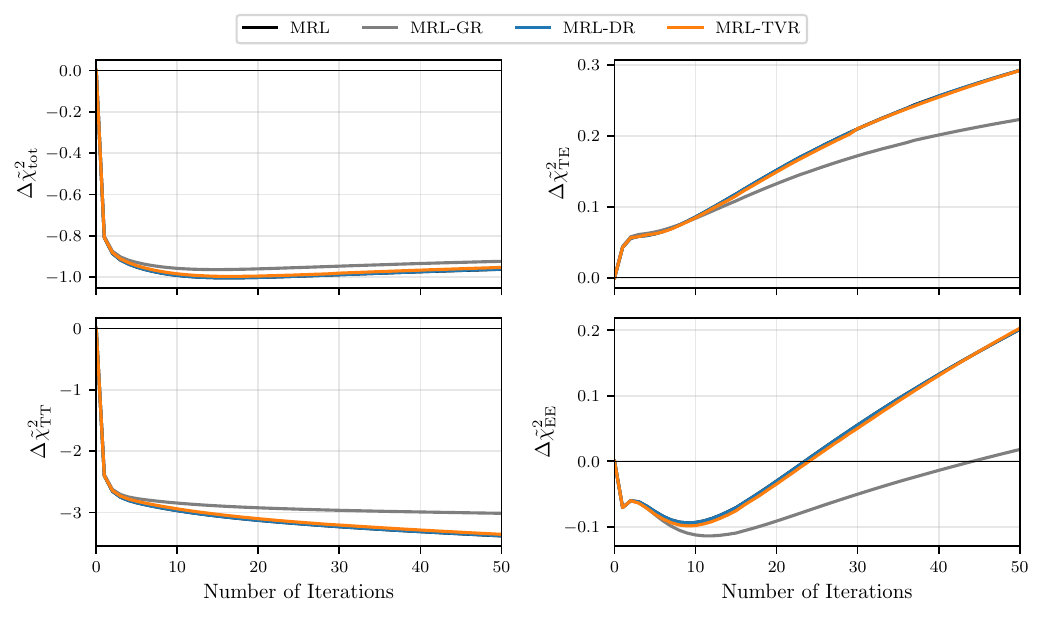}
        \caption{}
        \label{fig:COADACT_CamsBF-1b}
    \end{subfigure}

    \caption{(a) Primordial power spectra that minimize the total chi-square reconstructed employing the MRL algorithm with different regularization methods. The black dashed line represents the nearly scale-invariant power-law power spectrum. The range of $k$ in this panel is $[0.03 - 0.35]$. (b) Improvements obtained from reconstructed PPS in the chi-squared fit to data with every iteration for combined, TT, TE, and EE spectra. These results have been obtained from ACT data based on Planck PR4 CamSpec-NPIPE best-fit $\Lambda CDM$ background cosmology.}
    \label{fig:COADACT_CamsBF-1}
\end{figure}

\begin{figure}
    \centering

    \begin{subfigure}{0.7\textwidth}
        \centering
        \includegraphics[width=\textwidth]{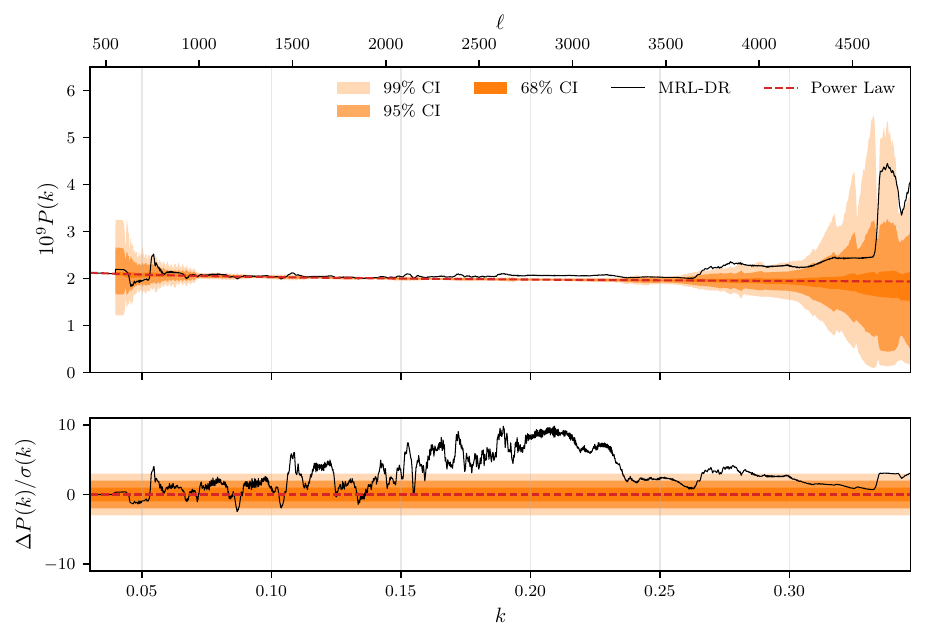}
        \caption{}
        \label{fig:COADACT_CamsBF-2a}
    \end{subfigure}

%    \vspace{0.5cm} % space between figures

    \begin{subfigure}{0.7\textwidth}
        \centering
        \includegraphics[width=\textwidth]{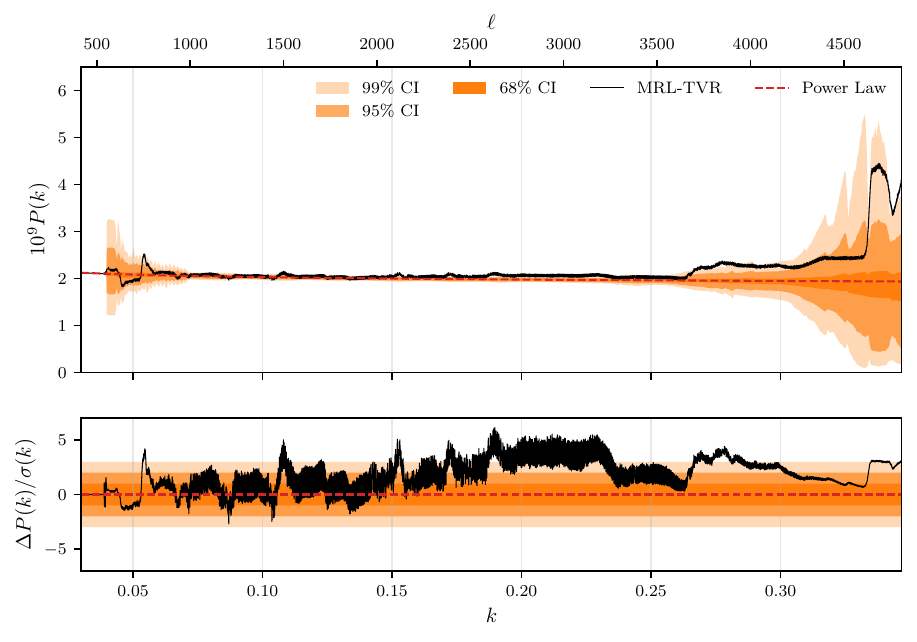}
        \caption{}
        \label{fig:COADACT_CamsBF-2b}
    \end{subfigure}

    \caption{(a) Simulation based error analysis on the power spectrum reconstructed using diffusive regularization method (MRL-DR). (b) Simulation based error analysis on the power spectrum reconstructed using total variation regularization method (MRL-TVR). Results in both panels are for ACT data and Planck PR4 CamSpec-NPIPE $\Lambda CDM$ best-fit cosmology. The range of $k$ in both panels is $[0.03 - 0.35]$.}
    \label{fig:COADACT_CamsBF-2}
\end{figure}

This section presents the reconstruction results from our third analysis with ACT-DR6 data. In this analysis, we assume the best-fit baseline $\Lambda CDM$ cosmology obtained from CamSpec-NPIPE Planck PR4 data as our background cosmology. The best-fit values of the six baseline $\Lambda CDM$ parameters are listed in Table~\ref{tab:bf_cosmo_all}. Our findings from this analysis are exhibited in Figs.~\ref{fig:COADACT_CamsBF-1} and~\ref{fig:COADACT_CamsBF-2}. The $\mathcal{P}_k$s we obtain from the MRL reconstruction are shown in Fig.~\ref{fig:COADACT_CamsBF-1a}. Here we exhibit $\mathcal{P}_k$s correspond to the iterations for which $\Delta\tilde{\chi}^2_{\mathrm{tot}}$s become minimum. %And different regularisation methods take different iteration steps to minimize the $\Delta\tilde{\chi}^2_{\mathrm{tot}}$. Here, we have done the same thing as discussed for Fig. \ref{fig:COADACT_ACTBF-2}, the only difference is that here we have considered the best-fit model from CamSpec data tabulated in Table \ref{tab:camspec_bf_cosmo} as our background cosmological model.
%As we find a bump in the reconstructed PPS around $k=0.20 \, \mathrm{Mpc}^{-1}$ (see Fig. \ref{fig:COADACT_P18BF-1a}) while reconstructing with the ACT-DR6 data and Planck 2018 PR3 background cosmology (refer to section \ref{subsubsec:case2_act}), a similar bump has been found when reconstructing with the ACT-DR6 data and CamSpec background cosmology (see Fig. \ref{fig:COADACT_CamsBF-1a}).
The ACT reconstruction under the CamSpec background cosmology contains a bump around $k=0.20 \, \mathrm{Mpc}^{-1}$ (Fig.~\ref{fig:COADACT_CamsBF-1a}), similar to that obtained under the Planck 2018 PR3 background (Fig.~\ref{fig:COADACT_P18BF-1a}). We estimate the statistical significance of this bump in Fig.~\ref{fig:COADACT_CamsBF-2}. Fig.~\ref{fig:COADACT_CamsBF-2a} shows a significant deviation from the power-law behaviour over $k \approx 0.15-0.24 \, \mathrm{Mpc}^{-1}$ for the MRL-DR reconstructed $\mathcal{P}_k$. However, this deviation is little less prominent in the MRL-TVR $\mathcal{P}_k$ because of the difference in regularization strategy. This deviation indicates towards an inconsistency between Planck PR4 CamSpec-NPIPE and ACT-DR6 datasets. It is evident from Figs. \ref{fig:COADACT_P18BF-2} and \ref{fig:COADACT_CamsBF-2} that the amount of deviation is more for the Planck PR4 CamSpec-NPIPE background than the Planck 2018 PR3 background. %This feature may cause the combined dataset of ACT and Planck PR4 to support a little higher value for spectral index compared to Planck-only or ACT-only scenario. In order to investigate this feature as well, we conduct another robust Bayesian analysis with a parametric form of the primordial power spectrum with two tilts (refer to Eq. \ref{eq:double_tilt_pk}) in the next section to quantify this observed inconsistency between the ACT-DR6 and Planck PR4 CamSpec-NPIPE datasets. 
We also explore this fixed-background diagnostic with robust Bayesian analyses in Section~\ref{sec:result_par}.

\subsection{Reconstruction from SPT data}\label{subsec:rec_spt}    
In this section, we present and discuss the results of the MRL reconstructions using SPT-3G D1 data. All results related to this reconstruction are shown in Figs.~\ref{fig:SPT_SPTBF-1}-\ref{fig:SPT_CamsBF-2}. For SPT also we consider the following three different background cosmologies: SPT, Planck 2018 PR3, and Planck PR4 CamSpec-NPIPE.
% \begin{itemize}
%     \item We first consider the baseline best-fit cosmology (see Table \ref{tab:bf_cosmo_all}) obtained from SPT-3G D1 data and based on this background cosmology we reconstruct PPS from SPT-3G D1 band power spectra data.
%     \item Next, we reconstruct PPS assuming the Planck PR3 baseline best-fit cosmology (refer to Table \ref{tab:bf_cosmo_all}) as the background cosmology.
%     \item Lastly, we perform the MRL reconstruction considering the best-fit baseline cosmology obtained from Planck PR4 CamSpec-NPIPE data as our background cosmology (shown in Table \ref{tab:bf_cosmo_all}).
% \end{itemize}
%For SPT data as well, we adopt the same values for the reconstruction tuning parameters, $\Sigma_{\mathrm{GR}}=0.03$, $\kappa_{\mathrm{DR}}=0.01$, and $\kappa_{\mathrm{TVR}}=0.008$ as assumed in the case of reconstruction with ACT data.
Now, let us analyse reconstruction results obtained for different background cosmologies one by one subsequently. %It is important to mention here that the results we found for SPT data have been obtained using the same methodology as adopted in case of ACT data discussed in section \ref{subsubsec:case1_act}. As in, we have adopted the same methods to conduct the error analysis and to compute the reduced chi-squared improvement. 

\subsubsection{Case I: SPT background cosmology}\label{subsubsec:case1_spt}
\begin{figure}
    \centering

    \begin{subfigure}{0.7\textwidth}
        \centering
        \includegraphics[width=\textwidth]{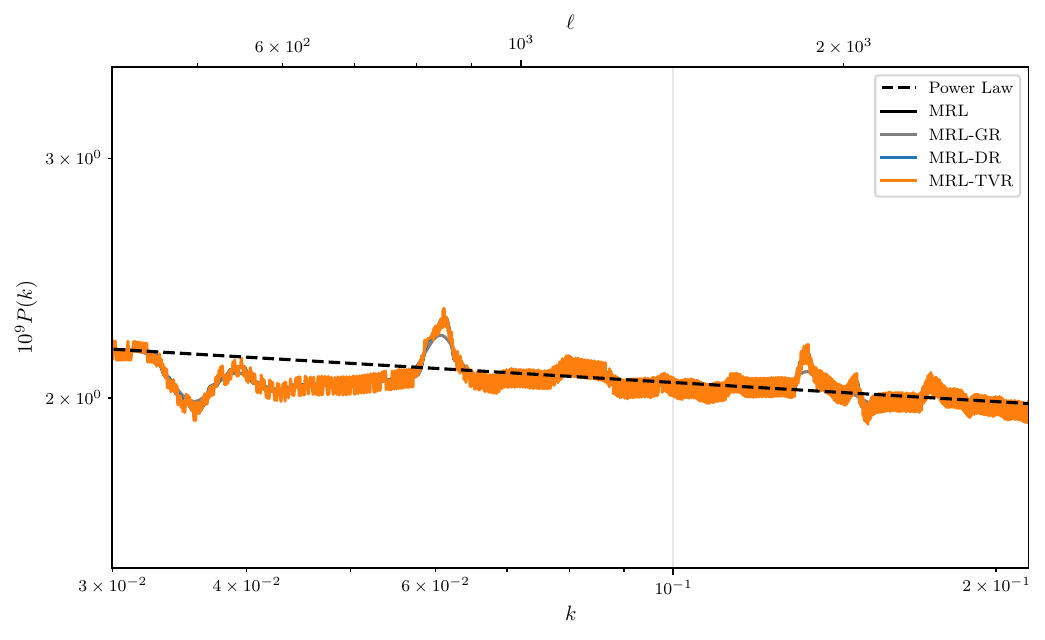}
        \caption{}
        \label{fig:SPT_SPTBF-1a}
    \end{subfigure}

%    \vspace{0.5cm} % space between figures

    \begin{subfigure}{0.7\textwidth}
        \centering
        \includegraphics[width=\textwidth]{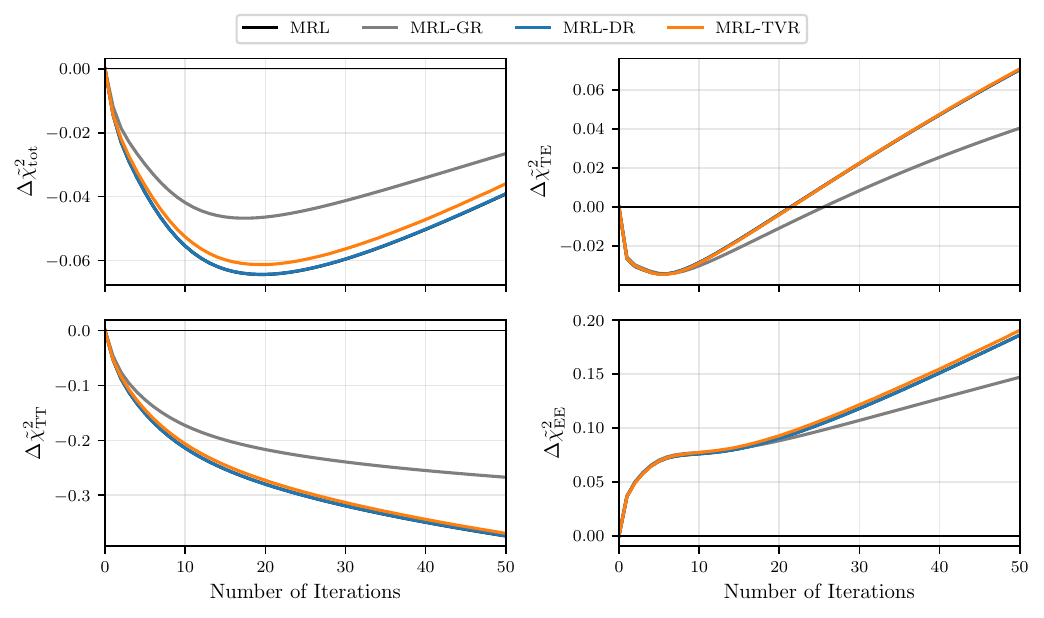}
        \caption{}
        \label{fig:SPT_SPTBF-1b}
    \end{subfigure}

    \caption{(a) Primordial power spectra that minimize the total chi-square reconstructed employing the MRL algorithm with different regularization methods. The black dashed line represents the nearly scale-invariant power-law power spectrum. The range of $k$ in this panel is $[0.03 - 0.214]$. (b) Improvements obtained from reconstructed PPS in the chi-squared fit to data with every iteration for combined, TT, TE, and EE spectra. These results have been obtained from SPT data based on SPT best-fit $\Lambda CDM$ background cosmology.}    \label{fig:SPT_SPTBF-1}
\end{figure}

\begin{figure}
    \centering

    \begin{subfigure}{0.7\textwidth}
        \centering
        \includegraphics[width=\textwidth]{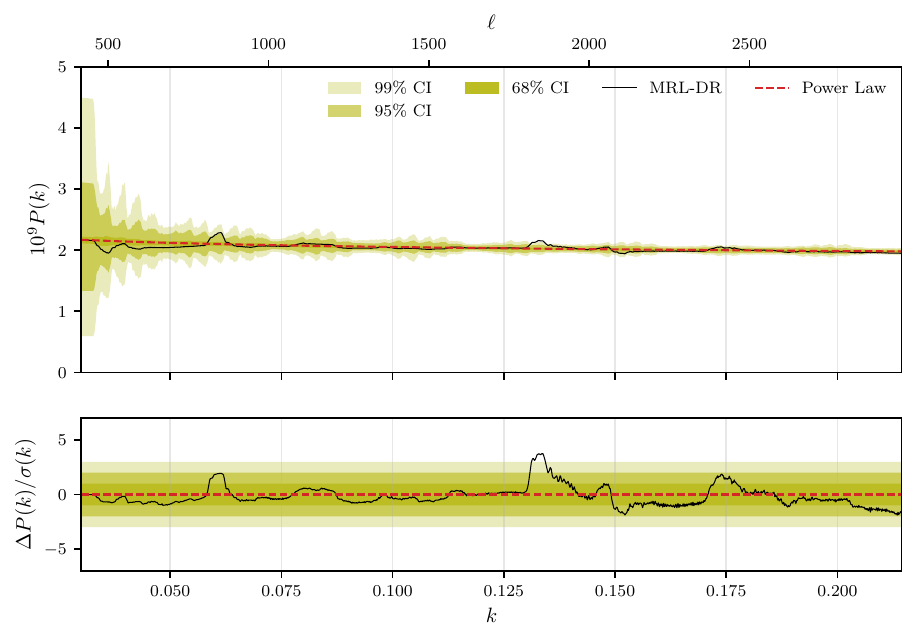}
        \caption{}
        \label{fig:SPT_SPTBF-2a}
    \end{subfigure}

%    \vspace{0.5cm} % space between figures

    \begin{subfigure}{0.7\textwidth}
        \centering
        \includegraphics[width=\textwidth]{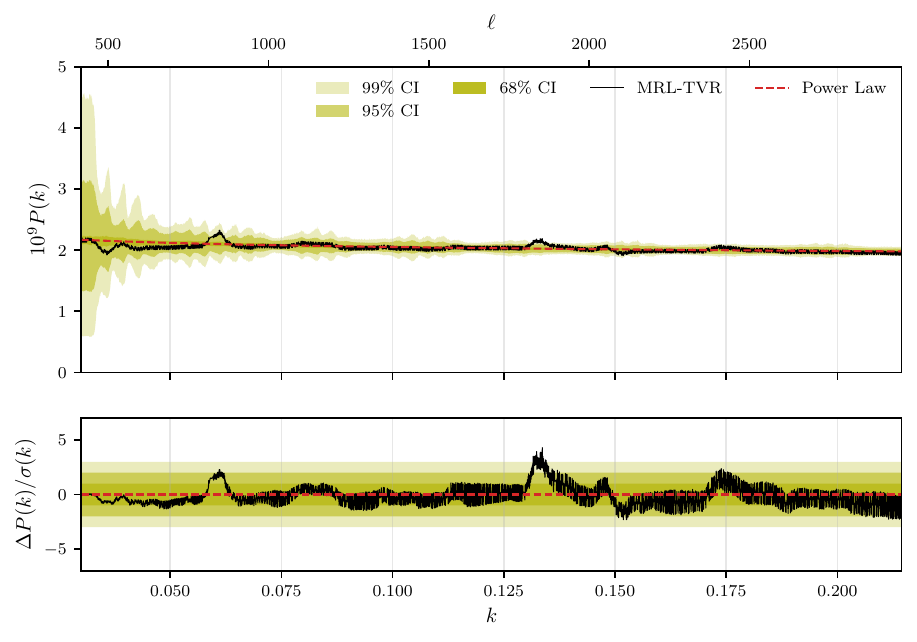}
        \caption{}
        \label{fig:SPT_SPTBF-2b}
    \end{subfigure}

    \caption{(a) Simulation based error analysis on the power spectrum reconstructed using diffusive regularization method (MRL-DR). (b) Simulation based error analysis on the power spectrum reconstructed using total variation regularization method (MRL-TVR). Results in both panels are for SPT data and SPT $\Lambda CDM$ best-fit cosmology. The range of $k$ in both panels is $[0.03 - 0.214]$.}
    \label{fig:SPT_SPTBF-2}
\end{figure}

In \texttt{Case I}, we discuss the results of the MRL reconstructions from SPT-3G D1 data assuming the six-parameter baseline best-fit cosmology obtained from SPT-3G D1 data as the background cosmology. The adopted best-fit values for this reconstruction are tabulated in Table~\ref{tab:bf_cosmo_all}. We obtain these best-fit values for the six cosmological parameters by minimizing the full posterior using the $\texttt{BOBYQA}$~\cite{Cartis_2021,cartis2018improvingflexibilityrobustnessmodelbased,Powell2009TheBA} method. We use the publicly available $\texttt{CAMB}$~\cite{Lewis:1999bs}\footnote{\url{https://github.com/cmbant/CAMB}} cosmology code with $\texttt{Cobaya}$~\cite{Torrado:2020dgo}\footnote{\url{https://github.com/CobayaSampler/cobaya}} sampler to find the best-fit values from the SPT-3G D1 data \footnote{\url{https://github.com/SouthPoleTelescope/spt_candl_data}}. %Here, we consider the baseline power-law primordial power spectrum while evaluating the best-fit cosmology.
We show all four MRL reconstructed $\mathcal{P}_k$s corresponding to different smoothing methods in Fig.~\ref{fig:SPT_SPTBF-1a}. The improvement in the reduced chi-square with every iteration for each method are depicted in Fig.~\ref{fig:SPT_SPTBF-1b}. %where we found that for TT data $\tilde{\chi}^2_{\mathrm{TT}}$ is improving all the way up to $50^{\mathrm{th}}$ iteration. However, in case of TE data $\tilde{\chi}^2_{\mathrm{TE}}$ becomes minimum at $5^{\mathrm{th}}$, $6^{\mathrm{th}}$, $5^{\mathrm{th}}$, and $5^{\mathrm{th}}$ iterations for methods MRL, MRL-GR, MRL-DR, and MRL-TVR, respectively. Similarly, for EE data, we have minimum $\tilde{\chi}^2_{\mathrm{EE}}$ at $1^{\mathrm{st}}$, $1^{\mathrm{st}}$, $1^{\mathrm{st}}$, and $1^{\mathrm{st}}$ iterations for MRL, MRL-GR, MRL-DR, and MRL-TVR methods, respectively. However, for EE data we are not having any improvements in reduced chi-square. The total $\tilde{\chi}^2_{\mathrm{tot}}$ becomes minimum at iterations $20$, $17$, $20$, and $20$ for MRL, MRL-GR, MRL-DR, and MRL-TVR methods, respectively. The achieved maximum improvements in total $\tilde{\chi}^2_{\mathrm{tot}}$ from methods MRL, MRL-GR, MRL-DR, and MRL-TVR are $-0.06437$, $-0.04675$, $-0.06435$, and $-0.06126$, respectively.
We perform the error analysis for this setup, the results of which are exhibited in Fig.~\ref{fig:SPT_SPTBF-2}. %, where we show results for two smoothing methods MRL-DR (\ref{fig:SPT_SPTBF-2a}) and MRL-TVR (\ref{fig:SPT_SPTBF-2b}).
The error analysis shows that the reconstructed power spectra are completely consistent with the power-law primordial power spectrum and we do not find any statistical evidence of the presence of any primordial features except at $k \approx 0.13 \, \mathrm{Mpc}^{-1}$. All the features that we observe for ACT data (Fig. \ref{fig:COADACT_ACTBF-2}) are absent in the PPS reconstructed from SPT data, and the feature at $k \approx 0.13 \, \mathrm{Mpc}^{-1}$ (Fig. \ref{fig:SPT_SPTBF-2}) found for SPT data is not present in the reconstructed PPS from ACT data. 

\subsubsection{Case II: Planck PR3 background cosmology}\label{subsubsec:case2_spt}
\begin{figure}
    \centering

    \begin{subfigure}{0.7\textwidth}
        \centering
        \includegraphics[width=\textwidth]{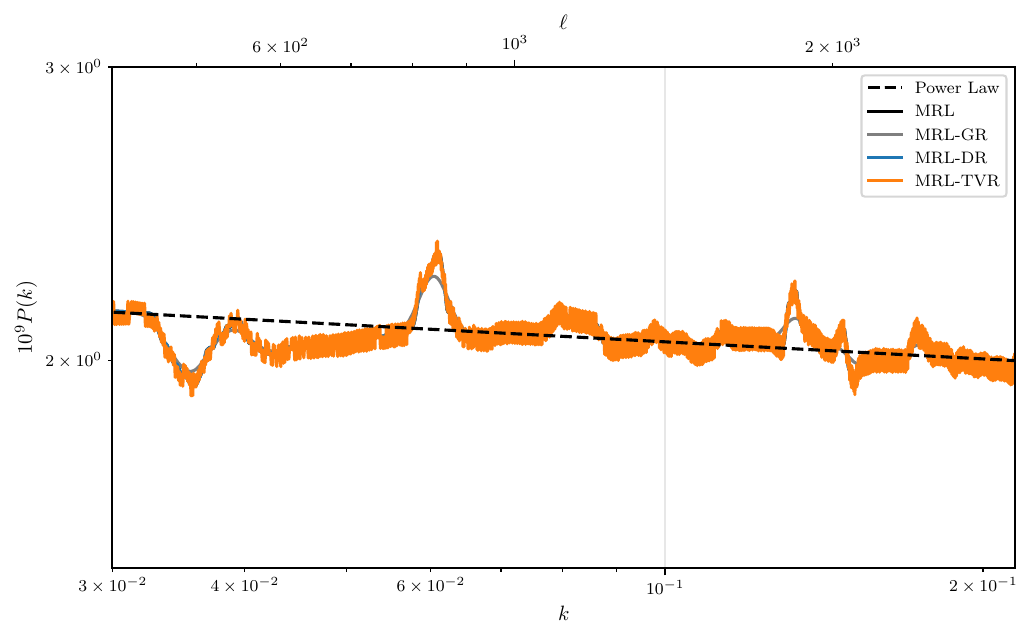}
        \caption{}
        \label{fig:SPT_P18BF-1a}
    \end{subfigure}

%    \vspace{0.5cm} % space between figures

    \begin{subfigure}{0.7\textwidth}
        \centering
        \includegraphics[width=\textwidth]{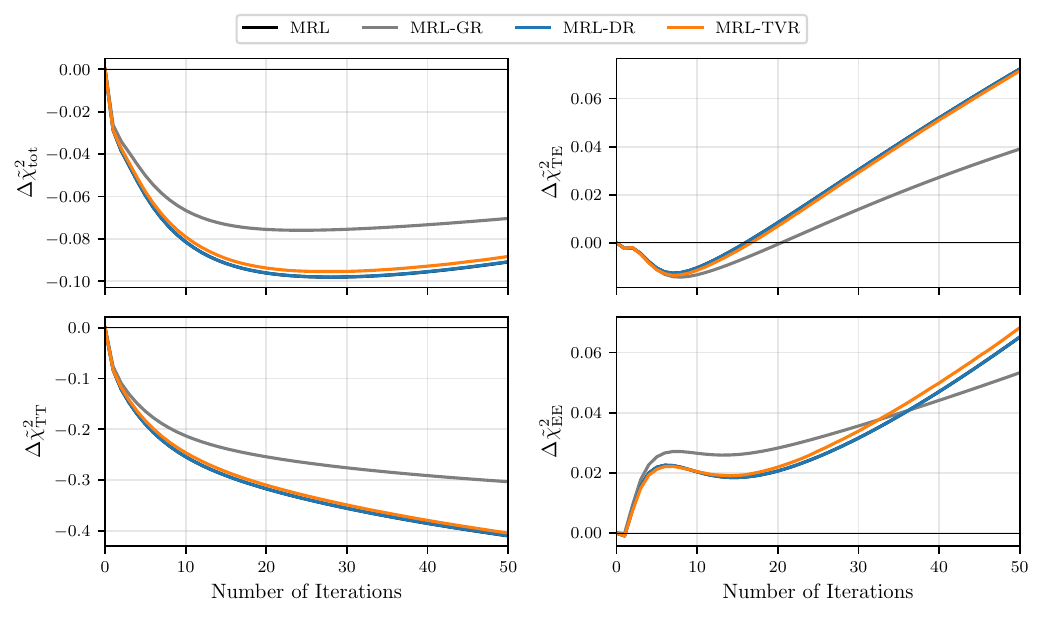}
        \caption{}
        \label{fig:SPT_P18BF-1b}
    \end{subfigure}

    \caption{(a) Primordial power spectra that minimize the total chi-square reconstructed employing the MRL algorithm with different regularization methods. The black dashed line represents the nearly scale-invariant power-law power spectrum. The range of $k$ in this panel is $[0.03 - 0.214]$. (b) Improvements obtained from reconstructed PPS in the chi-squared fit to data with every iteration for combined, TT, TE, and EE spectra. These results have been obtained from SPT data based on Planck PR3 best-fit $\Lambda CDM$ background cosmology.}
    \label{fig:SPT_P18BF-1}
\end{figure}

\begin{figure}
    \centering

    \begin{subfigure}{0.7\textwidth}
        \centering
        \includegraphics[width=\textwidth]{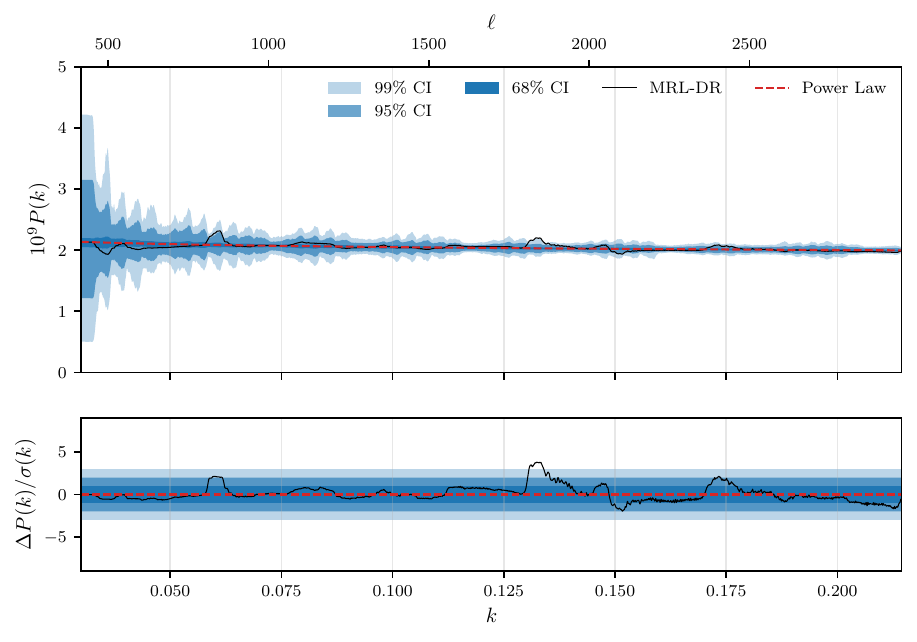}
        \caption{}
        \label{fig:SPT_P18BF-2a}
    \end{subfigure}

%    \vspace{0.5cm} % space between figures

    \begin{subfigure}{0.7\textwidth}
        \centering
        \includegraphics[width=\textwidth]{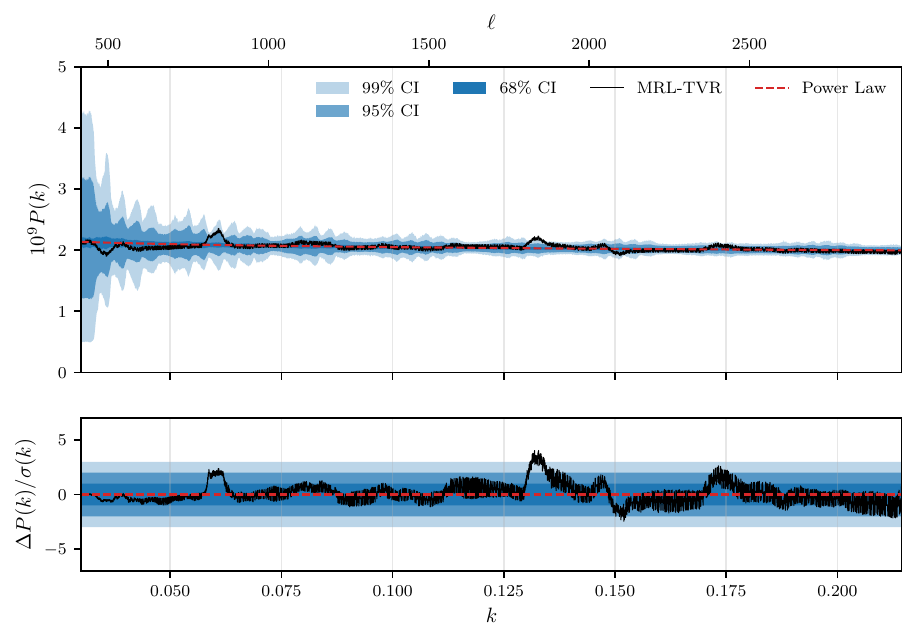}
        \caption{}
        \label{fig:SPT_P18BF-2b}
    \end{subfigure}

    \caption{(a) Simulation based error analysis on the power spectrum reconstructed using diffusive regularization method (MRL-DR). (b) Simulation based error analysis on the power spectrum reconstructed using total variation regularization method (MRL-TVR). Results in both panels are for SPT data and Planck PR3 $\Lambda CDM$ best-fit cosmology. The range of $k$ in both panels is $[0.03 - 0.214]$.}
    \label{fig:SPT_P18BF-2}
\end{figure}

In this section, we discuss the $\mathcal{P}_k$s reconstructed from SPT-3G D1 data for the Planck 2018 PR3 best-fit baseline cosmology (Table \ref{tab:bf_cosmo_all}). Figs. \ref{fig:SPT_P18BF-1a} and \ref{fig:SPT_P18BF-1b}, show the reconstructed power spectra and improvements in the fit statistic, respectively. 
%Our main motive in conducting this analysis is to make a consistency check between the Planck PR3 data and SPT-3G D1 data as we have done with the Planck PR3 data and ACT-DR6 data in section \ref{subsubsec:case2_act}. We intend to see whether the best-fit cosmology obtained from the Planck 2018 PR3 data is in agreement with the SPT data or not which we can be better understood from the error analysis.
The obtained plots from the error analysis performed for the MRL-DR and MRL-TVR power spectra are presented in Figs. \ref{fig:SPT_P18BF-2a} and \ref{fig:SPT_P18BF-2b}, which reveal that in contrast to ACT data (Section \ref{subsubsec:case2_act}), for SPT data, we do not find any such inconsistency between the Planck 2018 PR3 background and SPT-3G D1 data. They are in complete agreement with each other.   

\subsubsection{Case III: Planck PR4 background cosmology}\label{subsubsec:case3_spt}
\begin{figure}
    \centering

    \begin{subfigure}{0.7\textwidth}
        \centering
        \includegraphics[width=\textwidth]{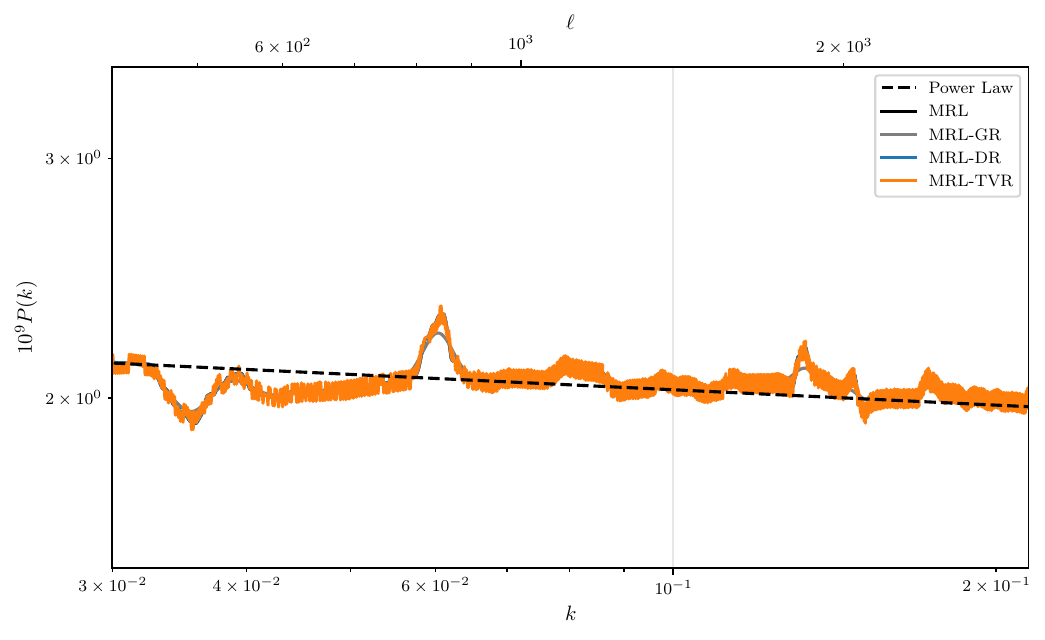}
        \caption{}
        \label{fig:SPT_CamsBF-1a}
    \end{subfigure}

%    \vspace{0.5cm} % space between figures

    \begin{subfigure}{0.7\textwidth}
        \centering
        \includegraphics[width=\textwidth]{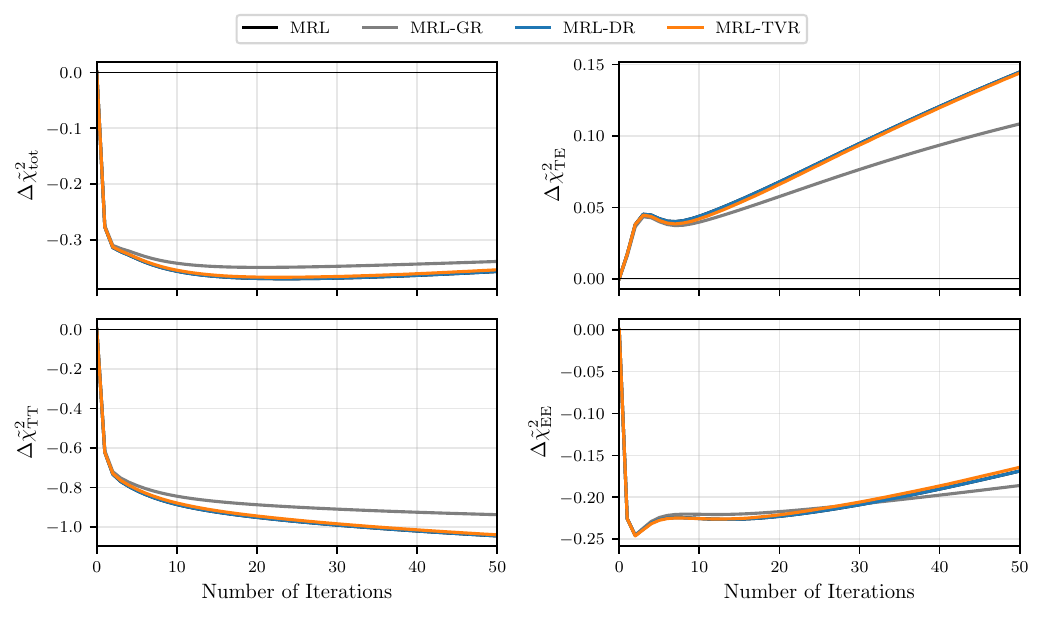}
        \caption{}
        \label{fig:SPT_CamsBF-1b}
    \end{subfigure}

    \caption{(a) Primordial power spectra that minimize the total chi-square reconstructed employing the MRL algorithm with different regularization methods. The black dashed line represents the nearly scale-invariant power-law power spectrum. The range of $k$ in this panel is $[0.03 - 0.214]$. (b) Improvements obtained from reconstructed PPS in the chi-squared fit to data with every iteration for combined, TT, TE, and EE spectra. These results have been obtained from SPT data based on Planck PR4 CamSpec-NPIPE best-fit $\Lambda CDM$ background cosmology.}
    \label{fig:SPT_CamsBF-1}
\end{figure}

\begin{figure}
    \centering

    \begin{subfigure}{0.7\textwidth}
        \centering
        \includegraphics[width=\textwidth]{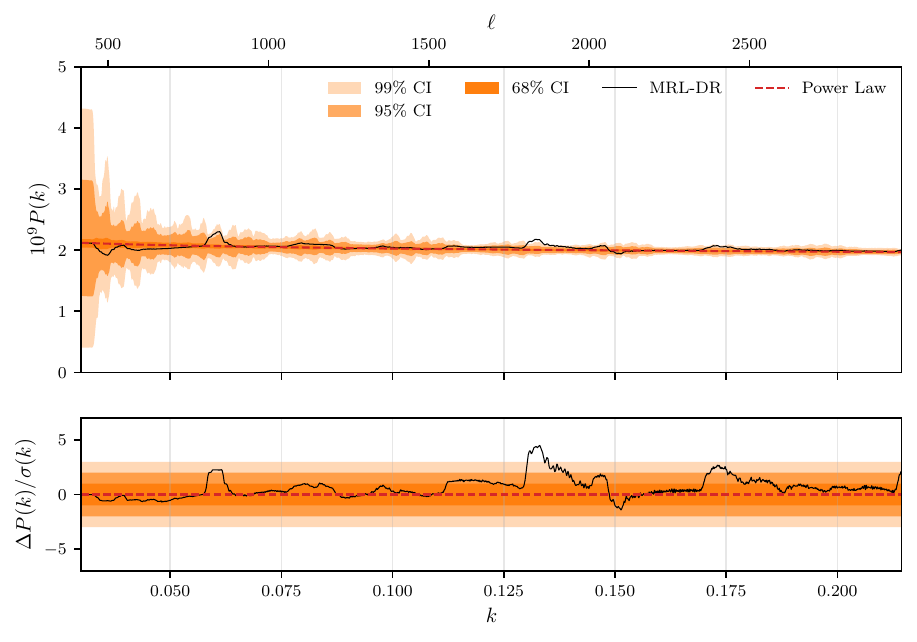}
        \caption{}
        \label{fig:SPT_CamsBF-2a}
    \end{subfigure}

%    \vspace{0.5cm} % space between figures

    \begin{subfigure}{0.7\textwidth}
        \centering
        \includegraphics[width=\textwidth]{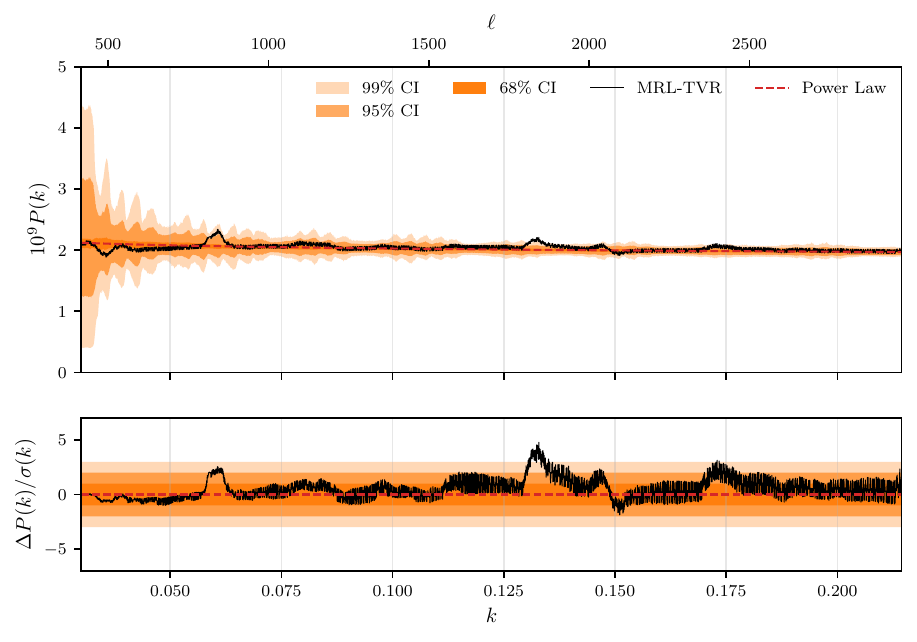}
        \caption{}
        \label{fig:SPT_CamsBF-2b}
    \end{subfigure}

    \caption{(a) Simulation based error analysis on the power spectrum reconstructed using diffusive regularization method (MRL-DR). (b) Simulation based error analysis on the power spectrum reconstructed using total variation regularization method (MRL-TVR). Results in both panels are for SPT data and Planck PR4 CamSpec-NPIPE $\Lambda CDM$ best-fit cosmology. The range of $k$ in both panels is $[0.03 - 0.214]$.}

    \label{fig:SPT_CamsBF-2}
\end{figure}

This third analysis is based on SPT-3G D1 data and the best-fit cosmology obtained from the Planck PR4 CamSpec-NPIPE data release (Table \ref{tab:bf_cosmo_all}). We present all our results of this analysis in Figs.~\ref{fig:SPT_CamsBF-1} and~\ref{fig:SPT_CamsBF-2}. % The purpose of this analysis is to test if there exists any tension between SPT-3G D1 and Planck PR4 CamSpec-NPIPE data. In reconstruction with ACT-DR6 data, we find that for the Planck PR4 CamSpec-NPIPE background cosmology reconstructions (\ref{fig:COADACT_CamsBF-2}) shows more departure from the single-tilt power-law shape than for the Planck PR3 background (\ref{fig:COADACT_P18BF-2}). Here also we see same pattern in the reconstructed PPS, for the Planck PR4 CamSpec-NPIPE background cosmology the power spectra~(\ref{fig:SPT_CamsBF-1a}) behave slightly more blue tilted than the power spectra reconstructed under the Planck PR3 background~(\ref{fig:SPT_P18BF-1a}). However, unlike ACT data for SPT these power spectra are well within the uncertainties as we can see from Fig. \ref{fig:SPT_CamsBF-2}.
Same as the Planck PR3 background (Section \ref{subsubsec:case2_spt}), here also SPT-3G D1 data does not exhibit any inconsistent behaviour when used with the Planck PR4 background cosmology. From Fig. \ref{fig:SPT_CamsBF-2}, we find that for the Planck PR4 background the reconstructed $\mathcal{P}_k$s from SPT-3G D1 data corroborate the power-law behaviour of the primordial power spectrum.  

\subsection{Comparison of ACT, SPT and CamSpec datasets}\label{subsec:compare_act_spt_cam}
\begin{figure}
    \centering

    \begin{subfigure}{0.7\textwidth}
        \centering
        \includegraphics[width=\textwidth]{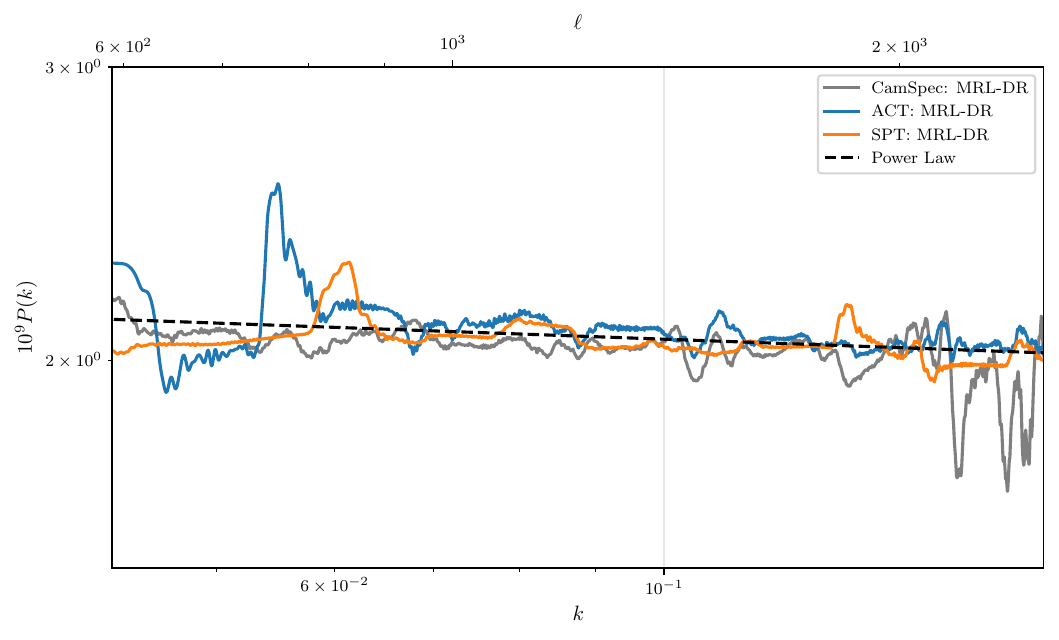}
        \caption{}
        \label{fig:comparison-a}
    \end{subfigure}

%    \vspace{0.5cm} % space between figures

    \begin{subfigure}{0.7\textwidth}
        \centering
        \includegraphics[width=\textwidth]{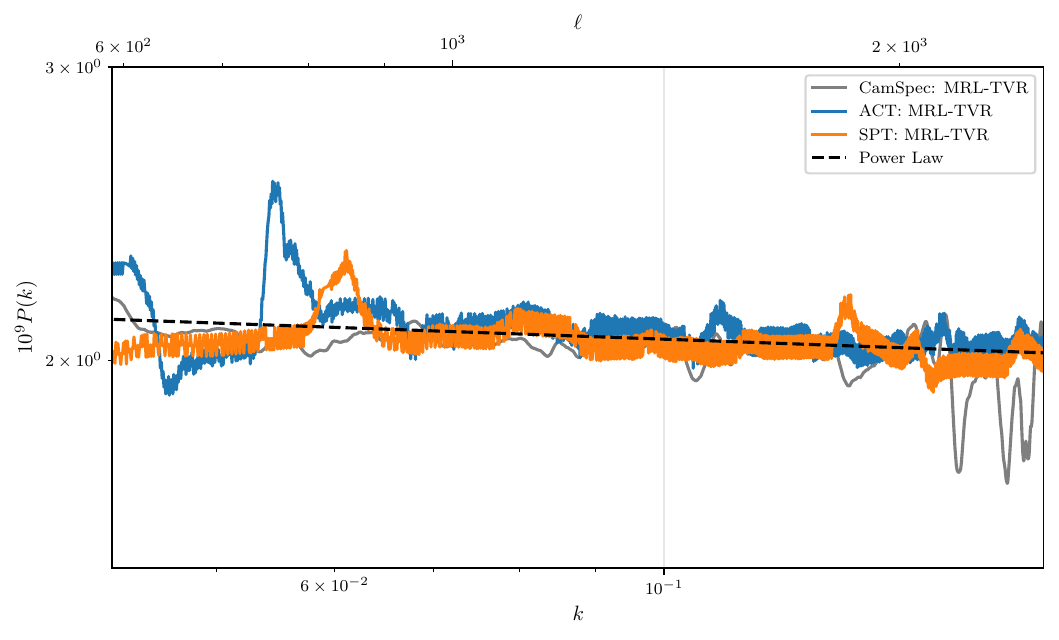}
        \caption{}
        \label{fig:comparison-b}
    \end{subfigure}

    \caption{(a) Comparison of power spectra reconstructed using diffusive regularization method (MRL-DR). (b) Comparison of power spectra reconstructed using total variation regularization method (MRL-TVR). These are the same MRL-DR and MRL-TVR power spectra shown in Figs. \ref{fig:COADACT_ACTBF-1a}, \ref{fig:SPT_SPTBF-1a}, and \ref{fig:Cams_CamsBF-1a} reconstructed from ACT, SPT, and CamSpec datasets, respectively, plotted for their overlapping multipole range $593-2500$. The black dashed line represents the nearly scale-invariant power-law power spectrum. To plot this power-law power spectrum, we consider the mean values of $\ln(10^{10}A_{\mathrm{s}})$ and $n_s$ reported by article~\cite{SPT-3G:2025bzu} in "TABLE I" estimated by considering SPT, ACT, and Planck data together (CMB-SPA).}
    \label{fig:comparison}
\end{figure}

In this section, we discuss the comparison of MRL-DR and MRL-TVR power spectra reconstructed from CamSpec, ACT and SPT data in Sections \ref{subsec:rec_cam}, \ref{subsubsec:case1_act}, and \ref{subsubsec:case1_spt}, respectively. We show the comparison in Fig. \ref{fig:comparison}. These are the same MRL-DR and MRL-TVR power spectra which are shown in Figs. \ref{fig:Cams_CamsBF-1a}, \ref{fig:COADACT_ACTBF-1a}, and \ref{fig:SPT_SPTBF-1a}, plotted for the common multipole range $\ell = 593-2500$. Since, the CamSpec reconstruction (Figs. \ref{fig:Cams_CamsBF-1a}) shows highly oscillatory behaviour, we use the Savitzky-Golay filter to smooth the spectrum and highlight the main features in the spectrum. This comparison allows us to probe any features that are common to these datasets even if they are statistically insignificant. The MRL algorithm has a strength that it allows us to reconstruct from each data point. As a result it picks all plausible features that are present in the data even if that are not statistically favoured by a full parameter estimation analysis as 'look-elsewhere effect' diminishes the global significance of detecting such features. Thus, if some features are repeatedly appearing in different individual observations then they may carry some physical importance apart from being only statistical signals or reconstruction artifacts since they are exhibiting cross-dataset consistency. We notice from Fig. \ref{fig:comparison} that certain features are present for all three datasets, while some are common to ACT and Planck at different scales. We also found that the reconstructed spectra from ACT and Planck closely follow each other, however SPT reconstructions do not exhibit such trend. Rather for SPT data we find at certain scales a completely out of phase behaviour with respect to either ACT or Planck. To measure these behaviours, we conduct the Pearson and Spearman correlation analyses with the reconstructed power spectra from CamSpec, ACT, and SPT datasets. The obtained results from this correlation analysis are tabulated in Table \ref{tab:correlation_results}. In this analysis, we consider two different $k$ ranges, full $0.043\leq k\leq0.180~{\rm Mpc}^{-1}$ and truncated $0.07\leq k\leq0.14~{\rm Mpc}^{-1}$. In this analysis, we consider the weighted correlation, and to compute the weight, we use our simulated power spectra from CamSpec, ACT, and SPT datasets which are shown in Figs. \ref{fig:Cams_CamsBF-2}, \ref{fig:COADACT_ACTBF-2}, and \ref{fig:SPT_SPTBF-2}. From the simulated power spectra, we have the point-error ($\sigma(k_i)$),  
\begin{equation}
\sigma(k_i) = \sqrt{\frac{1}{N_{\mathrm{sim}}-1}\sum_{j=1}^{N_{\mathrm{sim}}}\left[\mathcal{P_{\mathrm{sim}}}^{(j)}(k_i)-\mathcal{\bar{P}}_{\mathrm{sim}}(k_i)\right]^2},
\label{eq:point_error}
\end{equation}
where,
\[
\mathcal{\bar{P}}_{\mathrm{sim}}(k_i) = \frac{1}{N_{\mathrm{sim}}} \sum_{j=1}^{N_{\mathrm{sim}}} \mathcal{P_{\mathrm{sim}}}^{(j)}(k_i).
\]
The number of simulations $N_{\mathrm{sim}}= 1000$ in this analysis. We use the following definition for the weighted Pearson correlation,
\begin{equation}
r_w = \frac{\sum_i w_i (X_i-\bar{X}_w) (Y_i-\bar{Y}_w)}{\sqrt{\left[\sum_i w_i(X_i-\bar{X}_w)^2\right]
\left[\sum_i w_i(Y_i-\bar{Y}_w)^2\right]}},
\label{eq:weighted_pearson}
\end{equation}
where,
\[
\sigma_{XY}^2(k_i)
=
\sigma_X^2(k_i)
+
\sigma_Y^2(k_i),
\qquad
w_i=\frac{1}{\sigma_{XY}^2(k_i)},
\qquad
\bar{X}_w=\frac{\sum_i w_i X_i}{\sum_i w_i},
\qquad
\bar{Y}_w=\frac{\sum_i w_i Y_i}{\sum_i w_i}.
\]
Here, $X$ and $Y$ denote the CamSpec, ACT, SPT.
To evaluate Spearman's rank correlation we use weighted Pearson correlation to ranks. We evaluate correlation for both power spectra, reconstructed from the observed data as well as for power spectra reconstructed from the power-law based simulated $C_{\ell}$s (Section \ref{subsec:stat_sig}). From the correlation coefficients computed for the simulated $\mathcal{P_{\mathrm{sim}}}$s, we compute the 16th, 50th, and 84th percentiles, and then compare them with the correlation coefficients obtained for power spectra reconstructed from the observed datasets. And also compute the $p$ values, 
\[
p_{\mathrm{lower}} = \frac{1+N_{\mathrm{lower}}} {N_{\mathrm{sim}}+1},
\qquad
p_{\mathrm{upper}} = \frac{1+N_{\mathrm{upper}}}{N_{\mathrm{sim}}+1},
\]
here, $N_{\mathrm{lower}}$ and $N_{\mathrm{upper}}$ are denoting the numbers of simulations whose statistic are less than or equal to the observed statistic and greater than or equal to the observed statistic, respectively. This helps us understand that if the underlying power spectrum is power-law then given the same observed covariance and MRL reconstruction pipeline how likely the entire pipeline would produce a correlation higher or lower than the correlations obtained for the power spectra reconstructed from the observed datasets. The Table \ref{tab:correlation_results} shows that for CamSpec-ACT pair the correlation ($r_{\mathrm{obs}}$) for reconstructions from observed data significantly higher than the 50th percentile ($q_{50}$) for truncated $k$ range, which is true for both Pearson and Spearman. The CamSpec–SPT and ACT-SPT pairs show a mild negative correlation for the truncated region. We find the correlation between CamSpec and ACT beyond power-law and even at the feature level. %And it further establishes a consistency between CamSpec and ACT.
%From our reconstructed MRL-DR and MRL-TVR power spectra for CamSpec, ACT, and SPT datasets it is noticeable that for any feature that is common to any two datasets at some scale then another feature from the third dataset is appearing around that location with an out of phase behaviour and it is statistically marginalising the possibility of having such feature at that location. 

\begin{table*}[htbp]
\centering
\caption{Weighted Pearson and Spearman correlation results for primordial power spectra shown in Figs. \ref{fig:Cams_CamsBF-1a}, \ref{fig:COADACT_ACTBF-1a}, and \ref{fig:SPT_SPTBF-1a} reconstructed from CamSpec, ACT, and SPT datasets, respectively. The null distributions are obtained from 1000 power-law-based MRL simulations. Full and truncated stand for ranges $0.043\leq k\leq0.180~{\rm Mpc}^{-1}$ and $0.07\leq k\leq0.14~{\rm Mpc}^{-1}$, respectively. Here, $q_{16}$, $q_{50}$, and $q_{84}$ denote the 16th, 50th, and 84th percentiles of the corresponding null distribution.}
\label{tab:correlation_results}

%---------------------------------------------------------------
% Pearson
%---------------------------------------------------------------

\textbf{(a) Weighted Pearson correlation}

\vspace{2mm}

\begin{tabular}{llc cccc cc}
\hline
\hline
Method & Range & Pair &
$r_{\rm obs}$ & $q_{16}$ & $q_{50}$ & $q_{84}$
& $p_{\rm lower}$ & $p_{\rm upper}$ \\
\hline

MRLTV & Full & CamSpec--ACT
& 0.3920 & 0.1364 & 0.2561 & 0.3792 & 0.8691 & 0.1319 \\
MRLTV & Full & CamSpec--SPT
& -0.0054 & 0.1388 & 0.2935 & 0.4394 & 0.0370 & 0.9640 \\
MRLTV & Full & ACT--SPT
& 0.2970 & 0.1885 & 0.3330 & 0.4661 & 0.3996 & 0.6014 \\

MRLTV & Truncated & CamSpec--ACT
& 0.5487 & 0.0992 & 0.2715 & 0.4426 & 0.9620 & 0.0390 \\
MRLTV & Truncated & CamSpec--SPT
& -0.2770 & -0.0085 & 0.2495 & 0.4607 & 0.0220 & 0.9790 \\
MRLTV & Truncated & ACT--SPT
& -0.0476 & 0.0017 & 0.2333 & 0.4269 & 0.1129 & 0.8881 \\

\hline

regMRL & Full & CamSpec--ACT
& 0.4367 & 0.1578 & 0.2930 & 0.4256 & 0.8561 & 0.1449 \\
regMRL & Full & CamSpec--SPT
& -0.0088 & 0.1464 & 0.3033 & 0.4552 & 0.0330 & 0.9680 \\
regMRL & Full & ACT--SPT
& 0.3333 & 0.2401 & 0.4108 & 0.5670 & 0.3137 & 0.6873 \\

regMRL & Truncated & CamSpec--ACT
& 0.6202 & 0.1453 & 0.3577 & 0.5482 & 0.9251 & 0.0759 \\
regMRL & Truncated & CamSpec--SPT
& -0.3212 & -0.0006 & 0.2675 & 0.4908 & 0.0200 & 0.9810 \\
regMRL & Truncated & ACT--SPT
& -0.0766 & 0.0178 & 0.3449 & 0.5937 & 0.1009 & 0.9001 \\

\hline
\hline
\end{tabular}

\vspace{5mm}

%---------------------------------------------------------------
% Spearman
%---------------------------------------------------------------

\textbf{(b) Weighted Spearman correlation}

\vspace{2mm}

\begin{tabular}{llc cccc cc}
\hline
\hline
Method & Range & Pair &
$\rho_{\rm obs}$ & $q_{16}$ & $q_{50}$ & $q_{84}$
& $p_{\rm lower}$ & $p_{\rm upper}$ \\
\hline

MRLTV & Full & CamSpec--ACT
& 0.3958 & 0.1433 & 0.2849 & 0.4277 & 0.7692 & 0.2318 \\
MRLTV & Full & CamSpec--SPT
& 0.0225 & 0.1356 & 0.3282 & 0.4863 & 0.0609 & 0.9401 \\
MRLTV & Full & ACT--SPT
& 0.3045 & 0.2404 & 0.3904 & 0.5187 & 0.2887 & 0.7123 \\

MRLTV & Truncated & CamSpec--ACT
& 0.4942 & 0.1043 & 0.2760 & 0.4465 & 0.9061 & 0.0949 \\
MRLTV & Truncated & CamSpec--SPT
& -0.0595 & 0.0085 & 0.2649 & 0.4772 & 0.1049 & 0.8961 \\
MRLTV & Truncated & ACT--SPT
& -0.0247 & 0.0086 & 0.2355 & 0.4290 & 0.1279 & 0.8731 \\

\hline

regMRL & Full & CamSpec--ACT
& 0.4407 & 0.1858 & 0.3550 & 0.5238 & 0.6873 & 0.3137 \\
regMRL & Full & CamSpec--SPT
& 0.0583 & 0.1379 & 0.3451 & 0.5151 & 0.0809 & 0.9201 \\
regMRL & Full & ACT--SPT
& 0.3167 & 0.3276 & 0.5296 & 0.6645 & 0.1518 & 0.8492 \\

regMRL & Truncated & CamSpec--ACT
& 0.5427 & 0.1541 & 0.3883 & 0.5743 & 0.7792 & 0.2218 \\
regMRL & Truncated & CamSpec--SPT
& -0.0450 & 0.0178 & 0.2934 & 0.5164 & 0.1169 & 0.8841 \\
regMRL & Truncated & ACT--SPT
& -0.0964 & 0.0281 & 0.3686 & 0.6256 & 0.1009 & 0.9001 \\

\hline
\hline
\end{tabular}

\end{table*}

\subsection{Reconstruction from CamSpec and ACT data combination}\label{subsec:combine_act_cam}
\begin{figure}
    \centering

    \begin{subfigure}{0.7\textwidth}
        \centering
        \includegraphics[width=\textwidth]{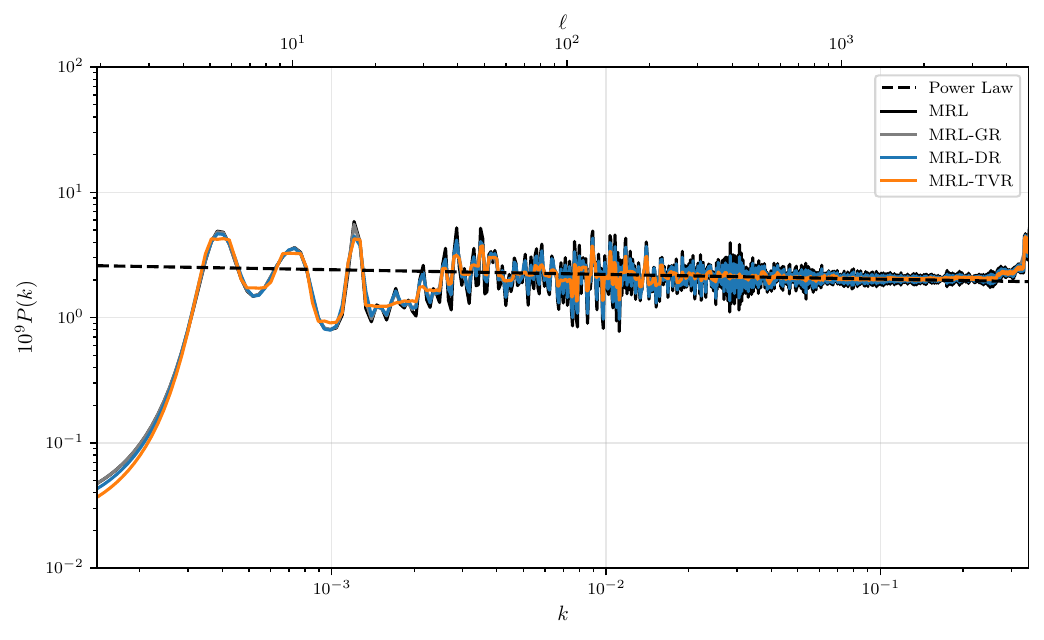}
        \caption{}
        \label{fig:ACT_Cams-1a}
    \end{subfigure}

%    \vspace{0.5cm} % space between figures

    \begin{subfigure}{0.7\textwidth}
        \centering
        \includegraphics[width=\textwidth]{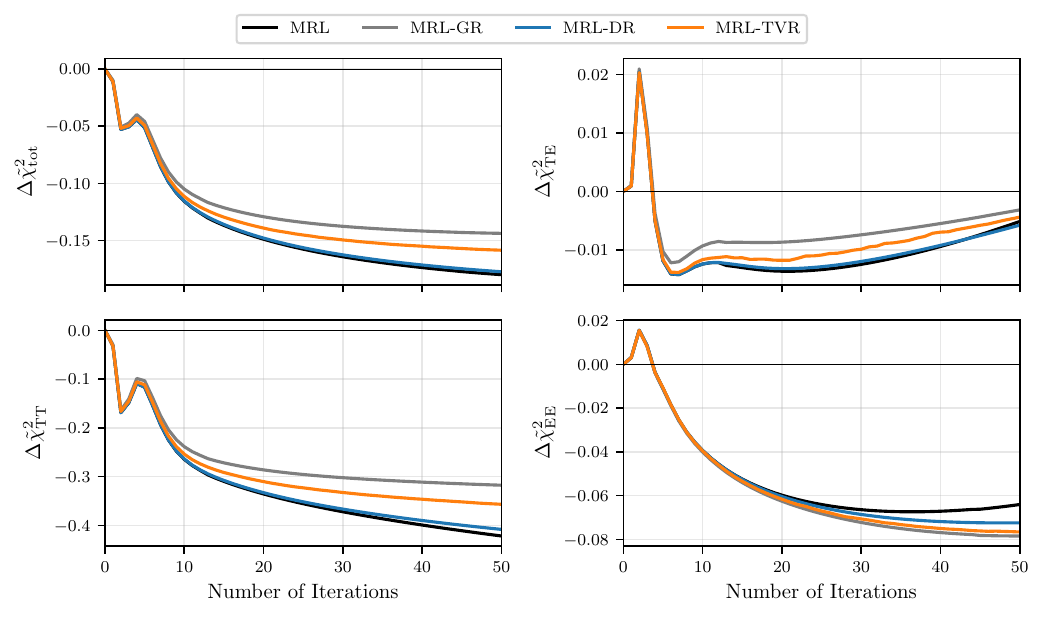}
        \caption{}
        \label{fig:ACT_Cams-1b}
    \end{subfigure}

    \caption{(a) Primordial power spectra that minimize the total chi-square reconstructed employing the MRL algorithm with different regularization methods. The black dashed line represents the nearly scale-invariant power-law power spectrum. The range of $k$ in this panel is $[0.00014 - 0.35]$. (b) Improvements obtained from reconstructed PPS in the chi-squared fit to data with every iteration for combined, TT, TE, and EE spectra. These results have been obtained from ACT and Planck PR4 CamSpec-NPIPE combined data based on Planck PR4 CamSpec-NPIPE best-fit $\Lambda CDM$ background cosmology.}
    \label{fig:ACT_Cams-1}
\end{figure}

In this section, we reconstruct the MRL power spectra for the combined datasets of Planck PR4 CamSpec-NPIPE and ACT-DR6. This reconstruction is to date the only analysis in the literature where the primordial power spectrum is reconstructed from the largest to the smallest accessible CMB scale, specifically for the multipole range $\ell=2 - 4810$. %To perform combined analysis with Planck and ACT data, in ACT paper~\cite{AtacamaCosmologyTelescope:2025blo}, they have chosen the Planck high-$\ell$ TT data up to $\ell < 1000$ and TE/EE polarization data up to  $\ell < 600$.
In this analysis, to combine ACT and CamSpec datasets, we evaluate the multipoles up to which the data uncertainties ($\Delta C_{\ell}$) in the CamSpec data are smaller than those in the ACT-DR6 data, and accordingly we use CamSpec data up to that multipole and ACT data for the remaining multipoles, for both the temperature and polarization spectra. %becomes smaller than the uncertainties ($\Delta C_{\ell}$) coming from CamSpec data for any given $\ell$, for each TT, TE, and EE data. 
We find following crossing multipoles $\ell^{\text{TT}}_{\text{cross}} = 1097$, $\ell^{\text{TE}}_{\text{cross}} = 992$ and $\ell^{\text{EE}}_{\text{cross}}=803$ beyond which ACT-DR6 data uncertainties become smaller than CamSpec. Thus, for our analysis, we consider CamSpec data for multipole ranges of $[2-1086]$, $[30-981]$, and $[30-792]$ for TT, TE, and EE spectra, respectively, and use ACT data for $\ell$ ranges of $[1097-4810]$, $[992-4810]$, and $[803-4810]$ for TT, TE, and EE spectra, respectively. %In this reconstruction, we have also followed the same strategy in combining CamSpec and ACT datasets as adopted in ACT analysis~\cite{AtacamaCosmologyTelescope:2025blo}.
The Planck PR4 CamSpec-NPIPE $\Lambda CDM$ best-fit cosmology listed in Table \ref{tab:bf_cosmo_all} has been used as the background cosmology for this reconstruction. %Fig.~\ref{fig:ACT_Cams-1} depicts the $\mathcal{P}_k$s reconstructed from the CamSpec and ACT combined dataset implementing the MRL algorithm. 
In Fig. \ref{fig:ACT_Cams-1a}, the reconstructed $\mathcal{P}_k$s using different smoothing prescriptions are illustrated, and in the bottom panel (Fig.~\ref{fig:ACT_Cams-1b}), the variation of improvement in the reduced chi-square $\Delta\tilde{\chi}^2_{\mathrm{XY}}$ with iteration number has been shown. Fig. \ref{fig:ACT_Cams-1b} reveals that for the combined dataset the $\tilde{\chi}^2_{\mathrm{tot}}$ is improving throughout 50 iterations and displays a behaviour almost identical to $\tilde{\chi}^2_{\mathrm{TT}}$, since improvement coming from TT data is more compared to TE/EE data. Furthermore, the reconstructed power spectra are improving the fit to TT, TE, and EE all three datasets. Otherwise it is usually seen in other cases like for ACT/SPT -only scenario (Sections \ref{subsec:rec_act} and \ref{subsec:rec_spt}) that the algorithm pushes the power spectrum to improve the fit to TT data at the cost of giving a poorer fit to TE and EE datasets. 
%And surprisingly, the reconstructed power spectra are giving better fit to combined EE data except a few early iterations (up to around 9 iterations) along with combined TT data.  

\subsection{Reconstruction from CamSpec and SPT data combination}\label{subsec:combine_spt_cam}
\begin{figure}
    \centering

    \begin{subfigure}{0.7\textwidth}
        \centering
        \includegraphics[width=\textwidth]{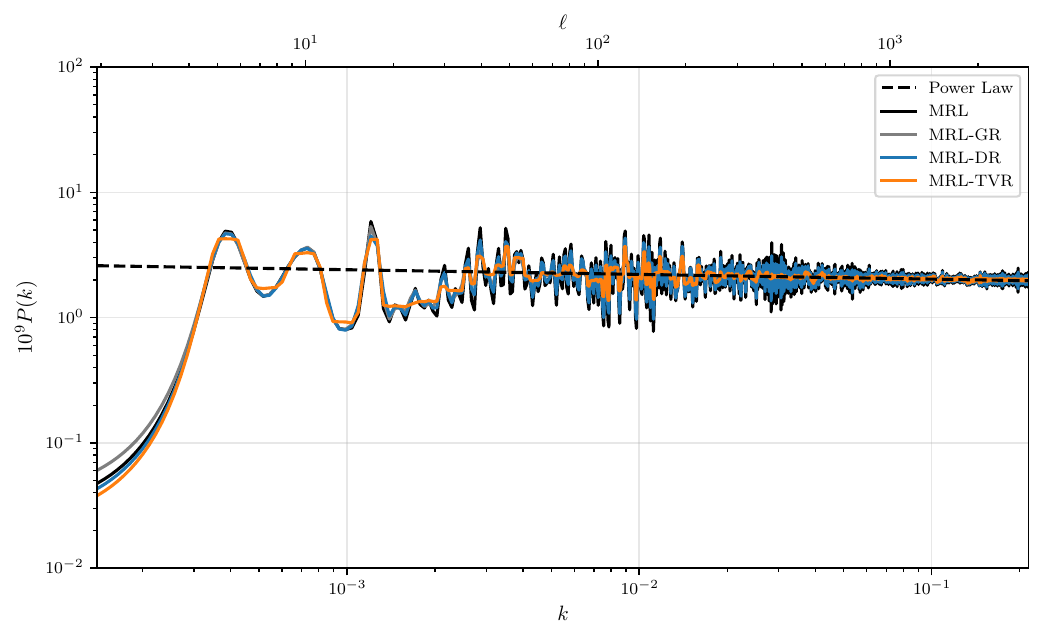}
        \caption{}
        \label{fig:SPT_Cams-1a}
    \end{subfigure}

%    \vspace{0.5cm} % space between figures

    \begin{subfigure}{0.7\textwidth}
        \centering
        \includegraphics[width=\textwidth]{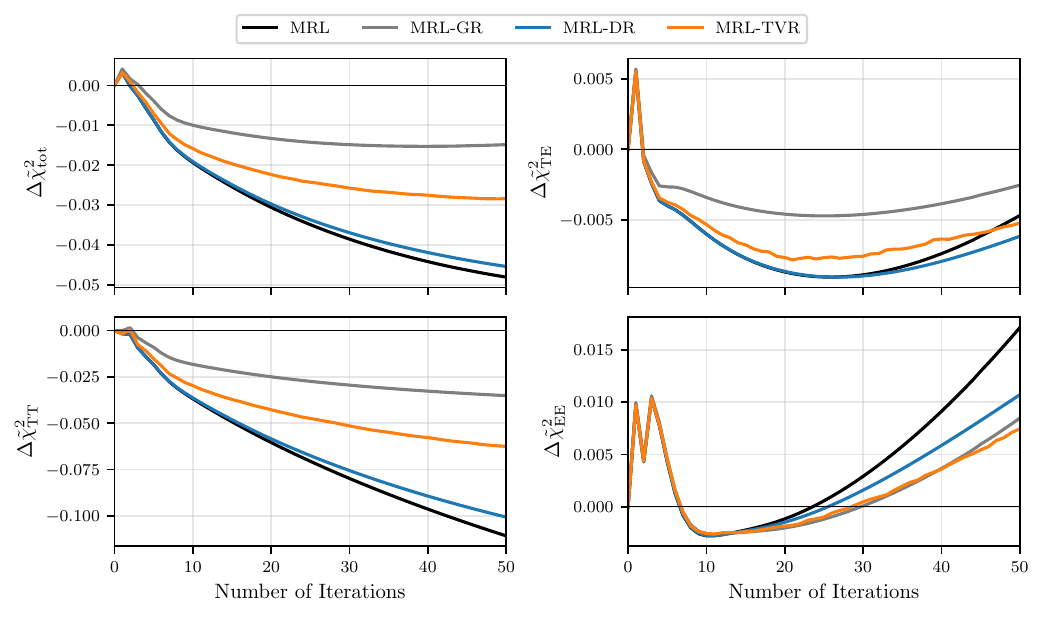}
        \caption{}
        \label{fig:SPT_Cams-1b}
    \end{subfigure}

    \caption{(a) Primordial power spectra that minimize the total chi-square reconstructed employing the MRL algorithm with different regularization methods. The black dashed line represents the nearly scale-invariant power-law power spectrum. The range of $k$ in this panel is $[0.00014 - 0.214]$. (b) Improvements obtained from reconstructed PPS in the chi-squared fit to data with every iteration for combined, TT, TE, and EE spectra. These results have been obtained from SPT and Planck PR4 CamSpec-NPIPE combined data based on Planck PR4 CamSpec-NPIPE best-fit $\Lambda CDM$ background cosmology.}
    \label{fig:SPT_Cams-1}
\end{figure}

Here we reconstruct the MRL power spectra for the combined datasets of Planck PR4 CamSpec-NPIPE and SPT-3G D1. This allows us to reconstruct the free-form power spectrum for the multipole range of $\ell=2 $ to $ \ell = 3000$. %using the MRL algorithm from the synergy of data coming from a space-based (Planck) and a ground-based (SPT) mission. 
For this analysis, we follow the same strategy adopted in Section~\ref{subsec:combine_act_cam}, selecting each dataset over the multipole range in which it provides greater constraining power. %We identify the $\ell$ range over which the uncertainties ($\Delta C_{\ell}$) in angular power spectra from CamSpec-NPIPE measurement are smaller than those of SPT-3G D1.
We identify that the uncertainties ($\Delta C_{\ell}$) in the TT, TE, and EE power spectra from the CamSpec-NPIPE measurement are smaller than those of SPT-3G D1 up to multipoles $2049$, $1399$, and $1049$, respectively. Accordingly, we use the CamSpec-NPIPE data for TT, TE, and EE spectra only for $\ell$ ranges $2-2049$, $30-1399$, and $30 - 1049$, respectively, and for the remaining multipole ranges $[2074.49-2974.46]$, $[1424.47-2974.47]$, and $[1074.46-2974.49]$, we use SPT-3G D1 data for TT, TE, and EE band power spectra, respectively. In this reconstruction, as the background cosmology, we consider the best-fit baseline $\Lambda CDM$ cosmology obtained from the Planck PR4 CamSpec-NPIPE data (Table \ref{tab:bf_cosmo_all}). %the CamSpec high-$\ell$ TT data up to $\ell < 1000$ and TE/EE polarization data up to  $\ell < 600$. In this reconstruction, we have also followed the same strategy in combining CamSpec and SPT datasets as adopted in ACT analysis~\cite{AtacamaCosmologyTelescope:2025blo}.
Fig. \ref{fig:SPT_Cams-1a} shows the MRL reconstructed $\mathcal{P}_k$s from this combined data for all adopted smoothing strategies, and Fig. \ref{fig:SPT_Cams-1b} describes the change in $\Delta\tilde{\chi}^2_{\mathrm{XY}}$ with increasing iterations. %Fig. \ref{fig:SPT_Cams-1b} reveals that the reconstructed $\mathcal{P}_k$s improve the fit for TT and TE data for all 50 iterations however, for EE data we get improvements in the fit up to around 24-29th iterations based on the smoothing method. 

%PPS combined dataset the $\tilde{\chi}^2_{\mathrm{tot}}$ is improving throughout 50 iterations and displays behaviour identical to $\tilde{\chi}^2_{\mathrm{TT}}$.
%And surprisingly, the reconstructed power spectra are giving better fit to combined EE data except a few early iterations (up to around 9 iterations) along with combined TT data. Otherwise it is usually seen in other cases like for ACT and SPT only scenarios (see Figs. \ref{fig:COADACTBF-1b} and \ref{fig:SPT-1b}) that the algorithm pushes the power spectrum to improve the fit to TT data at the cost of giving a poorer fit to TE and EE datasets.

\section{Methodology and datasets: parametric-form statistics}\label{sec:result_par}
In our reconstruction analysis for the non-parametric power spectrum with the MRL algorithm, we find that both the ACT-DR6 and SPT-3G D1 observations are consistent with the power-law primordial power spectrum (\ref{eq:dis_base_pk}) (Sections~\ref{subsubsec:case1_act} and~\ref{subsubsec:case1_spt}, respectively) when we use their respective $\Lambda CDM$ baseline best-fit cosmologies listed in Table \ref{tab:bf_cosmo_all}, as the background cosmology. However, when we consider the Planck 2018 PR3 and Planck PR4 CamSpec-NPIPE baseline best-fit models (Table \ref{tab:bf_cosmo_all}) as the background cosmology for reconstruction from ACT-DR6 and SPT-3G D1 data, it shows that SPT-3G D1 is completely consistent with Planck 2018 PR3 (Section~\ref{subsubsec:case2_spt}) and Planck PR4 CamSpec-NPIPE (Section~\ref{subsubsec:case3_spt}) baseline best-fit cosmologies but ACT-DR6 data do not corroborate the baseline best-fit cosmologies of Planck 2018 PR3 (Section \ref{subsubsec:case2_act}) and CamSpec-NPIPE (Section \ref{subsubsec:case3_act}) data. We intend to quantify this discordance between the Planck (both PR3 and PR4 data releases) and ACT-DR6 observations in probing the primordial power spectrum from the MRL reconstruction. Additionally, we also found in all our MRL reconstructions with ACT-DR6 data that at $k \approx 0.33-0.34  \, \mathrm{Mpc}^{-1}$ there exists a localized bump in the PPS. We also investigate the statistical significance of this observed bump along with probing the disagreement between the Planck and ACT datasets. We adopt a parametric primordial power spectrum to quantify the observed inconsistency in terms of change in the spectral tilt of the PPS. %This section deals with analysing ACT-DR6 data with a parametric form of the primordial power spectrum. The major goal of this analysis is to constrain the deviation of the spectral index indicated by the non-parametric analysis for the primordial power spectrum done in section \ref{sec:result_nonpar} and perform a consistency check between these two approaches (parametric and non-parametric) adopted in this work.

\subsection{Model}
To examine the statistical significance of the observed features in the MRL reconstructions from ACT-DR6 data, we adopt the following power-law power spectrum having two separate spectral indices before and after a particular junction point in the $k$ space which we call $k_{\mathrm{break}}$: 

\begin{equation}
\mathcal{P}^{\mathrm{double-tilt}}_{\mathcal{R}}\, (k) = A_s(k = k_{\mathrm{break}}) \times
\begin{cases}
\left(\dfrac{k}{k_{\mathrm{break}}}\right)^{n_{s1}-1}, & \text{if } k \le k_{\mathrm{break}}, \\[6pt]
\left(\dfrac{k}{k_{\mathrm{break}}}\right)^{n_{s2}-1}, & \text{if } k \ge k_{\mathrm{break}},
\end{cases}
\label{eq:double_tilt_pk}
\end{equation}
where $n_{s1}$ is the spectral tilt for scales before $k_{\mathrm{break}}$ and when $k$ crosses $k_{\mathrm{break}}$ then the spectral tilt is denoted by $n_{s2}$. It is important to mention here that we define the amplitude of the primordial power spectrum at the junction point $k_{\mathrm{break}}$, not at the pivot scale $k_{\mathrm{pivot}}=0.05 \, \mathrm{Mpc}^{-1}$. The difference between the two tilts is denoted by $\Delta n=n_{s2}-n_{s1}$. When $\Delta n=0$, we retrieve the single-tilt power-law power spectrum:

\begin{equation}
\mathcal{P}^0_{\mathcal{R}}\, (k) = A_s \left(\frac{k}{k_{\mathrm{pivot}}} \right)^{n_s-1}.
\label{eq:base_pk}
\end{equation}
Note here that the Eqs. (\ref{eq:dis_base_pk}) and (\ref{eq:base_pk}) are the same equation but for discrete values of $k$ we use the notation $\mathcal{P}^0_k$ to represent the baseline power-law power spectrum, so parameters $A_s$ and $n_s$ here carry the same definition as described in Eq. (\ref{eq:dis_base_pk}). 
The marginalized constraint on $\Delta n$ decides which one is the statistically allowed model for the primordial power spectrum, if it rejects $\Delta n=0$ with a strong statistical significance then it hints towards a double-tilt model as the statistically favoured model compared to a single-tilt one.

\subsection{Priors}
%The priors we consider for this analysis have been provided in Table \ref{tab:model_priors}. In Table \ref{tab:lcdm_prior}, we present the adopted priors considered for the baseline $\Lambda CDM$ parameters, and in Table \ref{tab:double_tilt_prior}, we tabulate the priors used for the parameters of the double-tilt model (\ref{eq:double_tilt_pk}). We use the same priors for the four background cosmological parameters as shown in Table \ref{tab:lcdm_prior} while sampling for the double-tilt model for two different prior sets listed in Tables \ref{tab:double_tilt_prior_a} and \ref{tab:double_tilt_prior_b}. 
The priors for the cosmological parameters of the baseline $\Lambda CDM$ and double-tilt models adopted in this analysis are provided in Table \ref{tab:model_priors}. We consider uniform priors for all the cosmological parameters related to $\Lambda CDM$ and double-tilt models. For two separate analyses, we use two different priors for the parameter $k_{\mathrm{break}}$, one in logarithmic scale and another in linear scale.
\begin{table}[t]
\centering
\caption{Uniform priors adopted for the cosmological parameters of the baseline $\Lambda CDM$ and double-tilt models. The prior on $k_{\mathrm{break}}$ is considered in both logarithmic and linear parameterizations for different analyses.}
\label{tab:model_priors}
\renewcommand{\arraystretch}{1.1}
\setlength{\tabcolsep}{6pt}

\begin{adjustbox}{max width=\textwidth}
\begin{tabular}{|lc|lc|}
\hline
\multicolumn{2}{|c|}{$\boldsymbol{\Lambda CDM}$} &
\multicolumn{2}{c|}{\textbf{Double-tilt}} \\
\textbf{Parameter} & \textbf{Prior} &
\textbf{Parameter} & \textbf{Prior} \\
\hline
\hline
$\Omega_{\mathrm b}h^2$
& $\sim \mathcal{U}(0.017,\,0.027)$
&
$\log(10^{10}A_{\mathrm s})$
& $\sim \mathcal{U}(1.61,\,3.91)$ \\

$\Omega_{\mathrm c}h^2$
& $\sim \mathcal{U}(0.09,\,0.15)$
&
$n_{\mathrm s1}$
& $\sim \mathcal{U}(0.9,\,1.1)$ \\

$100\,\theta_{\mathrm{MC}}$
& $\sim \mathcal{U}(1.038,\,1.044)$
&
$n_{\mathrm s2}$
& $\sim \mathcal{U}(-7.5,\,7.5)$ \\

$\tau_{\mathrm{reio}}$
& $\sim \mathcal{U}(0.02,\,0.08)$
&
$\log_{10}(k_{\mathrm{break}})$
& $\sim \mathcal{U}(-2.3,\,-0.75)$ \\
---
& ---
&
$\log_{10}(k_{\mathrm{break}})$
& $\sim \mathcal{U}(-1.0,\,-0.75)$ \\
---
& ---
&
$k_{\mathrm{break}}$
& $\sim \mathcal{U}(0.25,\,0.32)$ \\

\hline
\end{tabular}
\end{adjustbox}
\end{table}

\subsection{Sampling strategy}
We use the publicly available cosmology code $\texttt{CAMB}$~\cite{Lewis:1999bs}\footnote{\url{https://github.com/cmbant/CAMB}}, and the $\texttt{Cobaya}$~\cite{Torrado:2020dgo}\footnote{\url{https://github.com/CobayaSampler/cobaya}} sampling interface in our analysis. We modify the $\texttt{CAMB}$ code to implement Eq. (\ref{eq:double_tilt_pk}) in the primordial power spectrum sector of the code. We use nested sampling to perform the parameter estimation for our model when we consider sampling all the nuisance parameters related to the adopted likelihood(s). However, when we do not sample the nuisance parameters then we conduct the Markov chain Monte Carlo (MCMC) sampling. To execute nested sampling we employ $\texttt{Polychord}$ sampler~\cite{Handley_2015,Handley:2015fda}\footnote{\url{https://github.com/PolyChord/PolyChordLite}}, and to perform MCMC sampling we use the internal default sampler of the $\texttt{Cobaya}$ code. 
%and keeping then fixed at their respective best-fit values obtained for the baseline $\Lambda CDM$ scenario  for threefold reasons: 1. we anticipate a complicated shape of the posterior, which includes multimodal behaviour, non-linear correlation among parameters while sampling the parameter $k_{\mathrm{break}}$; 2. We aim to conduct evidence analysis between single-tilt and double-tilt models for which $\texttt{Polychord}$ is best as it directly evaluates the Bayesian evidence ($Z$) while sampling; 3. Furthermore, as we want to stringently constrain posterior tails of parameters $n_{s1}$, $n_{s2}$, and $k_\mathrm{break}$ for which $\texttt{Polychord}$ is the safest choice. To execute nested sampling we employ $\texttt{Polychord}$ sampler.
In all our analyses, we consider the default setting of $\texttt{Cobaya}$ for the number of live points, which is $25d$ ($d$ stands for the number of dimension), and we use $0.005$ as precision criterion in the logarithm of evidence. For all MCMC analyses, we consider the Gelman-Rubin convergence criterion $R-1 < 0.005$.
%In this analysis, we use the $\texttt{Polychord}$ sampler through $\texttt{Polychord}$~\cite{Handley_2015,Handley:2015fda}\footnote{\url{https://github.com/PolyChord/PolyChordLite}}
\subsection{Datasets and likelihoods}
We use the $\text{Planck}$ 2018 PR3 (\texttt{plik})~\cite{Planck:2018vyg} and PR4 (\texttt{CamSpec-NPIPE})~\cite{Rosenberg:2022sdy} data along with the ACT-DR6 multifrequency~\cite{AtacamaCosmologyTelescope:2025blo} data. We provide in Table \ref{tab:likelihoods} a complete list of the specific likelihoods that we use in this work. In the first column of Table \ref{tab:likelihoods}, we provide the aliases adopted in this article in referring the used likelihoods. The second column of the Table \ref{tab:likelihoods} provides the details of the likelihoods. In the third column, the adopted multipole ranges related to selected likelihoods are tabulated. When we perform the parameter estimation with the ACT data only then in order to constrain the optical depth ($\tau_\mathrm{reio}$) and to alleviate its degeneracy with the power spectrum amplitude ($A_s$), we use the $\texttt{sroll2 prior}$ on $\tau_\mathrm{reio}$. Otherwise, when sampling together with Planck data (\texttt{plik} and \texttt{NPIPE}), we use Planck low-$\ell$ EE polarization data ($\texttt{SimAll}$). %Here we use the binned data for Planck likelihood.
%We vary all nuisance parameters associated with both Planck (plik and NPIPE) and ACT likelihood in our analyses with their default priors unless specifically mentioned for some specific runs. 

\begin{table}[t]
\centering
\caption{The Planck and ACT likelihoods used in this work, and the associated multipole ranges and adopted aliases.}
\label{tab:likelihoods}
\begin{adjustbox}{max width=\textwidth}
\begin{tabular}{|c|c|c|}
\hline
\textbf{Name} & \textbf{Likelihood} & \textbf{Multipole} \\
\hline
\hline
%P18 & plik TTTEEE high-$\ell$ & 2--2508 (TT) \\
% & + TT lowL Commander + EE SimAll~\cite{Planck:2018vyg} & 2--1996 (TEEE) \\
%\hline
P18-TT ($\ell \leq 1996$) & plik TT high-$\ell$ + TT lowL Commander + EE SimAll & 2--1996 \\
\hline
P18 ($\ell \leq 1996$) & plik T \& E high-$\ell$ + TT lowL Commander + EE SimAll & 2--1996 \\
\hline
CamSpec-TT ($\ell \leq 1996$) & NPIPE TT high-$\ell$ + TT lowL Commander + EE SimAll & 2--1996 \\
\hline
CamSpec ($\ell \leq 1996$) & NPIPE T \& E high-$\ell$ + TT lowL Commander + EE SimAll & 2--1996 \\
\hline
ACT-TT ($1997 \leq \ell \leq 3000$) & ACT-DR6 MFLike (only TT) & 1997--3000 \\
\hline
ACT ($1997 \leq \ell \leq 3000$) & ACT-DR6 MFLike (T \& E) & 1997--3000 \\
\hline
ACT-TT($\ell \leq 4810$) & ACT-DR6 MFLike (only TT) & 600--4810 \\
\hline
ACT ($\ell \leq 4810$) & ACT-DR6 MFLike (T \& E) & 600--4810 \\
\hline
\end{tabular}
\end{adjustbox}
\end{table}

\section{Result and analysis: parametric-form statistics}
In this section, we thoroughly examine all the results of parameter estimation obtained from different likelihoods (Table \ref{tab:likelihoods}). Here we present and discuss our results in three sections. In the first section, we discuss the results related to the combined dataset of Planck 2018 PR3 and ACT-DR6. Then in the next section, we discuss the results obtained from the Planck PR4 and ACT-DR6 data combination. Finally, the third section is dedicated to discussing the results obtained from the analyses performed using the ACT-DR6 data alone.  %We present the evidence and Bayes' factor obtained for different datasets adopted in this analysis, which quantifies how significantly double-tilt model is statistically favour or disfavoured in comparison to the single-tilt baseline power-law model. 
%\subsection{P18-TT ($\ell < 1996$)\,+\,ACT-TT ($1997 < \ell < 3000$):} 

\subsection{Planck PR3 and ACT-DR6 data combination}
In this analysis, we intend to explore the plausible signature of discrepancy between two datasets, viz., the Planck 2018 PR3 and ACT-DR6, that has been observed in Fig. \ref{fig:COADACT_P18BF-2} when reconstructing the primordial power spectra employing the MRL algorithm from the ACT-DR6 data based on the Planck PR3 $\Lambda CDM$ best-fit background cosmology (Section~\ref{subsubsec:case2_act}). 

\begin{itemize}[label=\ding{109}]
    \item \textbf{P18-TT ($\ell \leq 1996$)\,+\,ACT-TT ($1997 \leq \ell \leq 3000$):}

We investigate three different cases for this dataset. In \texttt{Case I}, we vary the parameter $k_\mathrm{break}$ for a broad prior range. In \texttt{Case II}, we consider a smaller prior range for the parameter $k_\mathrm{break}$ compared to the prior considered in \texttt{Case I}. In \texttt{Case III}, we keep the value of $k_\mathrm{break}$ fixed at $0.165 \, \mathrm{Mpc}^{-1}$. Below, we discuss in detail the inferences drawn from each of these analyses separately.  

    \begin{itemize}[label={\tiny$\blacksquare$}]
    
         \item \textbf{Case I:}
         
         In this analysis, we combine the Planck TT data with ACT-DR6 TT data (Table \ref{tab:likelihoods}). For Planck data, along with high-$\ell$ TT data we also use the Planck low-{$\ell$} TT and EE measurements. We truncate the Planck high-$\ell$ TT data at $\ell=1996$, and from $\ell=1997$ to $\ell=3000$, we consider the ACT-DR6 TT data. We adopt this multipole range to explore specifically the region of the reconstructed power spectrum in which the identified feature suggests a potential discrepancy between the Planck and ACT datasets. We find from Fig. \ref{fig:COADACT_P18BF-2} that the discrepancy appears at $k \approx [0.15-0.23]  \, \mathrm{Mpc}^{-1} $, and before and after this window for the rest of the $k$ range the ACT data is consistent with the Planck $\Lambda CDM$ best-fit background cosmology. Accordingly, we choose the prior for $k_\mathrm{break}$. The adopted priors for all the cosmological parameters are mentioned in Table \ref{tab:model_priors}. Here we use the prior $\log_{10}(k_{\mathrm{break}}) \sim \mathcal{U}(-2.3,\,-0.75)$ (Table \ref{tab:model_priors}) for $k_{\mathrm{break}}$. The results obtained from this analysis are shown in Fig.~\ref{fig:planck_act_tt_kbreak_full_prior}. %This choice has been made from the hint we get from the MRL constructed power spectrum shown in Fig. \ref{fig:COADACT_P18BF-2}. In Fig. \ref{fig:COADACT_P18BF-2}, from the error analysis of the reconstructed PPS we see a visible departure from the standard power-law behaviour.  
         We can see from Fig.~\ref{fig:planck_act_tt_kbreak_full_prior_posterior} that $n_{s2}$ starts deviating from $n_{s1}$ around $k_\mathrm{break}=0.13 \, \mathrm{Mpc}^{-1}$ and rejects $\Delta n=0$ with $2\sigma$ confidence level when $k_\mathrm{break}$ starts exceeding $0.15 \, \mathrm{Mpc}^{-1}$. The $n_{s2}$ shows preference for blue tilted region except for very large scales. We show the posterior of the double-tilt $\mathcal{P}^{\mathrm{double-tilt}}_{\mathcal{R}}\, (k)$ (\ref{eq:double_tilt_pk}) in Fig.~\ref{fig:planck_act_tt_kbreak_full_prior_fgivenx}. In this analysis, we sample all the nuisance parameters related both Planck and ACT likelihoods.

\begin{figure}[!htb]
    \centering

    \begin{subfigure}{0.7\textwidth}
        \centering
        \includegraphics[width=\textwidth]{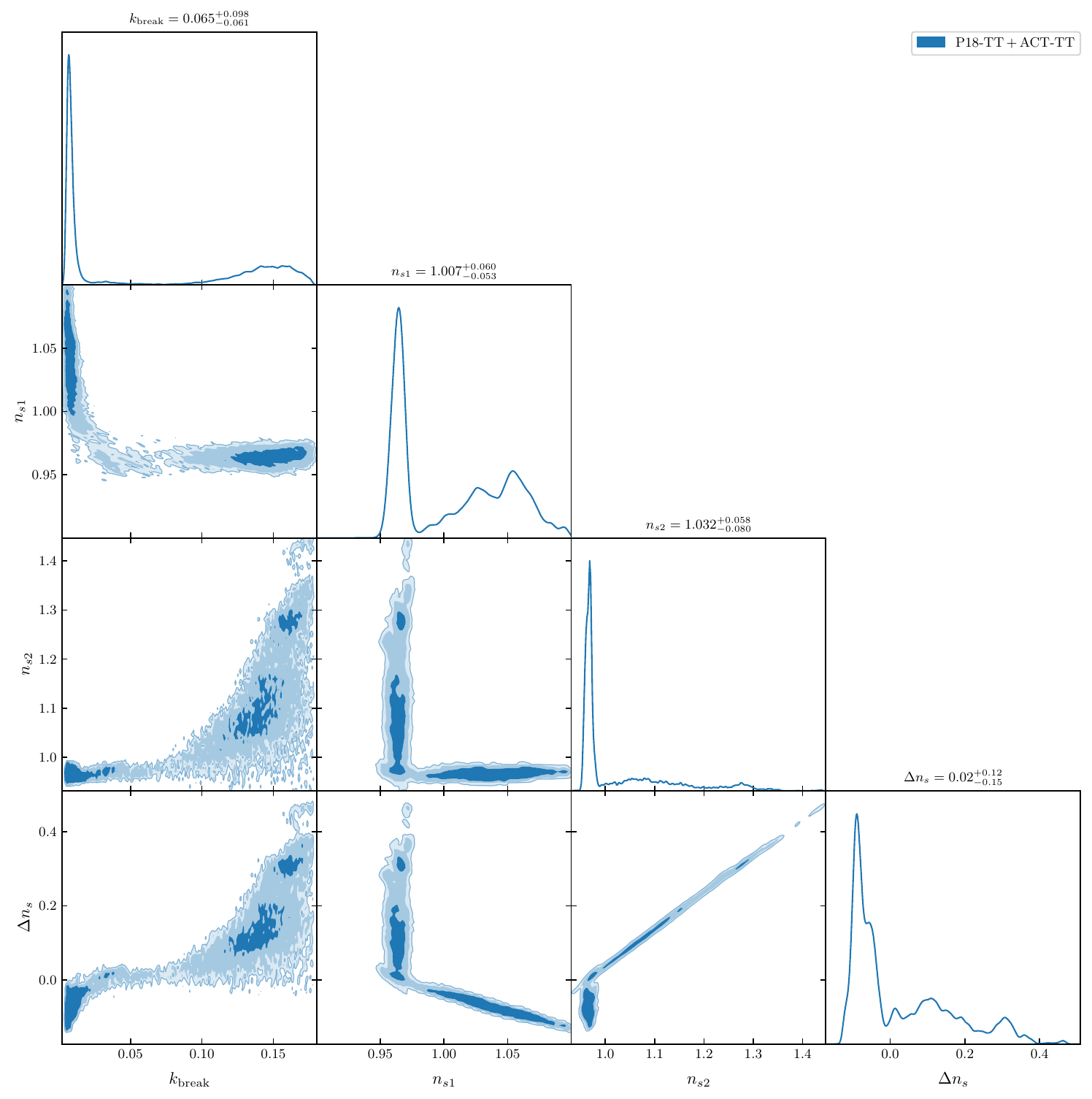}
        \caption{}
        \label{fig:planck_act_tt_kbreak_full_prior_posterior}
    \end{subfigure}

%    \vspace{0.3cm} % space between figures

    \begin{subfigure}{0.7\textwidth}
        \centering
        \includegraphics[width=\textwidth]{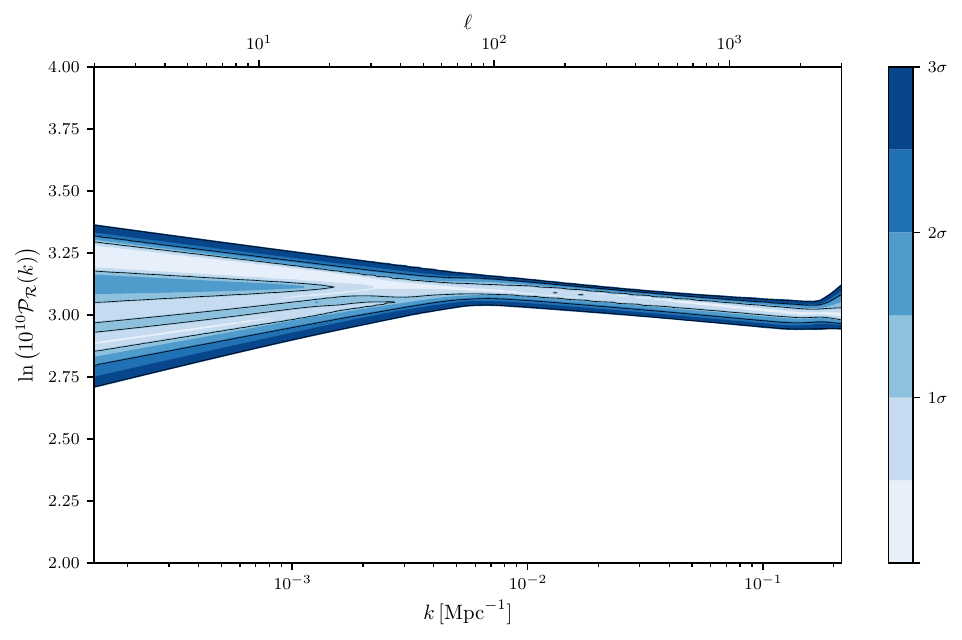}
        \caption{}
        \label{fig:planck_act_tt_kbreak_full_prior_fgivenx}
    \end{subfigure}

    \caption{The plots in both panels are for the prior $\log_{10}(k_{\mathrm{break}}) \sim \mathcal{U}(-2.3,\,-0.75)$. (a) $n_{\mathrm{s2}}$ vs $k_\mathrm{{break}}$ plot for Planck (2-1996) and ACT-DR6 multifrequency (1997-3000) combined TT data. Here, all the nuisance parameters related to both Planck and ACT likelihoods are sampled. (b) Plot of the double-tilt primordial power spectrum with its confidence intervals obtained from the combined Planck (2-1996) and ACT-DR6 multifrequency (1997-3000) TT dataset.}
    \label{fig:planck_act_tt_kbreak_full_prior}
\end{figure}

         \item \textbf{Case II:}   

          We conduct another analysis with keeping everything same as in \texttt{Case I} but with a reduced prior range of $k_{\mathrm{break}}$, $\log_{10}(k_{\mathrm{break}}) \sim \mathcal{U}(-1.0,\;-0.75)$ (Table \ref{tab:model_priors}). In this analysis, we sample all the nuisance parameters associated with both Planck and ACT likelihoods. The results of this analysis are shown in Fig. \ref{fig:pact_tt_ttteee_kbreak_lower_bound_0.1}. The constraints on $k_{\mathrm{break}}$, $n_{s1}$, $n_{s2}$, and $\Delta n_{s}$ are $0.148^{+0.024}_{-0.012}$, $0.9657 \pm 0.0054$, $1.115^{+0.079}_{-0.13}$, and $0.149^{+0.080}_{-0.13}$, respectively. This shows a deviation of $n_{s2}$ from $1$ by $1.1 \sigma$, and for $\Delta n_{s}$ a $1.42 \sigma$ deviation from $0$. The double-tilt primordial power spectrum with its confidence intervals obtained from this analysis is shown in Fig.~\ref{fig:Planck_ACT_TT_lower_bound_0.1_PC_fgivenx}.
          %The $ n_{s1}$ is consistent with Planck's constraint $n_s = 0.9626 \pm 0.0057 \, (68 \, \%  \, \text{CL})$ across the range of $\log_{10} k_\mathrm{break}$ except for the very large scales, which is an usual phenomenon observed when Planck TT only data is used~\cite{Hazra:2013nca}. From Planck 2018 data release, we have following estimation of spectral index for the baseline power-law power spectrum (\ref{eq:base_pk}) with single-tilt: $n_s = 0.9626 \pm 0.0057 \, (68 \, \%  \, \text{CL})$, for the Planck TT data in combination with low-{$\ell$} EE data, where Planck TT data considers multipole range of $2$ to $2508$. 

\begin{figure}[!htb]
    \centering
    \begin{subfigure}{0.7\textwidth}
        \centering    
        \includegraphics[width=\textwidth, keepaspectratio]{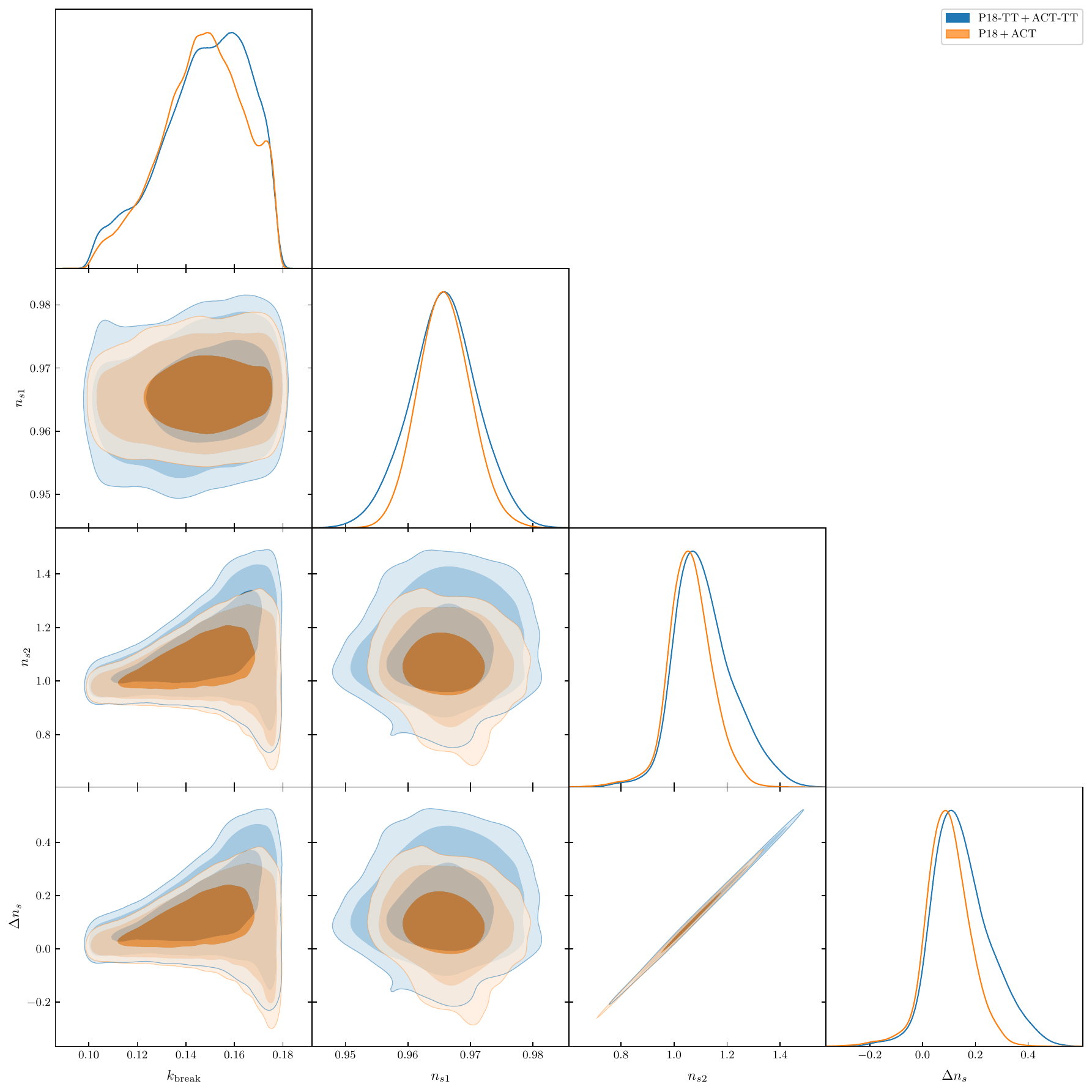}
        \caption{}
        \label{fig:pact_tt_ttteee_kbreak_lower_bound_0.1}
    \end{subfigure}

%    \vspace{0.3cm}

    \begin{subfigure}{0.49\textwidth}
        \centering
        \includegraphics[width=\textwidth, keepaspectratio]{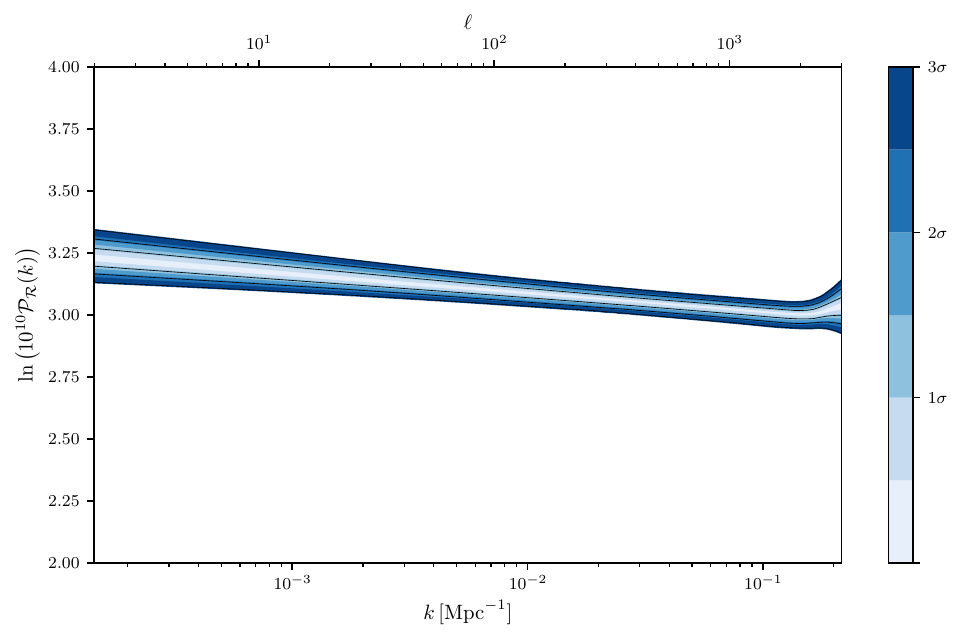}
        \caption{}
        \label{fig:Planck_ACT_TT_lower_bound_0.1_PC_fgivenx}
    \end{subfigure}
    \hfill
    \begin{subfigure}{0.49\textwidth}
        \centering
        \includegraphics[width=\textwidth, keepaspectratio]{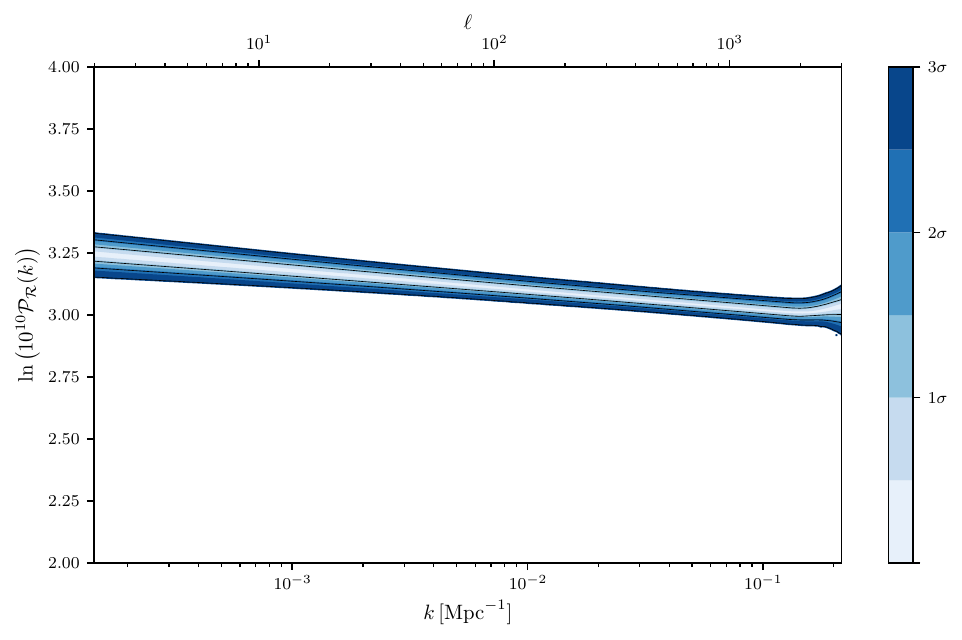}
        \caption{}
        \label{fig:Planck_ACT_lower_bound_0.1_PC_fgivenx}
    \end{subfigure}

    \caption{Marginalised posterior distributions of $n_{s2}$ versus $k_\mathrm{break}$ together with the double-tilt primordial power spectrum and its confidence intervals estimated from Planck (2-1996) and ACT-DR6 multifrequency (1997-3000) combined data. These plots are obtained for $\log_{10} k_\mathrm{break} \sim \mathcal{U} (-1.0, -0.75)$ and by varying all the nuisance parameters related to both Planck and ACT likelihoods. (a) Marginalised posterior distributions for TT and TTTEEE datasets. (b) Double-tilt primordial power spectrum for TT data combination. (c) Double-tilt primordial power spectrum for TTTEEE data combination.}
    \label{fig:Planck_ACT_TT_n_TTTEEE_lower_bound_0.1_PC_fgivenx}
\end{figure}
         
         \item \textbf{Case III:}
         
         We conduct another analysis with same datasets, to identify the exact deviation of $ n_{s2}$ from $ n_{s1}$, when the value of $k_\mathrm{break}$ is fixed at $0.165 \, \mathrm{Mpc}^{-1}$. We find from Figs. \ref{fig:planck_act_tt_kbreak_full_prior_posterior} and \ref{fig:pact_tt_ttteee_kbreak_lower_bound_0.1} that for $k_\mathrm{break} \gtrsim 0.16 \, \mathrm{Mpc}^{-1}$, $ \Delta n_{s}$ shows a considerable deviation from $0$, consequently, we set $k_\mathrm{break}$ at $0.165 \, \mathrm{Mpc}^{-1}$. To perform a consistent comparison throughout with other datasets in subsequent fixed $k_\mathrm{break}$ analyses, we consider $k_\mathrm{break}=0.165 \, \mathrm{Mpc}^{-1}$. We sample all the nuisance parameters of both Planck and ACT likelihoods in this analysis. We show the results obtained from this sampling in Fig. \ref{fig:all_pc_run_kbreak_fixed_0165} and Table~\ref{tab:double_tilt_constraints}. The estimation of $n_{s2}$ coming from this analysis is $ 1.17 \pm 0.12$, which implies that the parameter $n_{s2}$ is showing a $1.4167 \sigma$ deviation from $1$. We also have the constraints on the difference in spectral tilts ($ \Delta n_{s}) = 0.20 \pm 0.12$, which differs from $0$ by $1.67 \sigma$. Fig.~\ref{fig:Planck_ACT_kbreak_fix_0.165_TT_fgivenx} shows the double-tilt primordial power spectrum with its confidence intervals obtained from this analysis.
    
    \end{itemize}

    \item \textbf{P18 ($\ell \leq 1996$)\,+\,ACT ($1997 \leq \ell \leq 3000$):} 

    In this analysis, along with the high-$\ell$ temperature data we incorporate the high-$\ell$ polarization data as well for both Planck and ACT data. Here we explore four different scenarios. In \texttt{Case I}, we vary both, the parameter $k_\mathrm{{break}}$, and all the nuisance parameters associated with both the Planck and ACT likelihoods. In \texttt{Case II}, we consider everything same as in \texttt{Case I}, however, we assume a smaller prior range for $k_\mathrm{{break}}$ compared to \texttt{Case I}. In \texttt{Case III}, we keep the parameter $k_\mathrm{{break}}$ fixed at $0.165 \, \mathrm{Mpc}^{-1}$, but varying all the nuisance parameters. In \texttt{Case IV}, we keep both fixed, the parameter $k_\mathrm{{break}}$ at $0.165 \, \mathrm{Mpc}^{-1}$, and all the nuisance parameters at their respective best-fit values obtained for the $\Lambda CDM$ baseline cosmology. We analyse these three scenarios in detail subsequently.
    
    \begin{itemize}[label={\tiny$\blacksquare$}]
    
        \item \textbf{Case I:}
        
        In this data combination, as described in Table \ref{tab:likelihoods}, we consider the Planck TTTEEE high-$\ell$ data up to $\ell = 1996$ along with low-{$\ell$} TT and EE data, and for the multipole range $[1997,3000]$, we use the ACT TTTEEE multifrequency data. The results obtained for this data combination are shown in Fig. \ref{fig:pact_ttteee_kbreak_full_prior_ns1_vs_ns2}. The priors for the cosmological parameters used in this analysis are listed in Table \ref{tab:model_priors}. For $k_{\mathrm{break}} $, we use the prior $\log_{10}(k_{\mathrm{break}}) \sim \mathcal{U} (-2.3,\;-0.75)$. %In this analysis, we find $n_{s1} = 0.9654 \pm 0.0043 \, (68 \, \%  \, \text{CL})$, which is completely consistent with the bound reported by \cite{Planck:2018vyg} for the data combination of Planck (high-$\ell$ TTTEEE  + low-$\ell$ TT + low-$\ell$ EE) with $\ell$ range $[2,2508]$, $n_s = 0.9649 \pm 0.0044 \, (68 \, \%  \, \text{CL})$. %However, $n_{s2}$ shows both red and blue tilted behaviour.
        For $k_{\mathrm{break}} $ approximately up to $ 0.10 \, \mathrm{Mpc}^{-1}$, $\Delta n_s$ is consistent with null value ($\Delta n_s =0$), and beyond $ 0.10 \, \mathrm{Mpc}^{-1}$, we can see $\Delta n_s$ shows deviation from the null value by $1\sigma$, however stays within $2\sigma$ bounds. The posterior plot of the double-tilt power spectrum $\mathcal{P}^{\mathrm{double-tilt}}_{\mathcal{R}}\, (k)$ (\ref{eq:double_tilt_pk}) is exhibited in Fig. \ref{fig:pact_ttteee_kbreak_full_prior_fgivenx}. 

\begin{figure}[!htb]
    \centering

    \begin{subfigure}{0.7\textwidth}
        \centering
        \includegraphics[width=\textwidth]{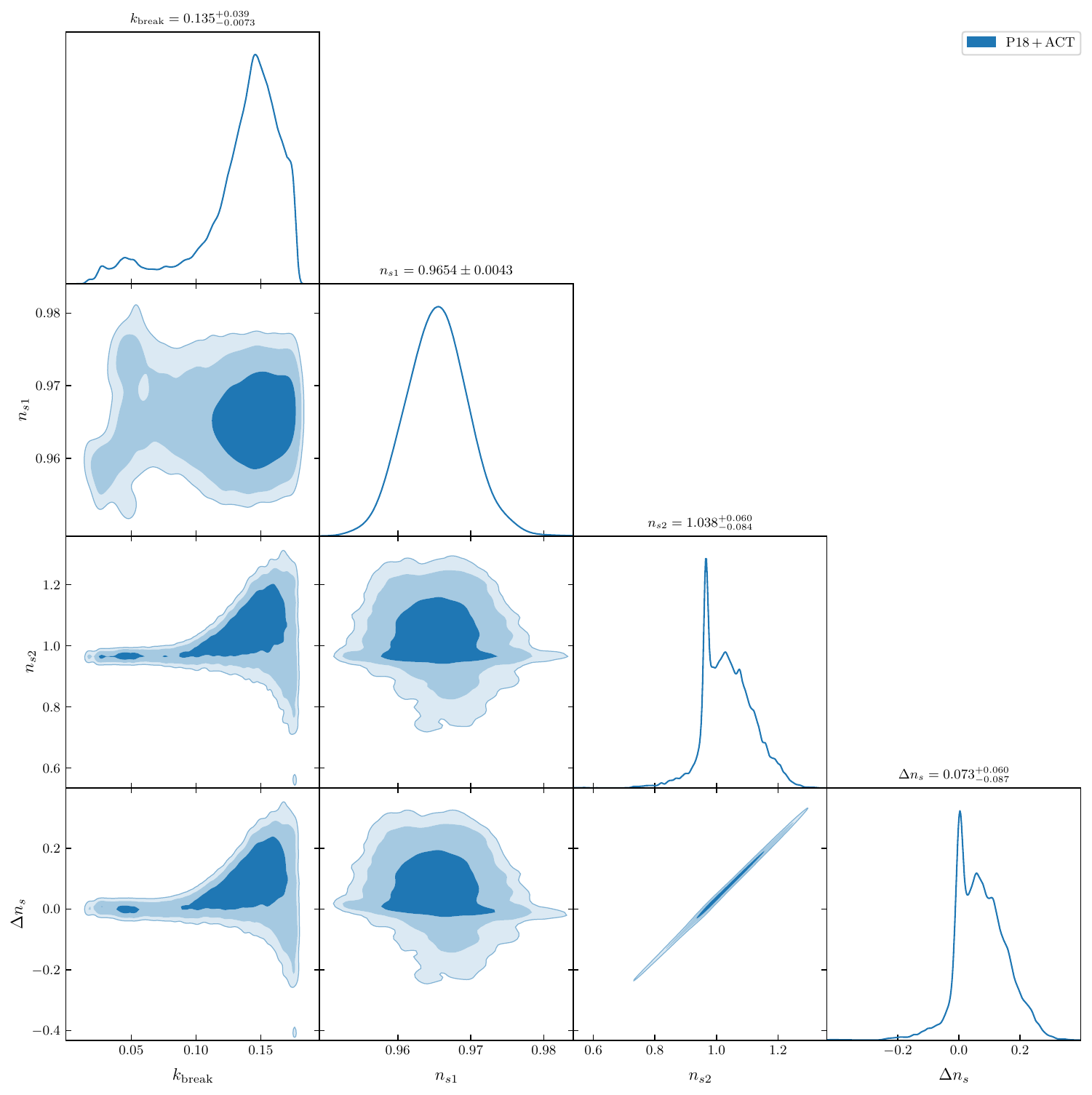}
        \caption{}
        \label{fig:pact_ttteee_kbreak_full_prior_ns1_vs_ns2}
    \end{subfigure}

%    \vspace{0.5cm} % space between figures

    \begin{subfigure}{0.7\textwidth}
        \centering
        \includegraphics[width=\textwidth]{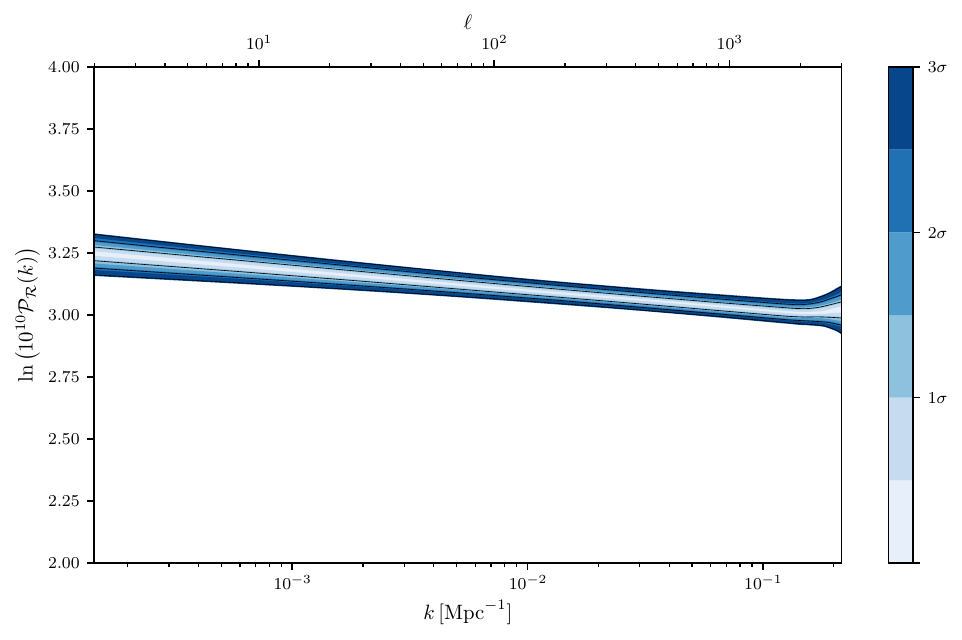}
        \caption{}
        \label{fig:pact_ttteee_kbreak_full_prior_fgivenx}
    \end{subfigure}

    \caption{The plots in both panels are for the prior $\log_{10} k_\mathrm{break} \sim \mathcal{U} (-2.3, -0.75)$. (a) $n_{\mathrm{s2}}$ vs $k_\mathrm{{break}}$ plot for Planck (2-1996) and ACT-DR6 multifrequency (1997-3000) combined TTTEEE data. Here, all the nuisance parameters related to both Planck and ACT likelihoods are sampled. (b) Plot of the double-tilt primordial power spectrum with its confidence intervals obtained from the Planck (2-1996) and ACT-DR6 multifrequency (1997-3000) TTTEEE data combination.}
    \label{fig:pact_ttteee_kbreak_full_prior}
\end{figure}

        \item \textbf{Case II:}
        
        In this analysis, we keep everything same as in \texttt{Case I}. Here we assume a truncated prior for $k_{\mathrm{break}}$, $\log_{10}(k_{\mathrm{break}}) \sim \mathcal{U} (-1.0,\;-0.75)$ (Table \ref{tab:model_priors}). We show the posteriors estimated from this analysis in Fig. \ref{fig:pact_tt_ttteee_kbreak_lower_bound_0.1}. The estimation of $k_{\mathrm{break}}$, $n_{s1}$, $n_{s2}$, and $\Delta n_{s}$ are $0.147^{+0.019}_{-0.015}$, $0.9659 \pm 0.0042$, $1.061^{+0.072}_{-0.083}$, and $0.095^{+0.073}_{-0.083}$, respectively. In this analysis, we find that the $n_{s2}$ and $\Delta n_{s}$ are $0.79\sigma$ and $1.22\sigma$ away from $1$ and $0$, respectively. The double-tilt primordial power spectrum with confidence intervals obtained from this analysis is shown in Fig. \ref{fig:Planck_ACT_lower_bound_0.1_PC_fgivenx}.

        \item \textbf{Case III:}
        
        We explore another scenario, where we keep the value of $k_\mathrm{break}$ fixed at $0.165 \, \mathrm{Mpc}^{-1}$, however sample all the nuisance parameters of Planck and ACT likelihoods. The obtained result from this sampling is depicted in Fig. \ref{fig:all_pc_run_kbreak_fixed_0165} and Table \ref{tab:double_tilt_constraints}. We find that at $k_\mathrm{break}=0.165 \, \mathrm{Mpc}^{-1}$, the posterior estimate of $n_{s2}$ and $\Delta n_s$ are $1.10^{+0.12}_{-0.098}$ and $0.132^{+0.12}_{-0.099}$, respectively. Thus, it shows nearly a $1\sigma$ deviation from $1$ and $0$ for $n_{s2}$ and $\Delta n_s$, respectively. We show the estimated posterior of the double-tilt primordial power spectrum in Fig.~\ref{fig:Planck_ACT_kbreak_fix_0.165_fgivenx}.
        
        \item \textbf{Case IV:}
        
        We also investigate the effect of simultaneously fixing $k_\mathrm{break}=0.165 \, \mathrm{Mpc}^{-1}$ and the nuisance parameters of both Planck and ACT likelihoods to their respective best-fit values based on the baseline $\Lambda CDM$ cosmology. We show the obtained constraints on spectral tilts in Fig. \ref{fig:all_mcmc_run_kbreak_fixed_0165_nui_fixed_bf} and Table \ref{tab:double_tilt_constraints}. In this case, we get slightly more statistically significant deviations for $n_{s2}$ and $\Delta n_s$ from $1$ and $0$, respectively. The spectral tilt $n_{s2}$ exhibits correlation with several nuisance parameters which include kinetic Sunyaev–Zel'dovich amplitude ($a_{\mathrm{kSZ}}$), thermal Sunyaev–Zel'dovich amplitude ($a_{\mathrm{tSZ}}$), tSZ shape parameter ($\alpha_{\mathrm{tSZ}}$), and Poisson point sources, all of them contribute at high-$\ell$ where $n_{s2}$ is effective, consequently fixing them at their best-fit values improves the bounds on $n_{s2}$. The posterior estimate obtained from this analysis for $n_{s2}$ and $\Delta n_s$ are $n_{s2} = 1.045\pm0.029$ and $\Delta n_s = 0.075\pm0.029$, which imply $1.55 \sigma$ and $2.59 \sigma$ deviations from $1$ and $0$, respectively. 
    
    \end{itemize}
    
\begin{figure}[!htb]
    \centering

    \begin{subfigure}{0.5\textwidth}
        \centering
        \includegraphics[width=\textwidth, keepaspectratio]{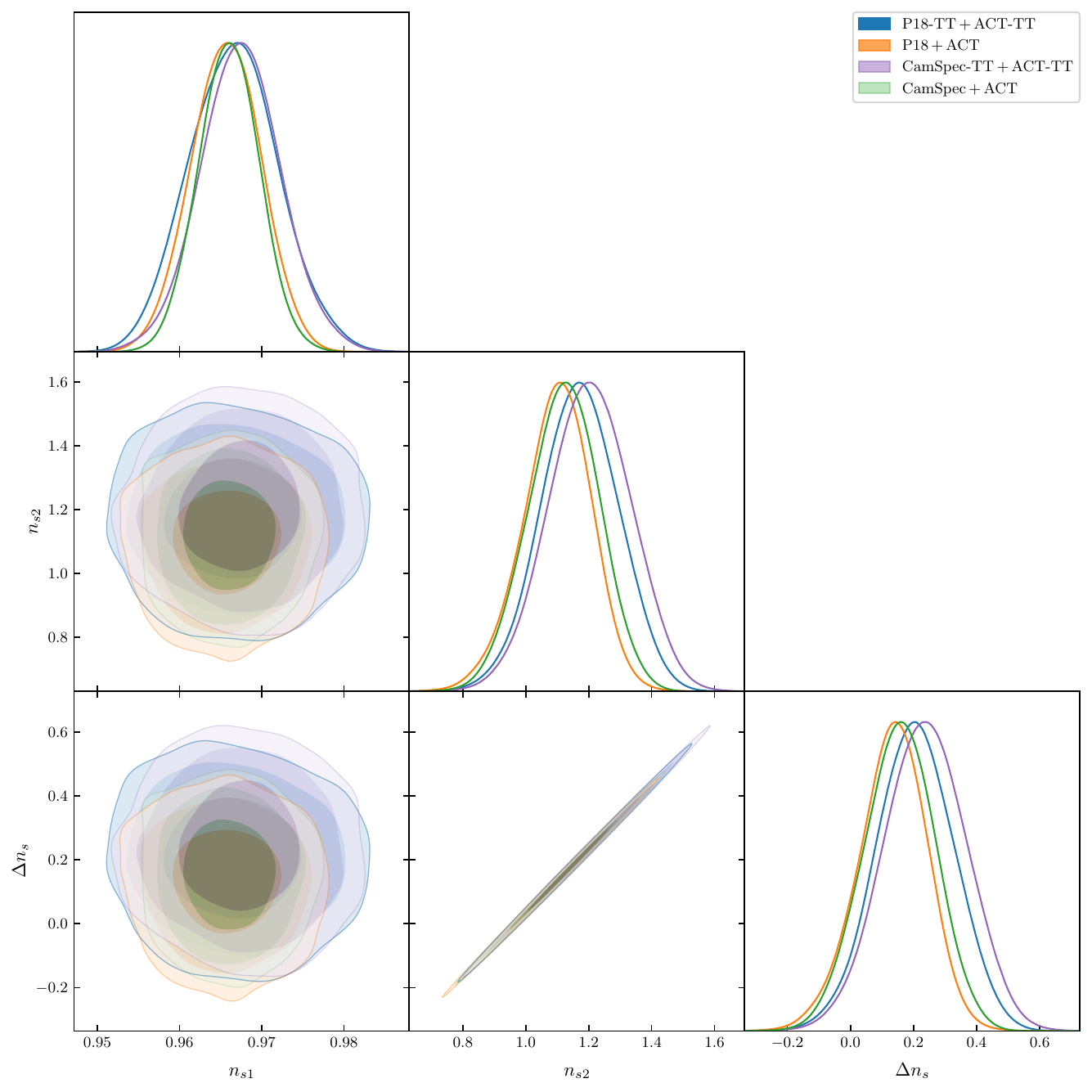}
        \caption{}
        \label{fig:all_pc_run_kbreak_fixed_0165}
    \end{subfigure}\hfill
    \begin{subfigure}{0.5\textwidth}
        \centering
        \includegraphics[width=\textwidth, keepaspectratio]{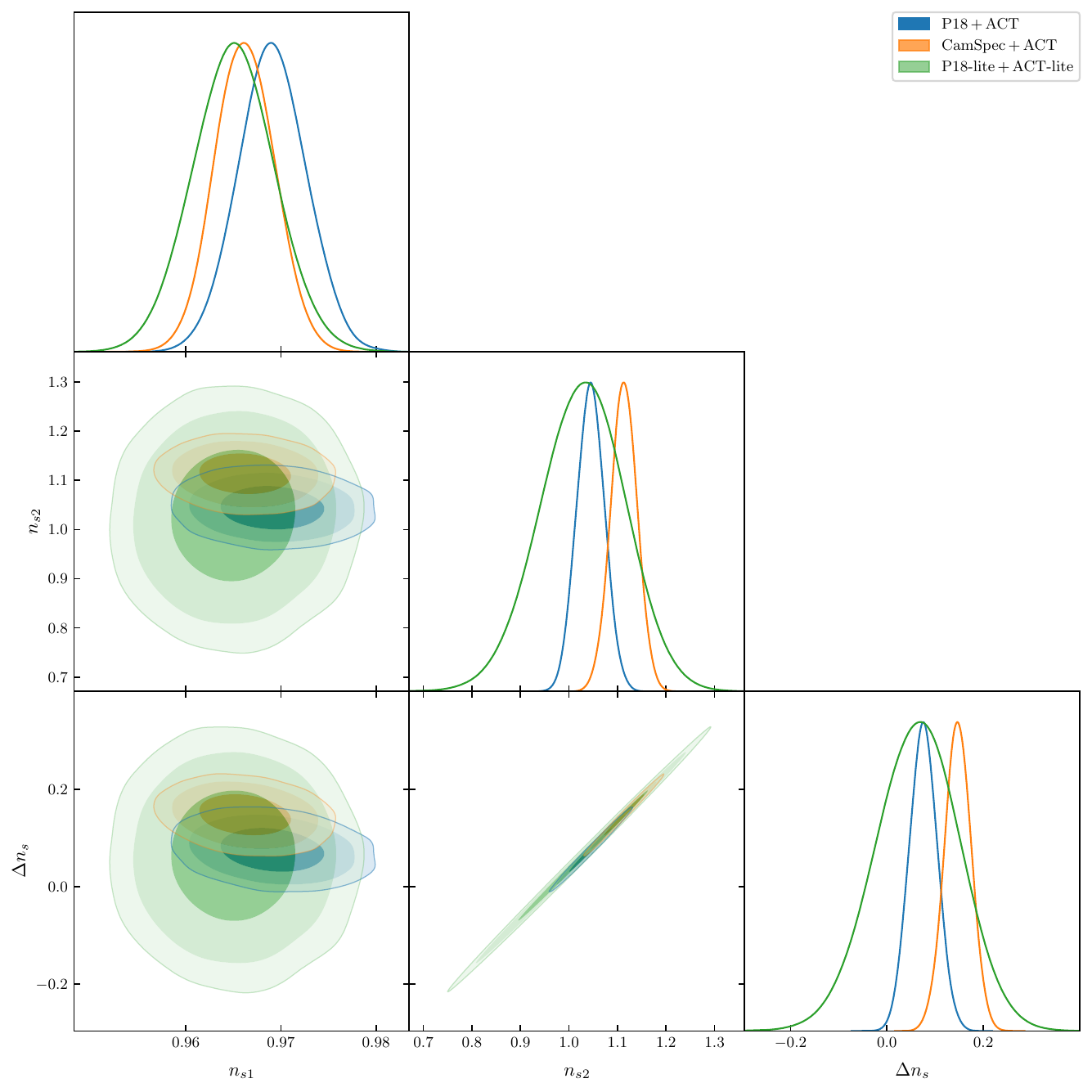}
        \caption{}
        \label{fig:all_mcmc_run_kbreak_fixed_0165_nui_fixed_bf}
    \end{subfigure}

    \caption{Marginalised posterior distributions of $n_{s1}$ versus $n_{s2}$, estimated from different likelihood combinations of Planck (PR3 and PR4 releases) (2-1996) and ACT (1997-3000) by setting $k_\mathrm{break}$ to $0.165 \, \mathrm{Mpc}^{-1}$. (a) Nuisance parameters of both Planck and ACT likelihoods are sampled. (b) Nuisance parameters of both Planck and ACT likelihoods kept fixed at their respective best-fit values based on $\Lambda CDM$ cosmology (except Planck-lite and ACT-lite likelihoods).}
    \label{fig:all_run_kbreak_fixed_0165}
\end{figure}

    \item \textbf{P18-lite ($\ell \leq 1996$)\,+\,ACT-lite ($1997 \leq \ell \leq 3000$):}
    
    In this analysis, we study the nuisance parameter marginalized likelihoods of Planck and ACT, viz., P18-lite ($\ell \leq 1996$) and ACT-lite ($1997 \leq \ell \leq 3000$). The obtained results for this combination of likelihoods are shown in Fig. \ref{fig:all_mcmc_run_kbreak_fixed_0165_nui_fixed_bf} and Table \ref{tab:double_tilt_constraints}. Here also we consider $k_\mathrm{break}=0.165 \, \mathrm{Mpc}^{-1}$. The posterior estimates of $n_{s2}$ and $\Delta n_s$ obtained from this combined likelihood are $1.028 \pm 0.088$ and $0.063 \pm 0.088$, respectively, indicating no statistically significant deviation from a single-tilt power-law primordial power spectrum.

\end{itemize}

\subsection{Planck PR4 and ACT-DR6 data combination}
In this section, we analyse the statistical significance of the feature identified in the reconstructed PPS for the fixed background cosmology inferred from the Planck PR4 data (Section~\ref{subsubsec:case3_act}). We then present and examine the results of the analyses performed using the combined Planck PR4 and ACT-DR6 datasets.

\begin{figure}[!htb]
    \centering

    \begin{subfigure}{0.49\textwidth}
        \centering
        \includegraphics[width=\linewidth]{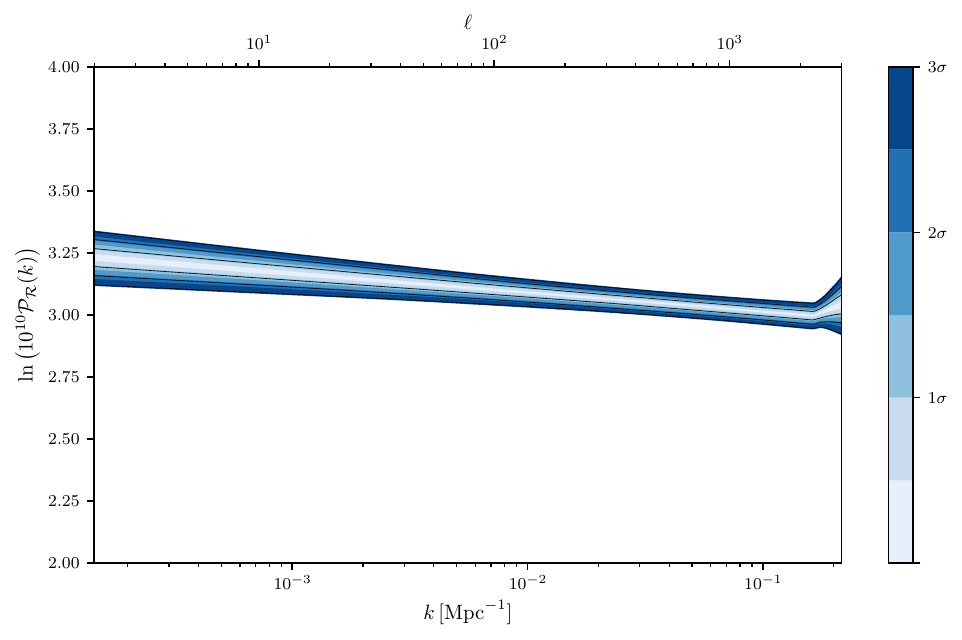}
        \caption{P18-TT and ACT-TT combined data}
        \label{fig:Planck_ACT_kbreak_fix_0.165_TT_fgivenx}
    \end{subfigure}
    \hfill
    \begin{subfigure}{0.49\textwidth}
        \centering
        \includegraphics[width=\linewidth]{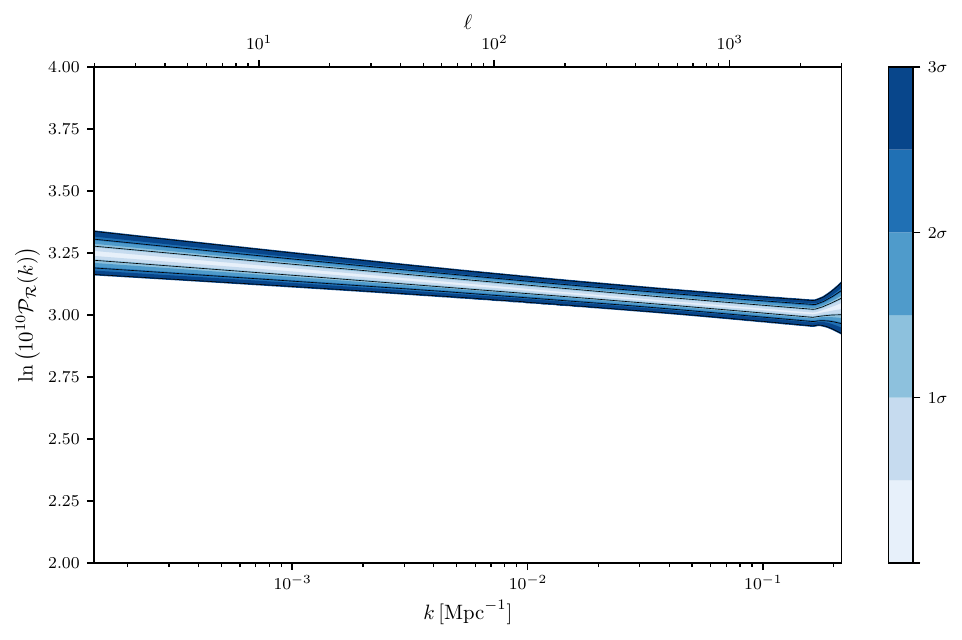}
        \caption{P18 and ACT combined data}
        \label{fig:Planck_ACT_kbreak_fix_0.165_fgivenx}
    \end{subfigure}

%    \vspace{0.5cm}

    \begin{subfigure}{0.49\textwidth}
        \centering
        \includegraphics[width=\linewidth]{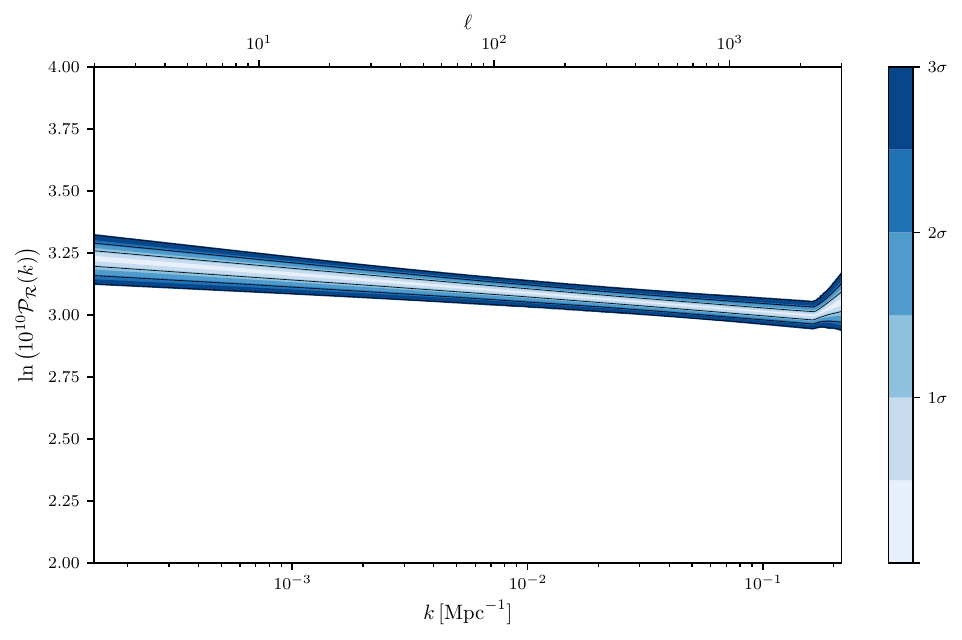}
        \caption{CamSpec-TT and ACT-TT combined data}
        \label{fig:NPIPE_ACT_kbreak_fix_0.165_TT_fgivenx}
    \end{subfigure}
    \hfill
    \begin{subfigure}{0.49\textwidth}
        \centering
        \includegraphics[width=\linewidth]{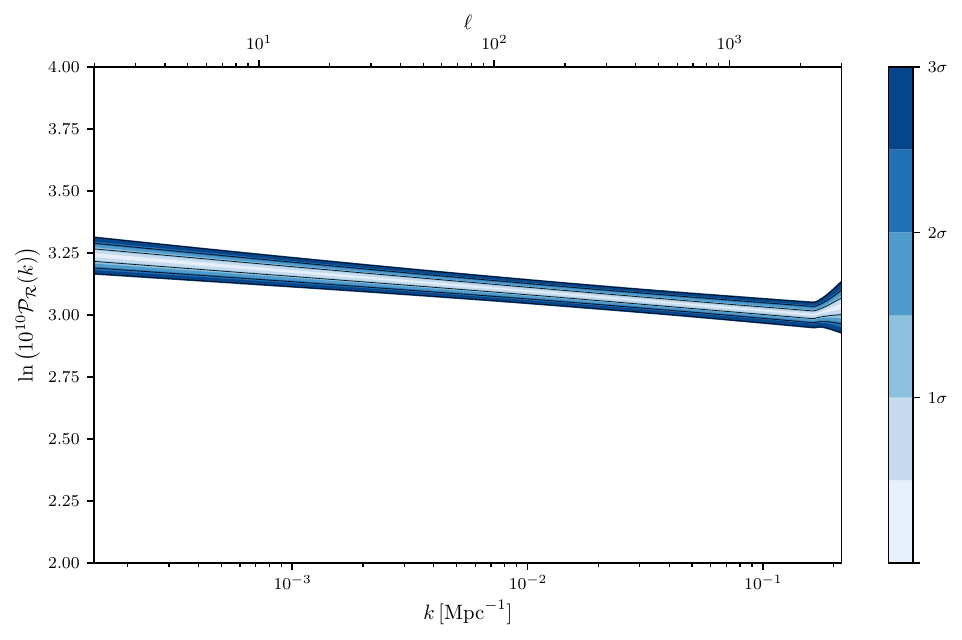}
        \caption{CamSpec and ACT combined data}
        \label{fig:NPIPE_ACT_kbreak_fix_0.165_fgivenx}
    \end{subfigure}

    \caption{Double-tilt primordial power spectrum with its confidence intervals obtained from different combinations of Planck (2-1996) and ACT-DR6 multifrequency (1997-3000) likelihoods. These plots are obtained for $k_\mathrm{break}=0.165 \, \mathrm{Mpc}^{-1}$ and by varying all the nuisance parameters related to both Planck and ACT likelihoods.}
    \label{fig:four_Planck_ACT_kbreak_fix_0.165_fgivenx}
\end{figure}

\begin{itemize}[label=\ding{109}]
    
    \item \textbf{CamSpec-TT ($\ell \leq 1996$)\,+\,ACT-TT ($1997 \leq \ell \leq 3000$):}
    
    We investigate here the combination of CamSpec-NPIPE and ACT multifrequency data. We consider only the TT data of both likelihoods in this sampling. The obtained results from this analysis are presented in Table \ref{tab:double_tilt_constraints} and Fig. \ref{fig:all_pc_run_kbreak_fixed_0165}. We vary all the nuisance parameters of both likelihoods in this analysis. Besides, we keep $k_\mathrm{break}$ fixed at $0.165 \, \mathrm{Mpc}^{-1}$. For this data combination, we infer $n_{s2} = 1.20 \pm 0.13$, which signifies a departure of $1.538 \sigma$ from '1'. The estimated mean of $n_{s2}$ shows a little more deviation from '1' compared to what we obtain from P18-TT+ACT-TT data, $n_{s2} = 1.17 \pm 0.12$ (Table \ref{tab:double_tilt_constraints}). The estimation of $\Delta n_s $ is $ 0.24 \pm 0.13$ for this dataset, and it shows a $1.846 \sigma$ departure from '0'. In contrast, for the TT-only case of the combined Planck PR3 and ACT-DR6 dataset, $\Delta n_s$ deviates from $0$ by $1.67 \sigma$. The posterior plot of the double-tilt primordial power spectrum for this dataset is shown in Fig.~\ref{fig:NPIPE_ACT_kbreak_fix_0.165_TT_fgivenx}.
    
    \item \textbf{CamSpec ($\ell \leq 1996$)\,+\,ACT ($1997 \leq \ell \leq 3000$):} 

    This analysis takes into consideration both high-$\ell$ temperature and polarization data. We investigate two individual cases for this dataset. In \texttt{Case I}, we keep the parameter $k_\mathrm{{break}}$ fixed at $0.165 \, \mathrm{Mpc}^{-1}$, but sample all the nuisance parameters associated with both the CamSpec and ACT likelihoods. In \texttt{Case II}, we keep both fixed, the parameter $k_\mathrm{{break}}$ at $0.165 \, \mathrm{Mpc}^{-1}$, and all the nuisance parameters at their respective best-fit values obtained for the $\Lambda CDM$ baseline cosmology. Subsequently, we present a detailed discussion on these two scenarios.
    
    \begin{itemize}[label={\tiny$\blacksquare$}]
    
        \item \textbf{Case I:}
        
        In this analysis the value of $k_\mathrm{break}$ is kept fixed at $0.165 \, \mathrm{Mpc}^{-1}$. However, all the nuisance parameters for both CamSpec and ACT likelihoods are sampled here. Our findings for this data combination are shown in Fig. \ref{fig:all_pc_run_kbreak_fixed_0165} and Table \ref{tab:double_tilt_constraints}. The obtained constraints on $n_{s2}$ and $\Delta n_s$ are $1.12 \pm 0.11$ and $0.15 \pm 0.11$, respectively. It shows that $n_{s2}$ and $\Delta n_s$ are consistent with $1$ and $0$ within $2\sigma$, respectively. We show the posterior plot of the double-tilt primordial power spectrum obtained for this analysis in Fig.~\ref{fig:NPIPE_ACT_kbreak_fix_0.165_fgivenx}.
        
        \item \textbf{Case II:}
        
        In this scenario, apart from keeping $k_\mathrm{break}$ fixed at $0.165 \, \mathrm{Mpc}^{-1}$, we also keep the values of all the nuisance parameters for both CamSpec and ACT likelihoods fixed at their respective best-fit values obtained for the $\Lambda CDM$ cosmology. We present our findings from this analysis in Fig. \ref{fig:all_mcmc_run_kbreak_fixed_0165_nui_fixed_bf} and Table \ref{tab:double_tilt_constraints}. From this analysis, we have $n_{s2}=1.113 \pm 0.027$ and $\Delta n_s = 0.147 \pm 0.028$, which signifies a $4.19 \sigma$ deviation from '1' and a $5.25 \sigma$ deviation from a single-tilt power-law power spectrum, respectively. For P18+ACT TTTEEE data with $k_\mathrm{break}$ and nuisance parameters fixed, we have $\Delta n_s = 0.075 \pm 0.029$  (Table \ref{tab:double_tilt_constraints}), and here for CamSpec+ACT TTTEEE data, we have $\Delta n_s = 0.147 \pm 0.028$, implying CamSpec data gives a higher statistical preference to a double-tilt power spectrum compared to Planck PR3 data.
        Since, for P18+ACT TTTEEE data, we have $n_{s2} = 1.045 \pm 0.029$ (Table \ref{tab:double_tilt_constraints}), which is $1.55\sigma$ away from $1$, whereas for CamSpec+ACT TTTEEE data, we find $4.19 \sigma$ deviation from $1$, thus it shows a blue tilted behaviour of the PPS at $k=0.165 \, \mathrm{Mpc}^{-1}$ is more preferred by CamSpec data. %Now the difference between two spectral indices indicating the inconsistency between two datasets, shows that for P18+ACT TTTEEE $\Delta n_s = 0.075 \pm 0.029$ and on the other hand for CamSpec+ACT TTTEEE $\Delta n_s = 0.147 \pm 0.028$, indicating that the mean difference is greater in case of CamSpec vs ACT ($0.147$) scenario than P18 vs ACT ($0.075$), however the uncertainties are similar ($0.029$ and $0.028$) in both datasets. Consequently, CamSpec+ACT TTTEEE is showing a $5.25\sigma$ deviation from a single-tilt primordial power spectrum and P18+ACT TTTEEE is exhibiting a $2.59\sigma$ deviation from a single-tilt primordial power spectrum.
        
    \end{itemize}

\end{itemize}

\begin{table*}[h]
\centering
\renewcommand{\arraystretch}{1.2}
\setlength{\tabcolsep}{5pt}
\caption{Estimated values of the spectral indices $n_{s1}$ and $n_{s2}$ obtained from various combinations of Planck (PR3 and PR4 releases) and ACT-DR6 multifrequency datasets at $k_\mathrm{break}=0.165 \, \mathrm{Mpc}^{-1}$. The Planck data are used at multipole $\ell \leq 1996$, and the ACT data are used for the multipole range $1997 \leq \ell \leq 3000$. The numbers reported within parentheses represent the statistical significance of the deviation of $n_{s2}$ and $\Delta n_s$ from $1$ and $0$, respectively.}
\label{tab:double_tilt_constraints}

\begin{adjustbox}{max width=\textwidth}
\begin{tabular}{|l|c|c|c|c|}
\hline
\textbf{Dataset} &
\textbf{Nuisance} &
$\mathbf{n_{s1}}$ &
$\mathbf{n_{s2}}$ &
$\mathbf{\Delta n_s}$ \\
\hline\hline

P18 + ACT &
Fixed &
$0.9691\pm0.0036$ &
$\begin{array}{c}
1.045\pm0.029\\
(1.55\sigma)
\end{array}$ &
$\begin{array}{c}
0.075\pm0.029\\
(2.59\sigma)
\end{array}$ \\
\hline

P18 + ACT &
Sampled &
$0.9657\pm0.0043$ &
$\begin{array}{c}
1.10^{+0.12}_{-0.10}\\
(0.92\sigma)
\end{array}$ &
$\begin{array}{c}
0.132^{+0.12}_{-0.10}\\
(1.21\sigma)
\end{array}$ \\
\hline

P18-TT + ACT-TT &
Sampled &
$0.9666\pm0.0054$ &
$\begin{array}{c}
1.17\pm0.12\\
(1.42\sigma)
\end{array}$ &
$\begin{array}{c}
0.20\pm0.12\\
(1.67\sigma)
\end{array}$ \\
\hline

P18-lite + ACT-lite &
Marginalized &
$0.9650\pm0.0043$ &
$\begin{array}{c}
1.028\pm0.088\\
(0.32\sigma)
\end{array}$ &
$\begin{array}{c}
0.063\pm0.088\\
(0.72\sigma)
\end{array}$ \\
\hline

CamSpec + ACT &
Fixed &
$0.9662\pm0.0032$ &
$\begin{array}{c}
1.113\pm0.027\\
(4.19\sigma)
\end{array}$ &
$\begin{array}{c}
0.147\pm0.028\\
(5.25\sigma)
\end{array}$ \\
\hline

CamSpec + ACT &
Sampled &
$0.9660\pm0.0037$ &
$\begin{array}{c}
1.12\pm0.11\\
(1.09\sigma)
\end{array}$ &
$\begin{array}{c}
0.15\pm0.11\\
(1.36\sigma)
\end{array}$ \\
\hline

CamSpec-TT + ACT-TT &
Sampled &
$0.9675\pm0.0050$ &
$\begin{array}{c}
1.20\pm0.13\\
(1.54\sigma)
\end{array}$ &
$\begin{array}{c}
0.24\pm0.13\\
(1.85\sigma)
\end{array}$ \\
\hline

\end{tabular}
\end{adjustbox}
\end{table*}

\subsection{ACT-DR6 data}

We conduct this analysis to examine another feature, a localized bump-like deformation identified at $ k \approx 0.33 -0.34  \, \mathrm{Mpc}^{-1}$, in the PPS when reconstructing from ACT-DR6 data using the MRL method (Section~\ref{subsubsec:case1_act}). %At $ k \approx 0.33  \, \mathrm{Mpc}^{-1}$, we find a bump-like deformation in the PPS for the ACT-DR6 data (see Fig. \ref{fig:COADACT_ACTBF-2}). 
We try to model this structure with a double-tilt power spectrum $\mathcal{P}^{\mathrm{double-tilt}}_{\mathcal{R}}\, (k)$ (\ref{eq:double_tilt_pk}). The prime objective of this analysis is to assess the statistical significance of the observed deformation and determine whether the ACT data favour a double-tilt power spectrum or remain consistent with the standard single-tilt power spectrum. To this end, we investigate the feature using the ACT-DR6 data alone. In this analysis, we take into consideration the TT-only, and the TTTEEE, both datasets individually. Below, we examine the results obtained from these analyses. %Since, we intended to show only that from statistical inference whether the found feature in the reconstructed power spectrum shows any statistical inconsistency in the value of $n_{s}$ from Planck reported value \cite{Planck:2018vyg} and favours blue tilt region or not.

\begin{itemize}[label=\ding{109}]

    \item \textbf{ACT-TT ($\ell \leq 4810$):}
    
    In this analysis, we consider only temperature data of ACT-DR6 for the multipole range $\ell=600-4810$. ACT does not measure below multipole $\ell=600$, consequently to constrain the optical depth ($\tau_\mathrm{reio}$), we use the $\texttt{sroll2 prior}$ for $\tau_\mathrm{reio}$. The priors of the cosmological parameters adopted in this analysis are listed in Table \ref{tab:model_priors}. We use the prior $\mathcal{U}  (0.25, 0.32)$ for $k_\mathrm{{break}}$. Here nuisance parameters of ACT likelihood are sampled. The results of this analysis are presented in Fig. \ref{fig:act_tt_n_ttteee_kbreak_full}, where we find that $n_{s2}$ is unconstrained ($n_{s2} < -1.40$). The estimated posterior of $n_{s2}$ indicates that the primordial power spectrum is consistent with a single-tilt power-law power spectrum. The posterior of the primordial power spectrum obtained from this analysis is shown in Fig.~\ref{fig:MRL_double_tilt_ACT_TT_PC_fgivenx}.
    
    %However, we found that in this scale $n_{s2}$ shows degeneracy with the nuisance parameter $a_{kSZ}$, hence by fixing $k_{break}$ at $k=0.25 \, \mathrm{Mpc}^{-1}$.  
    
    %In Fig. \ref{fig:COADACT-2}, we found that the reconstructed $\mathcal{P}_k$ is showing a bump near $k=0.33 \, \mathrm{Mpc}^{-1}$ when we considered ACT data up to $\ell= 4810$. In Fig. \ref{fig:COADACT-2}, we simulate 1000 mock $C_\ell$s from the power-law power spectrum ($\ref{eq:dis_base_pk}$) for ACT best-fit cosmology using same reconstruction algorithm to get confidence interval across scales $k$, where we found that around $k=0.33 \, \mathrm{Mpc}^{-1}$ the reconstructed $\mathcal{P}_k$ shows deviation from power-law behaviour falling outside $95 \%$ confidence interval.
\begin{figure}[!htb]
    \centering
    \begin{subfigure}{0.7\textwidth}
        \centering    
        \includegraphics[width=\textwidth, keepaspectratio]{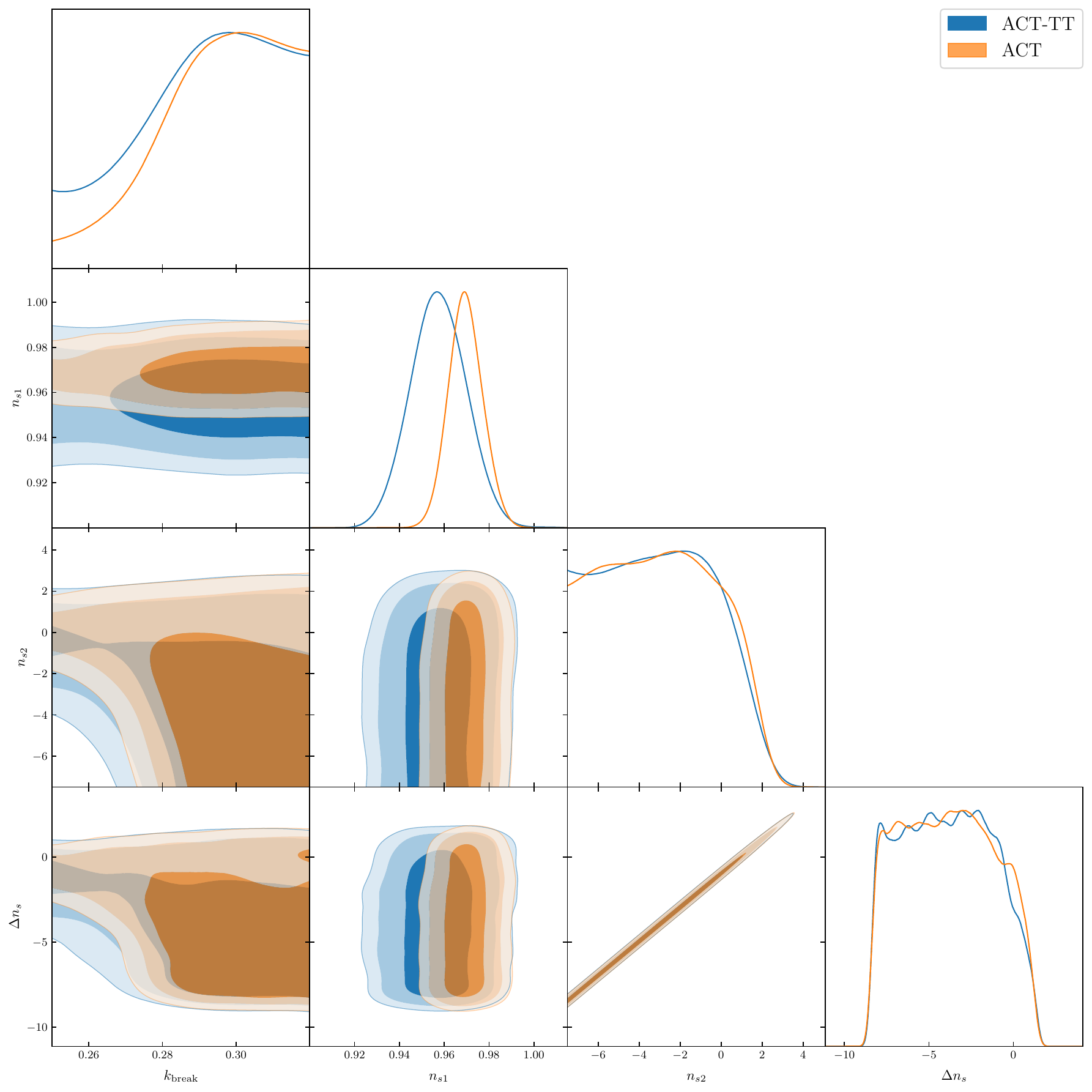}
        \caption{}
        \label{fig:act_tt_n_ttteee_kbreak_full}
    \end{subfigure}

%    \vspace{0.3cm}

    \begin{subfigure}{0.49\textwidth}
        \centering
        \includegraphics[width=\textwidth, keepaspectratio]{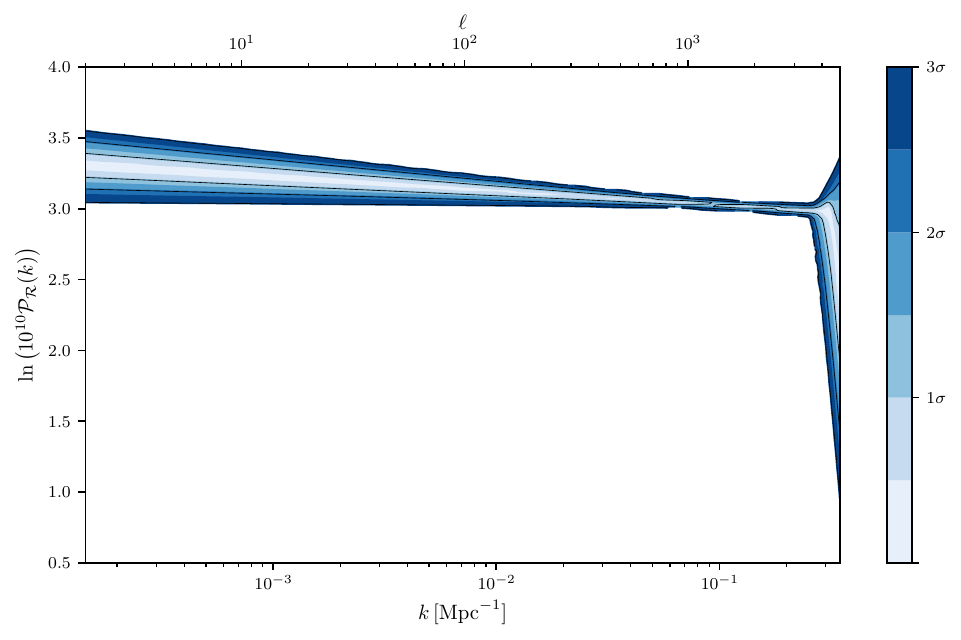}
        \caption{}
        \label{fig:MRL_double_tilt_ACT_TT_PC_fgivenx}
    \end{subfigure}
    \hfill
    \begin{subfigure}{0.49\textwidth}
        \centering
        \includegraphics[width=\textwidth, keepaspectratio]{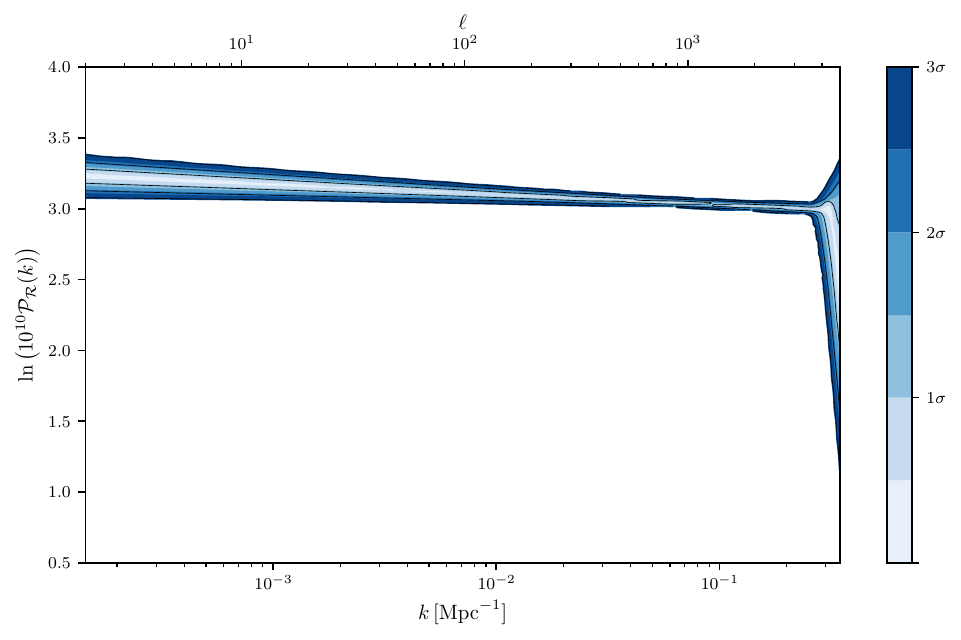}
        \caption{}
        \label{fig:MRL_double_tilt_ACT_PC_fgivenx}
    \end{subfigure}

    \caption{Marginalised posterior distributions of $n_{s2}$ versus $k_\mathrm{break}$ together with double-tilt primordial power spectrum and its confidence intervals estimated from the ACT-TT ($<4810$) and ACT-TTTEEE ($<4810$) datasets. The ACT likelihood nuisance parameters are sampled for both datasets. (a) Marginalised posterior distributions of $n_{s2}$ versus $k_\mathrm{break}$. (b) Double-tilt primordial power spectrum for TT data. (c) Double-tilt primordial power spectrum for TTTEEE data.}
    \label{fig:double_tilt-3}
\end{figure}

\begin{figure}[!htb]
\centering
\includegraphics[width=0.7\textwidth, keepaspectratio]{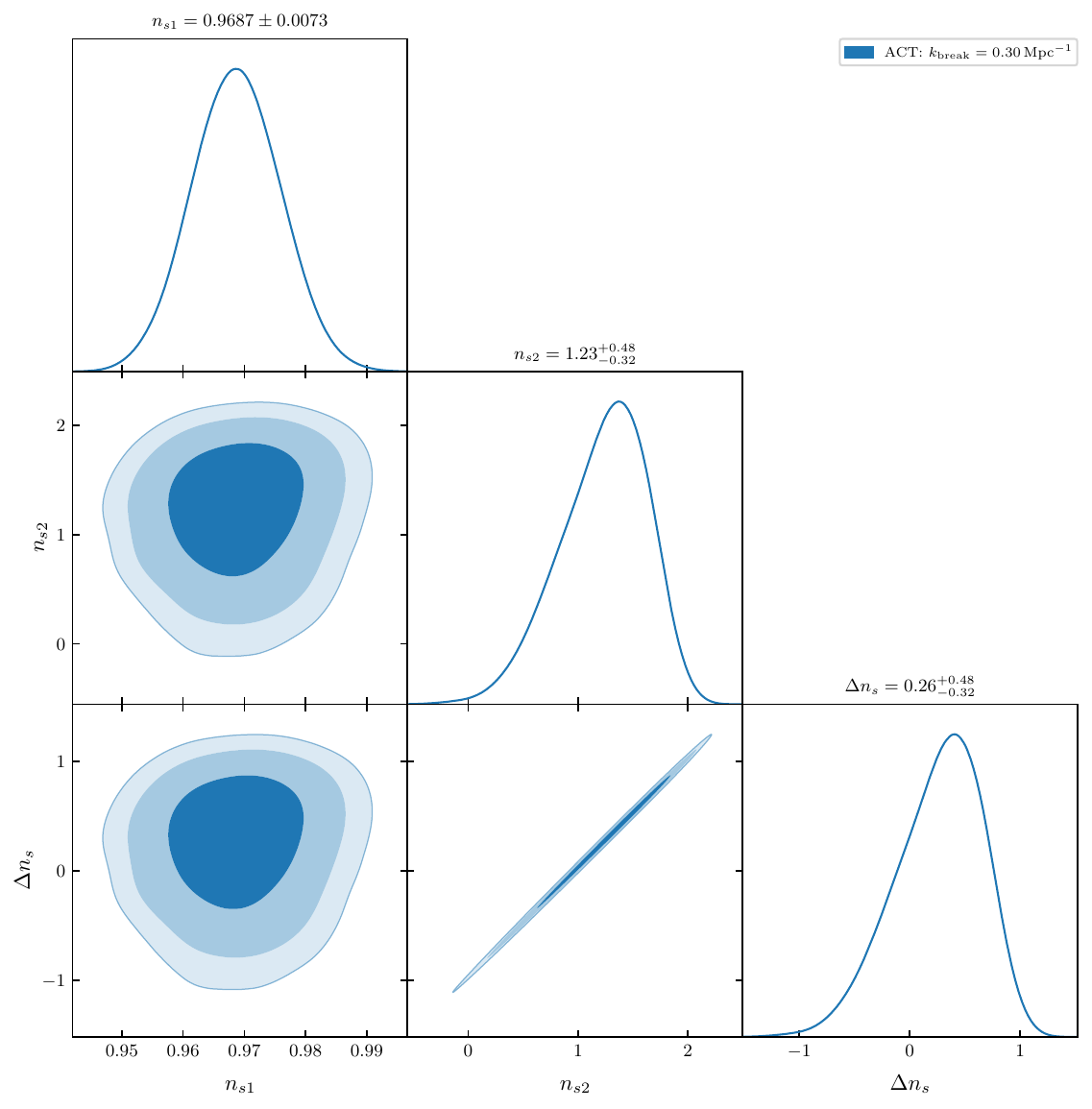}
\caption[]{Marginalised posterior distributions of $n_{s1}$ versus $n_{s2}$ estimated from ACT ($\ell< 4810$) data. This plot is obtained by setting $k_\mathrm{break}$ to $0.30 \, \mathrm{Mpc}^{-1}$ and by keeping the nuisance parameters related to ACT likelihood fixed at their respective best-fit values based on $\Lambda CDM$ cosmology.}
\label{fig:act_ttteee_kbreak_fixed_030_nui_fixed_bf}
\end{figure}

    \item \textbf{ACT ($\ell \leq 4810$):}
    \begin{itemize}[label={\tiny$\blacksquare$}]
        \item In this analysis, we consider both temperature and polarization data of ACT for the multipole range $[600,4810]$. We use the $\texttt{sroll2 prior}$ for the optical depth. We provide the priors adopted in this analysis for the cosmological parameters in Table \ref{tab:model_priors}. For the parameter $k_\mathrm{{break}}$, we consider the prior range $\mathcal{U} (0.25, 0.32)$. All the nuisance parameters of the ACT likelihood are varied here. We show the obtained results for this data in Figs. \ref{fig:act_tt_n_ttteee_kbreak_full} and \ref{fig:MRL_double_tilt_ACT_PC_fgivenx}. The constraint on $n_{s2}$ is $-2.9^{+2.3}_{-3.7}$, implying consistency with a single-tilt power-law power spectrum. 
        \item We also explore another scenario for this dataset. We fix all the nuisance parameters of the ACT TTTEEE likelihood at their respective best-fit values obtained for the baseline $\Lambda CDM$ model. Additionally, we set $k_{\mathrm{break}}$ to $0.30 \, \mathrm{Mpc}^{-1}$. In Fig. \ref{fig:act_ttteee_kbreak_fixed_030_nui_fixed_bf}, we present the results obtained from this analysis. The obtained constraints on $n_{s2}= 1.23^{+0.48}_{-0.32}$ and $\Delta n_{s} = 0.26 ^{+0.48}_{-0.32}$. Though the estimated mean of $n_{s2}$ is $>1$, it remains consistent with a single-tilt power-law power spectrum which is in complete agreement with what we found from the simulation-based error analysis in Section \ref{subsubsec:case1_act}. 
    \end{itemize}
     %However, in both the analyses, we find that $n_{s2}$ is showing degeneracy with the nuisance parameter $a_{\mathrm{kSZ}}$, hence for TTTEEE data we have performed another analysis by keeping $k_{\mathrm{break}}$ fixed at $0.25 \, \mathrm{Mpc}^{-1}$ and truncating the prior of $a_{\mathrm{kSZ}}$ from its default range $[0,10]$ to $[0,1]$, which allows us to constrain $n_{s2}$. In Fig. \ref{fig:double_tilt_pk-15}, we have presented our result for fixed $k_{\mathrm{break}}$  and reduced prior of $a_{\mathrm{kSZ}}$. From Fig. \ref{fig:double_tilt_pk-15}, we can find that the posterior of $n_{s2}$ is distributed on both sides of $1$: $n_{s2} = 0.57^{+ 0.69}_{-0.56}$, preferring both blue and red tilt.

\end{itemize}

\section{Conclusion and Outlook}
We have reconstructed the non-parametric primordial power spectra from the CamSpec, ACT-DR6, and SPT-3G D1 CMB datasets employing the MRL algorithm. We have also reconstructed free-form power spectra for the combined datasets of CamSpec and ACT/SPT. In all these reconstructions we have implemented three different regularization algorithms to get more physically viable reconstructions. %For both CamSpec-ACT and CamSpec-SPT data combinations, we have used a modified MRL algorithm to incorporate both unbinned and binned datasets.
In this work, both temperature and polarization spectra have been considered in reconstructing primordial power spectra for each observation, namely, CamSpec, ACT, and SPT and also for the combined datasets, making it the most complete analysis so far on non-parametric reconstruction of the primordial power spectrum from CMB data. Consideration of all TT, TE, and EE data allows us to reconstruct the power spectrum that fits all three angular power spectra together. We performed statistical test on the reconstructed power spectra to assess the significance of any observed features and to measure the goodness of fit to the data achieved by the reconstructed spectra. We conducted the $p$-value test by generating 1000 samples of $\mathcal{C}_{\ell}$ constructed from feature-free power-law power spectrum. 

In this work, apart from reconstructing model independent PPS from different CMB datasets individually and in combination, we have also investigated the consistency within and across different CMB missions. We have explored the consistency of ACT and SPT data with Planck data for its two data releases viz., PR3 and PR4.

We have also modelled any plausible disagreements with the single-tilt power-law power spectrum found in the reconstructed spectra with a parametric double-tilt power-law power spectrum and conducted the Bayesian analysis to estimate the statistical significance of the observed disagreements. 

The major contributions and key findings of this work are summarized below:

\begin{itemize}
    \item We reconstruct the primordial power spectrum in a model-independent manner using the CamSpec, ACT, and SPT datasets over the multipole ranges $2-2500$, $593-4810$, and $400<\ell<3000$, respectively. We also perform reconstructions using the combined CamSpec+ACT and CamSpec+SPT datasets, covering the multipole ranges $2-4810$ and $2-3000$, respectively. 
    \item We find no statistically significant features in the primordial power spectrum reconstructed from the CamSpec data. The reconstruction from ACT, however, exhibits hints of several local features that are neither present in the CamSpec nor in SPT reconstruction. Similarly, although a local feature is identified in the SPT reconstruction, the corresponding scale exhibits features that are out of phase in the ACT and Planck reconstructions. Consequently, we find no consistent evidence for the same feature across all three datasets and at the same time statistically significant in the error analysis tests. %all the plausible features found for ACT and SPT data, respectively, do not survive the look-else-where test. Therefore, our reconstructed power spectra from three datasets are consistent with a power-law power spectrum and rejects the presence of any viable features. This concludes that we do not find any evidence to reject the null hypothesis of a featureless power-law power spectrum.
    We therefore conclude that all three datasets are consistent with a featureless power-law primordial power spectrum and provide no statistically significant evidence to reject the null hypothesis.
    \item A comparison of the reconstructed primordial power spectra shows that the CamSpec and ACT reconstructions exhibit good overall agreement, differing only at a few localized scales and in their overall amplitude. %We find from the comparison analysis of reconstructed power spectra that the power spectra reconstructed from CamSpec and ACT data closely follow each other except for a few instances.
    \item We find no significant evidence of any tension between the Planck and ACT datasets. However, a marginal disagreement is observed. Quantified in terms of the inferred change in the spectral index, the discrepancy corresponds to a $1-5 \sigma$ deviation from a single-tilt power-law primordial power spectrum, depending on the choice of likelihood and on whether the nuisance parameters are varied or held fixed. The statistical significance of the disagreement increases when the nuisance parameters are fixed rather than marginalized over. The disagreement is more pronounced for the Planck PR4 CamSpec-NPIPE likelihood than for the Planck PR3 likelihood. % We find that Planck data shows inconsistency with ACT data. This disagreement is more for Planck PR4 CamSpec-NPIPE data compare to Planck PR3 data. The statistical significance of this inconsistency becomes stronger when we fix the nuisance parameters. The change in the spectral index quantifying the inconsistency, shows $1-5 \sigma$ deviation from a single-tilt power-law power spectrum depending upon the choice of likelihoods and whether we vary the nuisance parameters or keeping them fixed.
    \item Our Bayesian analysis shows that, although the constraints remain weak, the posterior distribution of $n_{s2}$ consistently favours values greater than unity across all Planck and ACT data combinations considered in this work.
\end{itemize}
%From ACT data we have found some features that were in favour of our alternative hypothesis, which we explored rigorously with parametric-form analysis. However, our statistical test supports the null hypothesis for both ACT and SPT data and also for CamSpec-ACT and CamSpec-SPT data combinations. 
%With the parametric-form analysis, we have investigated the inconsistency between the CamSpec and ACT-DR6 observations found from non-parametric-form analysis. To model the discrepancy in datasets we have considered a primordial power spectrum having two tilts operating at two different scales about a junction. We have performed the Bayesian analysis to probe the significance of the discrepancy with ACT-only data and with combined data of Planck and ACT. In this analysis, to investigate the origin of this tension we have sample all the nuisance parameters.  

%We find from the Bayesian analysis that even though the constrains are weak, still $n_{s2}$ prefers a value $>1$ persistently in all the data combinations of Planck and ACT taken into consideration in this work.

\section*{Acknowledgments}
The authors thank Wuhyun Sohn for important discussions and valuable comments. D.K.H. acknowledges financial support from the Indo-French Centre for the Promotion of Advanced Research (IFCPAR/CEFIPRA), New Delhi, India, through the Collaborative Scientific Research Programme (Project No. 6704-4, “Testing flavors of the early universe beyond vanilla models with cosmological observations”); and from the Anusandhan National Research Foundation (ANRF), Government of India, under the ARG MATRICS scheme (Grant No. ANRF/ARGM/2025/000941/TS; project “EPOCH: Exploring Primordial Origins in Cosmic Hierarchies”). A.S. acknowledges funding from the Korea Astronomy and Space Science Institute (KASI) through the project “Research on the Principles of the Accelerating Expansion of the Universe” (Project Code: 2026183201). A.S. is grateful for support by the Open KIAS Center at Korea Institute for Advanced Study. TS acknowledges support from the J. C. Bose Grant of ANRF, India.
%%%%%%%%%%%%%%%%%%%%%%%%%%%%%%%%%%%%%%%%%%%%%%%%%%%%%%%%%%%%%%%%%%%%%%%%%%%%%%%%%%%%%%%%%%%%%%%%%%%%%%%%%%%%%%%%%%%%%%%%%%%%%%%%%%%%%%%%%%%%%%%%%%%%%%%%%%

\bibliographystyle{unsrt}
\bibliography{references}
%%%%%%%%%%%%%%%%%%%%%%%%%%%%%%%%%%%%%%%%%%%%%%%%%%%%%

\end{document}